\PassOptionsToPackage{table}{xcolor}
\documentclass[
nofootinbib,
a4paper,
aps,
prx,
superscriptaddress,
reprint,
twocolumn,
preprintnumbers,
nobalancelastpage,
floatfix,
longbibliography,
10pt]{revtex4-2}
\usepackage{array}[=2016-10-06]
\usepackage{graphicx}
\usepackage{quantikz}
\usepackage{float}
\usepackage{makecell}
\usepackage[colorlinks=true, linkcolor=blue, citecolor=blue, urlcolor=blue]{hyperref}
\usepackage{multirow}
\usepackage{xcolor}
\usepackage{graphicx}

\newcommand{\symbolset}[1]{\mathbb{#1}}
\newcommand{\noisy}[1]{\tilde{#1}}
\newcommand{\doubleket}[1]{|#1\rangle\!\rangle}
\newcommand{\doublebra}[1]{\langle\!\langle #1 |}

\usepackage[utf8]{inputenc}

\usepackage{amsmath, amsthm, amsfonts, amssymb,amstext,cleveref}
\usepackage{mathtools}
\usepackage{bm}

\usepackage{float}
\usepackage{gensymb}

\usepackage{braket}
\usepackage{siunitx}
\DeclareUnicodeCharacter{2212}{\textminus}
\begin{document}

\title{Easier randomizing gates provide more accurate fidelity estimation}

\author{Debankan Sannamoth}
    \altaffiliation{These authors contributed equally}
 
\affiliation{Department of Applied Mathematics, University of Waterloo, Waterloo, Ontario N2L 3G1, Canada}
\affiliation{Institute for Quantum Computing, University of Waterloo, Waterloo, Ontario N2L 3G1, Canada}

\author{Kristine Boone}
    \altaffiliation{These authors contributed equally}
    \affiliation{Photonic Inc., Coquitlam, British Columbia, Canada}
\author{Arnaud Carignan-Dugas}
    \affiliation{Keysight Technologies Canada, Kanata, Ontario K2K 2W5, Canada}
\author{Akel Hashim}
    \affiliation{Department of Physics, University of California at Berkeley, Berkeley, CA 94720, USA}
    \affiliation{Applied Math and Computational Research Division, Lawrence Berkeley National Lab, Berkeley, CA 94720, USA}
\author{Irfan Siddiqi}
    \affiliation{Department of Physics, University of California at Berkeley, Berkeley, CA 94720, USA}
    \affiliation{Applied Math and Computational Research Division, Lawrence Berkeley National Lab, Berkeley, CA 94720, USA}
\author{Karl Mayer}
    \affiliation{Quantinuum, Broomfield, CO 80021, USA}

\author{Joseph Emerson}
   \thanks{Author contacts: \href{dsannamo@uwaterloo.ca}{dsannamo@uwaterloo.ca}, \href{jemerson@uwaterloo.ca}{jemerson@uwaterloo.ca}}
    \affiliation{Department of Applied Mathematics, University of Waterloo, Waterloo, Ontario N2L 3G1, Canada}
    \affiliation{Institute for Quantum Computing, University of Waterloo, Waterloo, Ontario N2L 3G1, Canada}
    \affiliation{Keysight Technologies Canada Inc., Mississauga, Ontario L5N 2M2, Canada}

\date{\today}

\begin{abstract}
Accurate benchmarking of quantum gates is crucial for understanding and enhancing the performance of quantum hardware. A standard method for this is interleaved benchmarking, a technique which estimates the error on an interleaved target gate by comparing cumulative error rates of randomized sequences implemented with the interleaved gate and without it. In this work, we show both numerically and experimentally that the standard approach of interleaved randomized benchmarking (IRB), which uses the multi-qubit Clifford group for randomization, can produce highly inaccurate and even physically impossible estimates for the error on the interleaved gate in the presence of coherent errors. Fortunately we also show that interleaved benchmarking performed with cycle benchmarking, which randomizes with single qubit Pauli gates, provides dramatically reduced systematic uncertainty relative to standard IRB, and further provides as host of additional benefits including data reusability. We support our conclusions with a theoretical framework for bounding systematic errors,  extensive numerical results comparing a range of interleaved protocols under fixed resource costs, and experimental demonstrations on three quantum computing platforms. 
\end{abstract}

\maketitle
\raggedbottom  

\section{INTRODUCTION}

Randomized benchmarking \cite{Emerson_2005, PhysRevA.80.012304,PhysRevLett.106.180504,PhysRevA.85.042311,PhysRevA.75.022314} is a fundamental method in quantum computing that now comprises a broad class of methods to estimate the error rates of quantum gates \cite{PRXQuantum.3.020357,PRXQuantum.6.030202}. 
While these methods are often applied to estimate error rates averaged over some set or group of elementary gates, in this work we focus on interleaved benchmarking protocols which aim to estimate the error on some specific gate of interest, typically a two-qubit entangling gate such as the CNOT. The primary example of such a protocol is interleaved randomized benchmarking (IRB) \cite{PhysRevLett.109.080505}, which is currently a standard benchmarking approach with widespread adoption across all leading experimental platforms \cite{PhysRevA.87.030301,Barends_2014,PhysRevLett.112.240504,PhysRevLett.122.200502,Muhonen_2015,Kawakami_2016,PhysRevLett.108.260503,Manetsch_2025,Weinstein_2023,PhysRevX.13.041052,Li_2023,wu2507simultaneous,q418-pydy,vbh4-lysv,norris2025performance,h7cv-xgw2,PhysRevX.11.021058} . The standard IRB scheme proposes implementing a reference randomized benchmarking (RB) sequence, that is, a standard RB experiment with the full Clifford group, as well as an interleaved sequence with the gate of interest interleaved between the randomizing gates. The ratio of the latter over the former is intended to provide an accurate estimate of the error on the interleaved gate alone.

In this work, we demonstrate with theoretical, numerical, and experimental evidence that the systematic uncertainty for IRB can lead to significant inaccuracy in practice, producing not just inaccurate but even physically impossible error-rate estimates for the interleaved gate in the presence of even moderate amounts of coherent error.  Fortunately, we also show that this problem can be effectively mitigated by selecting an easier-to-implement randomization group based on single-qubit gates with lower average error, notably the Pauli randomization used in cycle benchmarking (CB) \cite{Erhard_2019} and the local Clifford randomization \cite{Emerson_2007} used in variety of other benchmarking protocols ~\cite{Emerson_2007,mckay2311benchmarking}. Benchmarking protocols with these simpler randomization groups have now been well-studied theoretically \cite{Emerson_2007,mckay2311benchmarking,carignan2023error,beale_2020_3945250,CarignanDugas2024estimatingcoherent,Chen_2023,calzona2024multi,PRXQuantum.6.010310,Flammia_2020} and experimentally across a wide-variety of platforms \cite{Emerson_2007,mckay2311benchmarking,PhysRevX.11.041039,fazio2025characterizing,PhysRevResearch.6.L022067,haghshenas2025digital}. 
In spite of the above works exploring single qubit randomization with low average error rates, there is still widespread adoption of IRB 
which requires multi-qubit randomization with associated high average error rates. We show below that the dramatic inaccuracy of IRB is entirely avoidable, where the single-qubit randomization of cycle benchmarking provides a far more accurate alternative for interleaved benchmarking. CB also offers additional practical advantages at scale, in particular reusability of the measured Pauli-error-eigenvalues and robustness to crosstalk when benchmarking full simultaneous cycles, both of which are discussed in detail in Section ~\ref{Conclusion}.

Our analysis of the systematic error in interleaved benchmarking protocols builds on the systematic bounds developed in \cite{Carignan_Dugas_2019} which characterize how component level errors in a composition of gates can interfere coherently to give rise to significant variation in the cumulative error across a product of gates. In particular, we evaluate the significance of this effect in different interleaved benchmarking protocols, where we demonstrate that extremely large systematic uncertainties can arise in realistic contexts. Our results counter the (incorrect) intuition that randomized sequences typically suppress coherent errors, and so destructive and constructive interference should not be a concern under randomized benchmarking. This intuition fails because for interleaved benchmarking protocol the suppression of coherence only occurs between the composed errors from the interleaved gate and the randomizing gates; coherence is not suppressed between each of these gates separately. 
Our work also counters the intuition that coherent  (cancellation or amplification) effects for gate errors are actually rare and thus unlikely to be significant in practice. We show below that this is decidedly not the case: under typical experimental conditions, we provide both numerical and experimental evidence indicating that coherence effects can be significant enough to produce physically impossible estimates, in particular, negative error probabilities. 

The manuscript is structured as follows. First, in Section \ref{Background}, we provide conceptual and theoretical background and establish terminology for the set of interleaved benchmarking  protocols that we evaluate. In Section \ref{Systematic bounds}, we discuss the details about systematic bounds. In Section \ref{XRB bounds}, we discuss how performing extended randomized benchmarking (XRB) to extract knowledge of the unitarity of the error along with estimates from protocols like Haar-RB and IRB helps in providing a tighter bound as compared to the systematic bounds. In Section \ref{Results}, we show numerical and experimental results across three different quantum computing platforms comparing different interleaved protocols in support of our conclusions. In Appendix \ref{Optimal gauge}, we provide a framework for constructing the self-consistent gauge under which a theoretical value can be rigorously compared to the protocol estimator, to address the gauge ambiguity of RB protocols. Finally, we conclude this work in Section \ref{Conclusion} with discussions and outlook.

\begin{figure*}[t]
  \centering
  \includegraphics[width=0.8\textwidth]{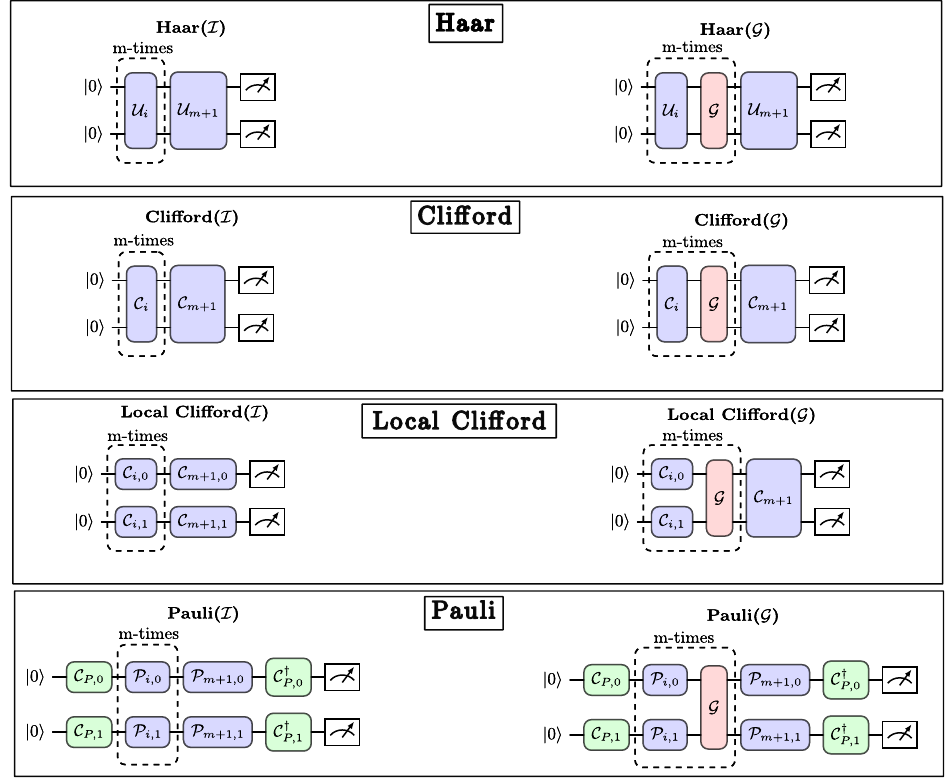}

  \caption{\footnotesize
     Circuits for interleaved protocol pairs for each of four different protocols with distinct randomizing/twirling groups which we label: Haar, Clifford, Local Clifford and Pauli; see also their descriptions in Table \ref{tab:TwirlingGroups}. In each pair, the interleaved and reference protocols are denoted by the twirling group name followed by ($\mathcal{G}$) and  by ($\mathcal{I}$), respectively,  which specifies where the gate $\mathcal{G}$ or no gate (viz $\mathcal{I}$) is interleaved in between the twirling gates. Please note that the notation $(\mathcal{I})$ does not mean that there is an identity gate or delay inserted in lieu of the interleaved gate, but denotes that the reference protocol is identical to the interleaved protocol except that it has no interleaved gate.  For \textit{Haar}, the twirling gates $\mathcal{U}_i$ are drawn uniformly at random from $\mathbb{SU}(4)$ according to the Haar measure and $\mathcal{U}_{m+1}$ denotes the correction gate $\mathbb{SU}(4)$ required in motion reversal circuits per the original RB protocol \cite{Emerson_2005}. For \textit{Clifford}, the twirling gates $\mathcal{C}_i$ are drawn at random from $\mathbb{C}_2$ where $\mathbb{C}_2$ is the 2-qubit Clifford group and $\mathcal{C}_{m+1}$ is the correction gate \cite{PhysRevLett.106.180504}. For \textit{Local Clifford}, twirling gates are drawn uniformly at random from the tensor product of 1-qubit Clifford groups $\mathbb{C}_1^{\otimes2}$ with  $\mathcal{C}_{i,j}\in \mathbb{C}_1$ at time step $i$ on qubit $j$, $\mathcal{C}_{m+1,j}$ are the single-qubit correction gates and $\mathcal{C}_{m+1}$ is the 2-qubit correction gate required for motion reversal of the interleaved sequence \cite{PhysRevLett.123.030503,Polloreno_2025,mckay2311benchmarking}. For \textit{Pauli}, twirling gates are drawn uniformly at random from the tensor product of 1-qubit Pauli group $\mathbb{P}_1^{\otimes2}$ with $\mathcal{P}_{i,j}\in \mathbb{P}_1$ at time step $i$ on qubit $j$, where in these circuits $\mathcal{C}_{P,j}$ are local Clifford operators are used only for generating the state  preparation and measurement in the +1 eigen state of a random Pauli $P$, which is the cycle benchmarking scheme proposed in \cite{Erhard_2019}.} 

  \label{fig:Circuits_for_all_protocols}
\end{figure*}

 \section{Background: Interleaved Protocols and Randomization via Twirling Groups}
\label{Background}

Interleaved benchmarking protocols aim to estimate the error on an individual gate (or cycle) of interest, rather than an average over a set of gates, and thus comprise two distinct experiments: a ``reference'' sequence consisting of a randomized benchmarking experiment, and an ``interleaved'' sequence which inserts a gate of interest between the random gates in the reference sequence. 

Each reference sequence is a randomized benchmarking experiment which is an implementation of a sequence of randomly drawn gates that yields a figure of merit associated with the average error rate over that set of gates \cite{Emerson_2005}. If the set forms a group, which is the case for all the protocols we consider below, then the implemented sequence of random gates can be shown to correspond to a sequence of twirls \cite{PhysRevA.75.022314}. For this reason, the randomizing gates are often assumed to be a \textit{twirling group}, although notably this condition is not fulfilled in the single qubit randomized benchmarking scheme proposed by Knill et al \cite{PhysRevA.77.012307}; see \cite{PhysRevA.99.032329} for a discussion. When the twirling group is a unitary 2-design  (over all qubits) \cite{Emerson_2005,PhysRevA.80.012304}, and the error model is sufficiently gate independent, then the error on each gate is reduced to a depolarizing channel, and the error rate is a unitarily invariant quantity  associated with the process fidelity of the error model \cite{Emerson_2005,PhysRevA.80.012304,PhysRevLett.106.180504,PhysRevA.85.042311}. The gate-independence condition can be relaxed and the same conclusions still hold for all practical purposes\footnote{Under gate-dependent error models, the above conclusions hold except for the caveat that the error model corresponds to the average error `between the gates' of the randomized sequence  \cite{Wallman_2018} rather than the error `on' each gate, resolving the concern raised in \cite{PhysRevLett.119.130502}. See Appendix for further details.}. In this work, we restrict our focus to two-qubit systems (generalizations to n-qubit protocols are discussed in the conclusion) and evaluate four different twirling groups, which are summarized in Table \ref{tab:TwirlingGroups} and Fig.~\ref{fig:Circuits_for_all_protocols} in order of decreasing group size and complexity : (Row 1) `Haar' randomization based on the original RB protocol \cite{Emerson_2005,PhysRevA.75.022314}; (Row 2) `Clifford' randomization based on the standard RB protocol \cite{PhysRevA.80.012304,PhysRevLett.106.180504,PhysRevA.85.042311,PhysRevA.75.022314}, which is the standard approach for interleaved benchmarking; (Row 3) `Local Clifford' randomization which was proposed in \cite{Emerson_2007} and has found renewed interest recently \cite{mckay2311benchmarking,proctor2022establishing}; and (Row 4) `Pauli' randomization which is leveraged in the cycle benchmarking \cite{Erhard_2019} and cycle error reconstruction \cite{carignan2023error} protocols. Note that for the final two rows in the table we are considering reference protocols which are not unitary 2-designs across the qubits. In particular, 
the `Local Clifford' protocol is a randomized benchmarking experiment with a group that only forms a unitary 2-design over the individual qubits (Row 3) and the `Pauli' protocol (Row 4) is  a randomized benchmarking experiment which only forms a unitary 1-design over the individual qubits. In these two protocols, the experimental decay nonetheless produces an estimator of the average process fidelity, as described in the final column. The circuits for the benchmarking experiments under each of the reference protocol groups are shown in the first column of Fig.~\ref{fig:Circuits_for_all_protocols}.

The interleaved sequences interleave a gate of interest $\mathcal{G}$ between the randomizing (i.e., twirling) gates of the associated reference protocol. The circuits for interleaved benchmarking experiments under each of the reference protocol groups are shown in the second column of Fig. ~\ref{fig:Circuits_for_all_protocols}. We have adopted a compact notation where we specify the reference and interleaved protocols with the same label but denote the interleaved gate  $\mathcal{G}$ in parentheses, and denote the lack of an interleaved as an identity gate $\mathcal{I}$ in parentheses,  highlighting that the reference protocol is equivalent to the corresponding interleaved protocol when the interleaved gate  is the identity gate. To be clear, this notation does not imply inserting an identity gate or a delay equivalent to the identity channel in lieu of the interleaved gate, but simply means that the reference protocol is associated with the absence of an interleaved gate. 

The composition of the interleaved gate with an adjacent randomizing gate is called the `dressed gate'. The interleave sequence actually directly estimates the error on this composition and the error rate on the interleaved gate is then inferred via an estimator, as discussed in the next section.  That is, the interleaved protocol estimates an average error rate associated with the  cumulative error on the dressed gate, and the process infidelity of the interleaved gate is then inferred by comparing the two measured error rates, as described in the next section.
The central problem that we describe and solve in this paper is that the inferred error on the interleaved gate can vary unpredictably and dramatically when the errors on the randomizing gate and interleaved gate are coherent, but this can be mitigated when the error rate of the reference randomizing group is significantly lower than the interleaved gate. 

While IRB using the  two-qubit `Clifford' group is the standard approach and cycle benchmarking using the single qubit `Pauli' group is the leading alternative, we also have proposed an interleaved benchmarking protocol  based on `Haar' randomization in Row 1 for two reasons: (i) the Haar twirl is the only twirl that naturally allows for interleaved gates that are arbitrary non-Clifford gates, and thus enables cross-platform comparisons that include platforms using non-Clifford gates such as Google's fSim gate \cite{PhysRevLett.125.120504}, and (ii) considering the `Haar' twirl (which requires ~3 CNOTs on average compared to ~1.5 CNOTs for `Clifford')  also allows us to explore inaccuracy as a function of varying circuit complexity. Relatedly, two-qubit cross-entropy benchmarking (XEB) protocols \cite{Wu_2021,arute_quantum_2019,PhysRevLett.125.120504,simakov2026decayratesinterleavedbenchmarking} used for gates such as Google's fSim can be viewed as estimating the dressed cycle infidelity with easy single-qubit randomization, so extracting a bare two-qubit gate infidelity involves an analogous reference infidelity subtraction step. We also included `Local Clifford' due to the recent benchmarking proposal in \cite{merkel2025clifford}. `Local Clifford' has a larger group size ($24^n$, which can be reduced to $12^n$) than the `Pauli' group ($4^n$), and can have greater circuit complexity depending on the compilation scheme and primitive gate set. In this work we considerd a compiling scheme such that these two single qubit randomization groups have the same circuit complexity (see \ref{subsection:numerical results} for further details).

\begin{table*}[t]
\centering
\renewcommand{\arraystretch}{1.9} 
\setlength{\tabcolsep}{5pt}

\begin{tabular}{|>{\centering\arraybackslash}p{2cm}|
                >{\centering\arraybackslash}p{1.2cm}|
                >{\centering\arraybackslash}p{2.2cm}|
                >{\centering\arraybackslash}p{3.0cm}|
                >{\centering\arraybackslash}p{2.6cm}|
                >{\centering\arraybackslash}p{1.8cm}|
                >{\centering\arraybackslash}p{2.5cm}|}
\hline
\multicolumn{3}{|c|}{\textbf{Twirling/Randomizing Group}} &
\multicolumn{2}{c|}{\textbf{Interleaved Protocol Pair}} &
\textbf{Initial State(s)} &
\textbf{Estimator for Process Infidelity} \\
\cline{1-5}
\textbf{Label} & \textbf{Symbol} & \textbf{Description} &
\textbf{Reference} & \textbf{Interleaved} & & \\
\hline

Haar &
$\mathbb{SU}(4)$ &
\parbox[c]{2.2cm}{\centering \vspace{1mm }\strut Haar-random 2-qubit \\ unitaries\vspace{1mm} \strut} &
\parbox[c]{3.0cm}{\centering \strut Haar($\mathcal{I}$) \\ (= Haar RB \cite{Emerson_2005}) \strut} &
\parbox[c]{2.6cm}{\centering \strut Haar($\mathcal{G}$) \\ (new) \strut} &
\parbox[c]{2.1cm}{\centering \strut $|0\rangle^{\otimes 2}$ \strut} &
\parbox[c]{2.5cm}{\centering \strut $\hat{\epsilon} = \epsilon$ \cite{Emerson_2005} \strut} \\
\hline
Clifford
 &
$\mathbb{C}_2$ &
\parbox[c]{2.2cm}{\centering \strut Random 2-qubit \\ Clifford gates \strut} &
\parbox[c]{3.0cm}{\centering \vspace{3.2mm }\strut Clifford($\mathcal{I}$) \\ (= Standard RB \cite{PhysRevLett.106.180504}) \strut} &
\parbox[c]{2.6cm}{\centering \strut Clifford($\mathcal{G}$) \\ (= IRB \cite{PhysRevLett.109.080505}) \strut} &
\parbox[c]{2.1cm}{\centering \strut $|0\rangle^{\otimes 2}$ \strut} &
\parbox[c]{2.5cm}{\centering \strut $\hat{\epsilon} = \epsilon$ \cite{Emerson_2005,PhysRevA.80.012304,PhysRevLett.106.180504,PhysRevA.85.042311,PhysRevLett.109.080505} \strut} \\
\hline

Local Clifford &
$\mathbb{C}_1^{\otimes2}$ &
\parbox[c]{2.2cm}{\centering \strut Random 1-qubit \\ Clifford gates \strut} &
\parbox[c]{3.0cm}{\centering  \vspace{3mm}\strut Local Clifford($\mathcal{I}$) \\ (= Simultaneous single-qubit RB \cite{PhysRevLett.109.240504}) \strut} &
\parbox[c]{2.6cm}{\centering \strut Local Clifford($\mathcal{G}$) \\ (= Direct RB \\ \cite{PhysRevLett.123.030503,Polloreno_2025,mckay2311benchmarking}) \strut} &
\parbox[c]{2.1cm}{\centering \strut $|0\rangle^{\otimes 2}$ \strut} &
\parbox[c]{2.5cm}{\centering \strut $\hat{\epsilon} \approx \epsilon$ \\ asymptotically \\ \cite{PhysRevLett.123.030503,Polloreno_2025} \strut} \\
\hline

\parbox[c]{2.1 cm}{\centering \vspace{2mm}\strut Pauli \\ (=CB \cite{Erhard_2019}) \vspace{2mm}\strut} &
$\mathbb{P}_1^{\otimes2}$ &
\parbox[c]{2.2cm}{\centering \strut Random Pauli \\ gates \strut} &
\parbox[c]{3.0cm}{\centering \strut Pauli($\mathcal{I}$) \\ (= CB with $\mathcal{I}$) \strut} &
\parbox[c]{2.8cm}{\centering\vspace{0.8mm} \strut Pauli($\mathcal{G}$) \\ (= CB with $\mathcal{G}$) \strut} &
\parbox[c]{2.1cm}{\centering \strut Pauli eigenstates \strut} &
\parbox[c]{2.5cm}{\centering \strut $\hat{\epsilon} = \epsilon + \mathcal{O}(\epsilon^2)$ \\ \cite{Erhard_2019} \strut} \\
\hline

\end{tabular}

\caption{Summary of the interleaved protocol pairs for different twirling groups, the circuits for which are shown in Fig.~\ref{fig:Circuits_for_all_protocols}. Protocols are listed in order of decreasing circuit complexity (and group size) for the randomizing gates, with the first two rows requiring multi-qubit (entangling gates) and the last two rows requiring only single qubit gates. See text and caption to Fig.~\ref{fig:Circuits_for_all_protocols} for further details.}
\label{tab:TwirlingGroups}

\end{table*}

\section{Theory}

The central quantity of interest is the process fidelity of an implemented gate, either averaged over a set of gates or associated with a target gate of interest, relative to its ideal counterpart. The process fidelity gives the same information as the average gate fidelity which only differ by dimensional factors \cite{Nielsen_2002}. Both of these fidelities are standard figures of merit because they can be defined as invariant quantities associated with the average overlap over a natural set of operators \cite{Polar}.
 
The process infidelity of a quantum error channel $\mathcal{E}$ is most easily computed from the Pauli Transfer Matrix, which is the Liouville representation of $\mathcal{E}$ in the Pauli basis and is denoted by $[\mathcal{E}]$, such that 
\begin{equation}
    \epsilon^{\mathcal{E}}=1-\frac{\mathrm{Tr}[\mathcal{E}]}{d^2}
    \label{eq:abstract infidelity}
\end{equation}
where $d$ is the dimension of the Hilbert space. 

In the context of our work, we are interested in the process infidelity of the error map $\mathcal{E}$ that is associated with an ideal unitary gate $U$ relative to its imperfect implementation,  which can be written as $\mathcal{E}_U= U \circ \mathcal{E}$. So we 
let \( \epsilon^\mathcal{E\circ F} \) denote the process infidelity associated with the error on the \textit{dressed gate}, which consists of a twirling gate combined with the interleaved gate,  and \( \epsilon^\mathcal{E} \) denote the process infidelity associated with the errors on the cycles of randomizing gates.

Interleaved protocols aim to estimate the process infidelity of the error on the interleaved gate alone, but this cannot be uniquely determined from  \( \epsilon^\mathcal{E\circ F} \) and  \( \epsilon^\mathcal{E} \) for general error models. 
Thus interleaved protocols require an estimator for process infidelity of the error on interleaved gate \( \epsilon^\mathcal{F} \), which can be obtained by relating  $ \epsilon^\mathcal{E\circ F} $ with  $ \epsilon^\mathcal{E} $ as follows
\begin{equation}
    {\hat{\epsilon}^\mathcal{F} = 1 - \frac{1 - \epsilon^\mathcal{E\circ F}}{1 - \epsilon^\mathcal{E}}}.
    \label{eq:ratio of two fidelities}
\end{equation}
For depolarizing errors, this estimator becomes exact as the error rate approaches zero, but for more general errors, with a coherent component, this estimator is subject to significant inaccuracy, and the extent of this inaccuracy is the main focus of our study. We note here that this estimator formula differs (albeit negligibly) from the estimator in the original interleaved proposal in \cite{PhysRevLett.109.080505}, which is expressed in terms of the decay parameter `$p$' for randomized benchmarking. Our formulation of the estimator is more general as it allows for a unified approach that applies to a general class of interleaved benchmarking schemes with twirling groups other than Clifford, including cases, in particular the \textit{Pauli} case for cycle benchmarking, in which no single decay parameter exists. In Appendix \ref{process infidelity calculation of intleaved gate}, we show that the difference between the two approaches is asymptotically negligible in the low error regime, and practically negligible differences under typical error rates for the numerical and experimental data in this paper.  

\subsection{Systematic Uncertainty for Interleaved Benchmarking}
\label{Systematic bounds}
The process infidelity of the interleaved gate is commonly estimated from the ratio of the estimators of the interleaved and reference experiments, as in Eqn.~\ref{eq:ratio of two fidelities}. This ratio produces a point estimate for the bare target-gate infidelity which would be subjected to its own statistical uncertainty due to finite sampling. However, even if the reference and dressed infidelity estimates were known with infinite precision, the bare-gate infidelity $\epsilon^{\mathcal F}$ would not, in general be uniquely determined by $\epsilon^{\mathcal E}$ and $\epsilon^{\mathcal E\circ\mathcal F}$. The resulting non-identifiability gives rise to a systematic interval of bare-gate infidelities consistent with the same measured data.\\
 The reason for this systematic uncertainty as established in \cite{Carignan_Dugas_2019} is that the infidelity of a composite channel cannot  be inferred uniquely from the infidelities of its components, because process infidelity contains only limited information about the error channel. 
 For fixed values of the reference infidelity $\epsilon^{\mathcal{E}}$ and the measured dressed-gate infidelity $\epsilon^{\mathcal{E}\circ\mathcal{F}}$, there is generally a range of bare-gate infidelities $\epsilon^{\mathcal{F}}$ consistent with the same experimental data. This range of possible bare-gate infidelities is quantified by the following bound \cite{Carignan_Dugas_2019}
\begin{align}
    &\left| -\epsilon^\mathcal{F} + \epsilon^\mathcal{E\circ F}
    + \epsilon^\mathcal{E}
    - 2 \epsilon^\mathcal{E\circ F} \epsilon^\mathcal{E} \right| \notag \\
    &\quad \leq
    2 \sqrt{(1 - \epsilon^\mathcal{E\circ F})
    (1 - \epsilon^\mathcal{E})
    \epsilon^\mathcal{E\circ F}
    \epsilon^\mathcal{E}} .
    \label{eq: sys bounds}
\end{align}
which is equivalently expressed as
\begin{subequations}
\label{eq:systematic-bound}
\begin{align}
\epsilon^{\mathcal{F}}
&\ge
\epsilon^{\mathcal{E\circ F}}
+ \epsilon^{\mathcal{E}}
- 2\,\epsilon^{\mathcal{E\circ F}}\epsilon^{\mathcal{E}}
\nonumber\\
&\quad
- 2\sqrt{
    (1-\epsilon^{\mathcal{E\circ F}})
    (1-\epsilon^{\mathcal{E}})
    \epsilon^{\mathcal{E\circ F}}
    \epsilon^{\mathcal{E}}
    },
\\[4pt]
\epsilon^{\mathcal{F}}
&\le
\epsilon^{\mathcal{E\circ F}}
+ \epsilon^{\mathcal{E}}
- 2\,\epsilon^{\mathcal{E\circ F}}\epsilon^{\mathcal{E}}
\nonumber\\
&\quad
+ 2\sqrt{
    (1-\epsilon^{\mathcal{E\circ F}})
    (1-\epsilon^{\mathcal{E}})
    \epsilon^{\mathcal{E\circ F}}
    \epsilon^{\mathcal{E}}
    } .
\end{align}
\end{subequations}
In the regime of small error rate, the scale of difference between the upper bound and lower bound of the in Eqn.~\ref{eq:systematic-bound} goes as $4\sqrt{\epsilon^\mathcal{E\circ\mathcal{F}}\epsilon^\mathcal{E}}$, which is the width of the systematic uncertainty interval. The theoretical value of the bare-gate process infidelity $\epsilon^{\mathcal F}$ is guaranteed to fall within the width of interval in Eqn.~\ref{eq:systematic-bound}. Therefore this interval should be as seen as the physically allowed range of bare-gate infidelities consistent with the measured reference and interleaved experiments. On the contrary, the conventional ratio estimator yields only one point which may produce unrealistic values, such as negative values of the estimated infidelities for destructive interference of coherent errors, which does not mean that the real infidelity is negative but only means that the point estimator is not sufficient to specify the infidelity. In our numerical analysis, we find that even if the ratio estimator produces an unphysical result, the theoretical process infidelity still falls within the range defined by Eqn.~\ref{eq:systematic-bound}.
\subsubsection{Systematic uncertainty versus statistical uncertainty}
\label{statistical-vs-systematic}
We emphasize that Eqn.\ref{eq:systematic-bound} is \emph{not}  a statistical error bar on the ratio estimator. Rather, it is a systematic interval for the physically possible theoretical value of the bare-gate process infidelity $\epsilon^{\mathcal F}$, given the measured dressed-channel and reference-channel infidelities. It is useful to distinguish three conceptually different mechanisms that can affect the inferred interleaved-gate infidelity which are summarized in Table~\ref{tab:mechanisms}. The first two of these mechanisms are statistical in nature and arise from finite sampling of shots, finite numbers of random sequences, and the propagation of fit uncertainty through the ratio estimator. Both of these contributions can be made arbitrarily small with increasing resources. The third mechanism, however, systematic in nature and remains even in the limit of infinite data. It is therefore an \emph{asymptotic accuracy floor} intrinsic to the interleaved protocol and the only adjustable parameter in this floor is the error on the randomizing layer $\epsilon^{\mathcal{E}}$ , as reducing it tightens the floor as $\sqrt{\epsilon^{\mathcal{E}}}$. This can be a dominant source of uncertainty and is the critical sense supporting our claim that easier (low error) randomization provides more accurate fidelity estimation of the target-gate.  

\smallskip
\noindent\textbf{}

\begin{table*}[t]
\centering
\begingroup
\renewcommand{\arraystretch}{1.45}
\setlength{\tabcolsep}{5pt}
\setlength{\extrarowheight}{2pt}
\footnotesize

\newcommand{\pcell}[2]{%
\parbox[c]{#1}{%
\centering
\vspace{2mm}
#2
\par\vspace{2mm}
}%
}

\begin{tabular}{|>{\centering\arraybackslash}m{0.75cm}|
                >{\centering\arraybackslash}m{3.0cm}|
                >{\centering\arraybackslash}m{2.6cm}|
                >{\centering\arraybackslash}m{2.0cm}|
                >{\centering\arraybackslash}m{5.0cm}|
                >{\centering\arraybackslash}m{1.7cm}|}
\hline

\pcell{0.75cm}{\textbf{Label}} &
\multicolumn{2}{c|}{\pcell{5.6cm}{\textbf{Mechanism}}} &
\pcell{2.0cm}{\textbf{Nature}} &
\pcell{5.0cm}{\textbf{Finite $s$ scaling}} &
\pcell{1.7cm}{\textbf{$s \to \infty$}} \\
\hline

\pcell{0.75cm}{1}
&
\multicolumn{2}{>{\centering\arraybackslash}m{5.6cm}|}
{\pcell{5.6cm}{Propagation of statistical uncertainty of the reference sequence infidelity estimator $\hat\epsilon^{\mathcal E}$ through the ratio estimator used to infer $\hat\epsilon^{\mathcal F}$}}
&
\pcell{2.0cm}{Derived\\ statistical\\ effect}
&
\pcell{5.0cm}{
$\displaystyle
\mathrm{Var}[\hat\epsilon^{\mathcal F}]
\propto
\mathrm{Var}[\hat\epsilon^{\mathcal E}]
\sim
\frac{1}{s}$

\vspace{1.5mm}

$\displaystyle
\mathrm{SE}[\hat\epsilon^{\mathcal F}]
\sim
\frac{1}{\sqrt{s}}$
}
&
\pcell{1.7cm}{vanishes}
\\
\hline
\multirow{2}{0.75cm}{\pcell{0.75cm}{2}}
&
\multirow{2}{3.0cm}{\pcell{3.0cm}{Sequence-to-sequence variance of the true sequence fidelity $F_{m,i}$ over different randomizations}}
&
\pcell{2.6cm}{2A: coherent errors}
&
\multirow{2}{2.0cm}{\pcell{2.0cm}{Statistical}}
&
\pcell{5.0cm}{
$\displaystyle 
\sigma_m^2 := \mathrm{Var}_i(F_{m,i}) = \mathcal{O}(m^2 r^2)$ \cite{Wallman_2016}

\vspace{1.5mm}

$\displaystyle
\mathrm{SE}[\widehat F_m]
=
\frac{\sigma_m}{\sqrt{s}}
=
\mathcal{O}\!\left(\frac{mr}{\sqrt{s}}\right)$
}
&
\multirow{2}{1.7cm}{\pcell{1.7cm}{vanishes}}
\\
\cline{3-3}\cline{5-5}

&
&
\pcell{2.6cm}{2B: stochastic errors}
&
&
\pcell{5.0cm}{
$\displaystyle 
\sigma_m^2 := \mathrm{Var}_i(F_{m,i}) =\mathcal{O}(m r^2)$ \cite{Wallman_2016}

\vspace{1.5mm}

$\displaystyle
\mathrm{SE}[\widehat F_m]
=
\frac{\sigma_m}{\sqrt{s}}
=
\mathcal{O}\!\left(\frac{\sqrt{m}r}{\sqrt{s}}\right)$
}
&
\\
\hline

\pcell{0.75cm}{3}
&
\multicolumn{2}{>{\centering\arraybackslash}m{5.6cm}|}
{\pcell{5.6cm}{Systematic interval of Eqs.~\ref{eq:systematic-bound}: non-identifiability due to coherent interference in the dressed-gate composition}}
&
\pcell{2.0cm}{{Systematic}}
&
\pcell{5.0cm}{
$\displaystyle
\sim
4\sqrt{\epsilon^{\mathcal E \circ\mathcal{F}}\epsilon^{\mathcal E}}
\quad \text{(leading order)}$
}
&
\pcell{1.7cm}{{does not}\\ {vanish}}
\\
\hline

\end{tabular}

\caption{\footnotesize
Table distinguishing statistical uncertainty from systematic uncertainty in the inference of the bare-gate infidelity in interleaved benchmarking; $s$ denotes the number of independently sampled randomizations, which in our setting is equivalent to taking each shot to be a new randomization.. For a fixed sequence length $m$, $F_{m,i}$ denotes the true sequence fidelity of the $i$th random circuit instance, and $
    \sigma_m^2 := \mathrm{Var}_i(F_{m,i})
$
denotes the sequence-to-sequence variance over randomizations which is often more pronounced for coherent errors than stochastic error but remains a finite-sample statistical effect that vanishes as $s\to\infty$. $\mathrm{SE}$ denotes the standard statistical error of the corresponding averaged
estimator. }
\label{tab:mechanisms}
\endgroup
\end{table*}

\subsection{Unitarity and XRB bounds}
\label{XRB bounds}

One can reduce the size of the systematic uncertainty identified in Eqn.~\ref{eq:systematic-bound} by performing an independent measurement of the degree of coherence of the method via protocols such as extended-RB (XRB)\cite{Wallman_2015}. It is a modification to RB style experiments where correction operation at the end of the sequence is replaced with state tomography. The basic idea is that coherent errors preserve the purity of the state, where purity is defined as $\text{Tr}(\rho^2)$. For pure states, $\text{Tr}(\rho^2)=1$ while for mixed state $\text{Tr}(\rho^2)<1$. Starting with a pure state, the coherent errors perform a rotation of the pure state on the Bloch sphere, keeping it a pure state. The quantity of interest that we measure using XRB is called \textit{unitarity} which is a quantifier of the coherence of the error acting on the gates of the experiments. Mathematically, we can define \textit{unitarity} as

\begin{align}
    u(\mathcal{E})=\frac{1}{d-1}\int d\psi||\mathbf{n}(\mathcal{E}
    |\psi\rangle\langle\psi|)-\mathbf{n}(\mathcal{E}(I/d))||^2
\end{align}
where $\mathbf{n}$ denotes the generalized $d$-dimensional Bloch vector and $||\mathbf{n}(\rho)||^2$ denotes the Euclidean norm. We can instantly see a direct relationship between purity of a state and the Euclidean norm of the Bloch vector, which for a single-qubit is  
\begin{equation}
\begin{split}
\text{Tr}(\rho^2) &= \frac{1}{4} \Big[\text{Tr}(I_2) + 2\text{Tr}(n_x X + n_y Y + n_z Z) \\
&\quad + \text{Tr}((n_x^2 + n_y^2 + n_z^2) I_2)\Big] \\
&= \frac{1}{2}\left[1 + \|\mathbf{n}(\rho)\|^2\right]
\end{split}
\end{equation}
Another equivalent definition of unitarity is 
\begin{align}
u(\mathcal{E})=\frac{1}{d^2-1}\text{Tr}[\mathcal{E}_u^\dagger\mathcal{E}_u]
\end{align}
where $\mathcal{E}_u$ denotes the $(d^2-1) \times (d^2-1)$ unital block of the CPTP map $\mathcal{E}$ in the Liouville representation.

For an in-depth description of the protocol, see \cite{Wallman_2015}. \\
Once we have the \textit{unitarity}, we can  provide a tighter bound on the systematic uncertainty in Eqn.~\ref{eq: sys bounds}. For distinguishability from systematic bounds, we would call these bounds \textit{XRB bounds,}  which are given as
\begin{equation}
\Big|p(\mathcal{Y})-\frac{p(\mathcal{XY})p(\mathcal{X})}{u(\mathcal{X})}\Big| \leq \sqrt{1-\frac{(p(\mathcal{X}))^2}{u(\mathcal{X})}}\sqrt{1-\frac{(p(\mathcal{XY}))^2}{u(\mathcal{X})}}
\end{equation}
where $p$ is the depolarizing parameter in RB experiments. The parameters  $p(\mathcal{X}), p(\mathcal{XY})$, and  $u(\mathcal{X})$ that appear in the bound can be obtained directly from the measurements of the decay parameters associated with the reference, interleaved, and XRB experiments, respectively, without any further assumptions. From Ref.~\cite{Carignan_Dugas_2019}, one notes that the worst case limit is when $u(\mathcal{X}) = 1$, which means that the channel $\mathcal{X}$ is a unitary transformation . While for the depolarizing limit where $u(\mathcal{X}) = p(\mathcal{X})^2$, the bound collapses in first order to $\epsilon^{\mathcal{E}} + \epsilon^{\mathcal{F}}$, which corresponds to the first-order assumption in the usual ratio estimator in Eqn.~\ref{eq:ratio of two fidelities}. This indicates that assuming the composite infidelity to be equal to the linear sum of the individual infidelities is justified when the errors are stochastic. The unitarity is therefore essential because it quantifies the coherent component of the noise and determines how much the systematic interval can tighten relative to the worst-case bound. \\

\subsection{Defining the relevant fidelity based on the effective error model between the gates}
\label{Gauge}

In the context of all numerical studies of benchmarking experiments, there is an ambiguity in computing the fidelity of the operationally relevant error model \cite{PhysRevLett.119.130502,Carignan_Dugas_2018,Wallman_2018}.  This arises because an average fidelity computed naively based on the average `error-on-the-gate' may not be relevant when there can be systematic cancellation of error between a gate and the subsequent gate. This is sometimes stated as a dependence on a choice of gauge, where the appropriate gauge  itself depends on the randomizing group. Thus, to define the relevant fidelity in a rigorous way, a method is required to identify the self-consistent gauge (SCG) associated with each randomizing group.  To compare the  accuracy of these benchmarking protocols to a rigorous theoretical fidelity in numerical studies, we also introduce a protocol for finding the gauge associated with any gate-dependent error model for each randomized benchmarking sequence. We develop this framework in the Appendix \ref{Optimal gauge}, and provide a comparison of the SCG fidelity for the `error-between-the-gates' and the theoretical fidelity associated with the errors `on' the gates in the results. As a corollary, we find that for realistic error models and randomizing groups of interest, the differences between SCG fidelities and the theoretical fidelity is negligible.

\section{Results}
\label{Results}
\begin{figure}
    \centering
    \includegraphics[width=0.48\textwidth]{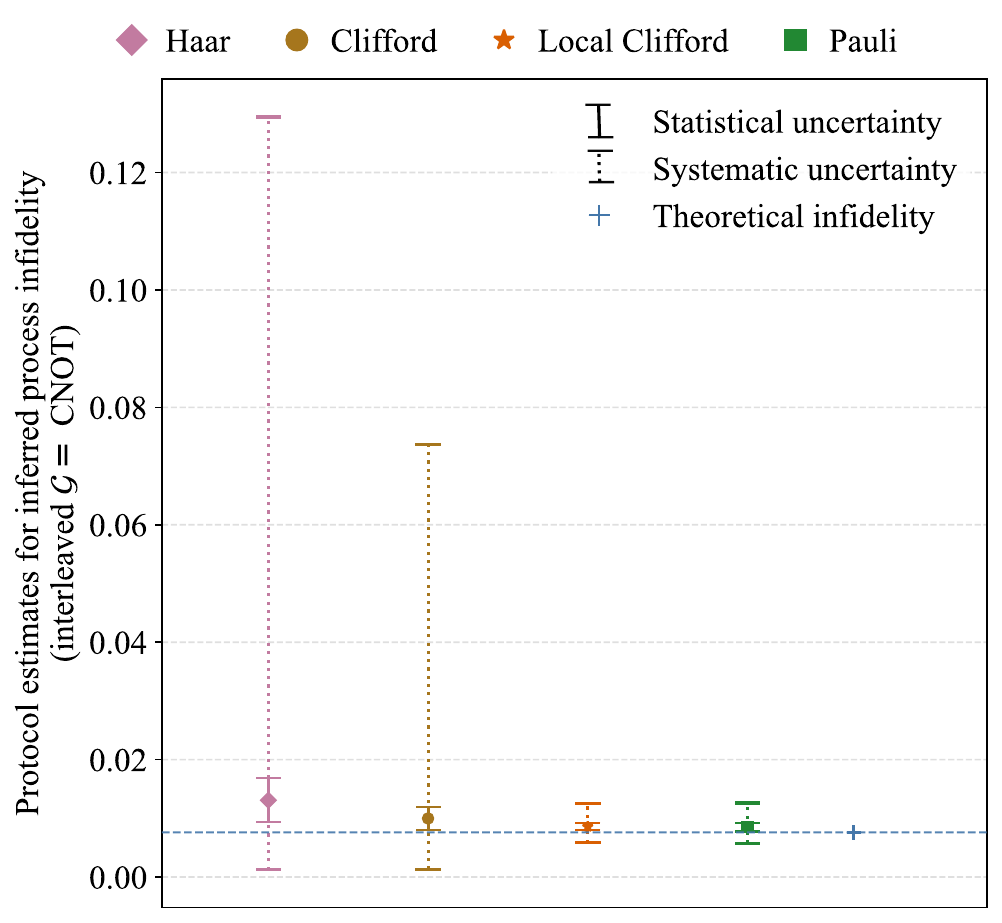} 
    \caption{\footnotesize Comparison of systematic uncertainty as a function of the different twirling groups defined in Table \ref{tab:TwirlingGroups}, showing that Pauli and Local Clifford twirling groups have the smallest systematic uncertainties, while keeping the number of shots fixed.  The error model comprises coherent error on the two-qubit gates of the form $e^{i\theta_2 ZZ}$ with $\theta_2=10^{\circ} $ and coherent error on the single-qubit gates of the form $e^{i\theta_1 Z}$ with $\theta_1=1^{\circ}$. The theoretical infidelity is the obtained using Eqn.~\ref{eq:abstract infidelity} under a naive (identity) choice of gauge. Even though Local Clifford group is larger than Pauli group, due to our chosen decomposition and error model on single-qubit gates  as detailed in \ref{subsection:numerical results}, both have almost same error magnitude and provides almost the same accuracy in the estimate. }
    \label{fig:sys uncertainty as function of twirling group}
\end{figure}
\begin{figure}
    \centering
    \includegraphics[width=0.48\textwidth]{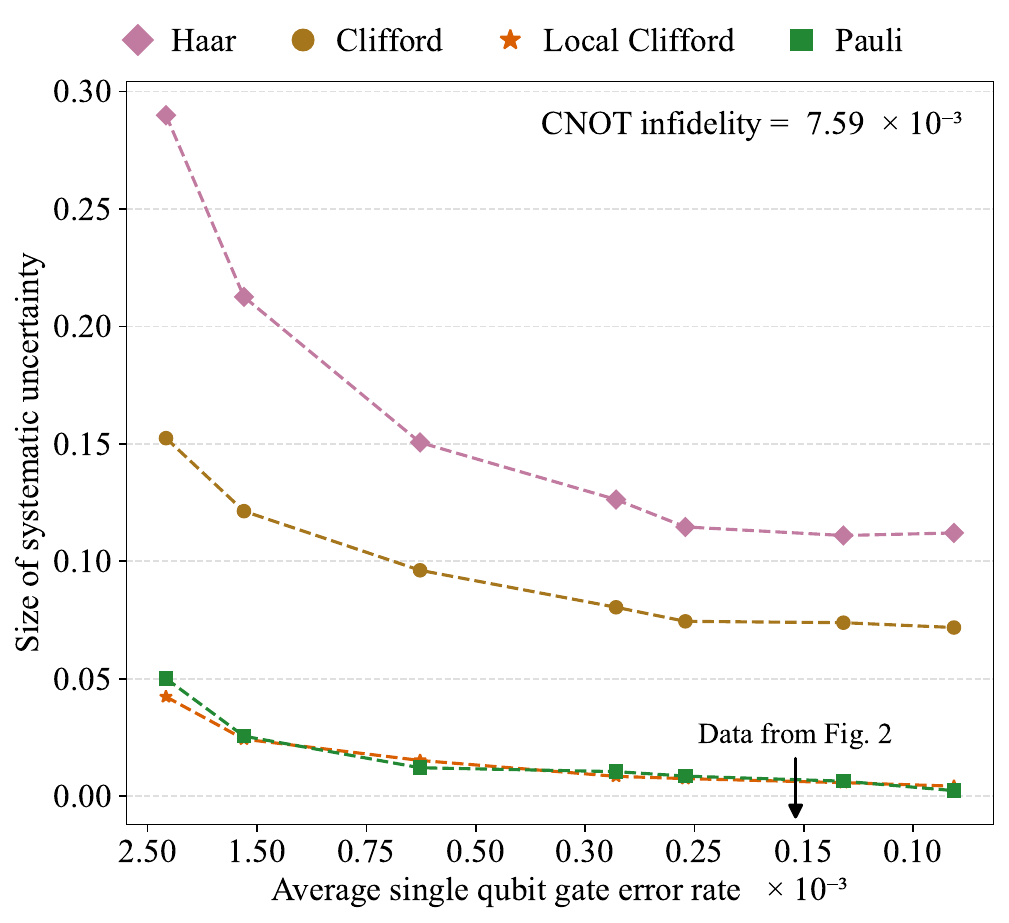} 
    \caption{\footnotesize Numerical plot showing the size of the systematic uncertainty for varying average single-qubit gate error rate. The error model is the same as the one used in Fig.~\ref{fig:sys uncertainty as function of twirling group} with $\theta_2$ fixed at $10^{\circ}$ and varying $\theta_1$ so as to vary the average single-qubit gate error rate along the x-axis. The other simulation parameters are the same as Fig.~\ref{fig:sys uncertainty as function of twirling group} which is a typical instance of this plot. }
    \label{fig:rel error vs ratio}
\end{figure}
\begin{figure}
    \centering
    \includegraphics[width=0.48\textwidth]{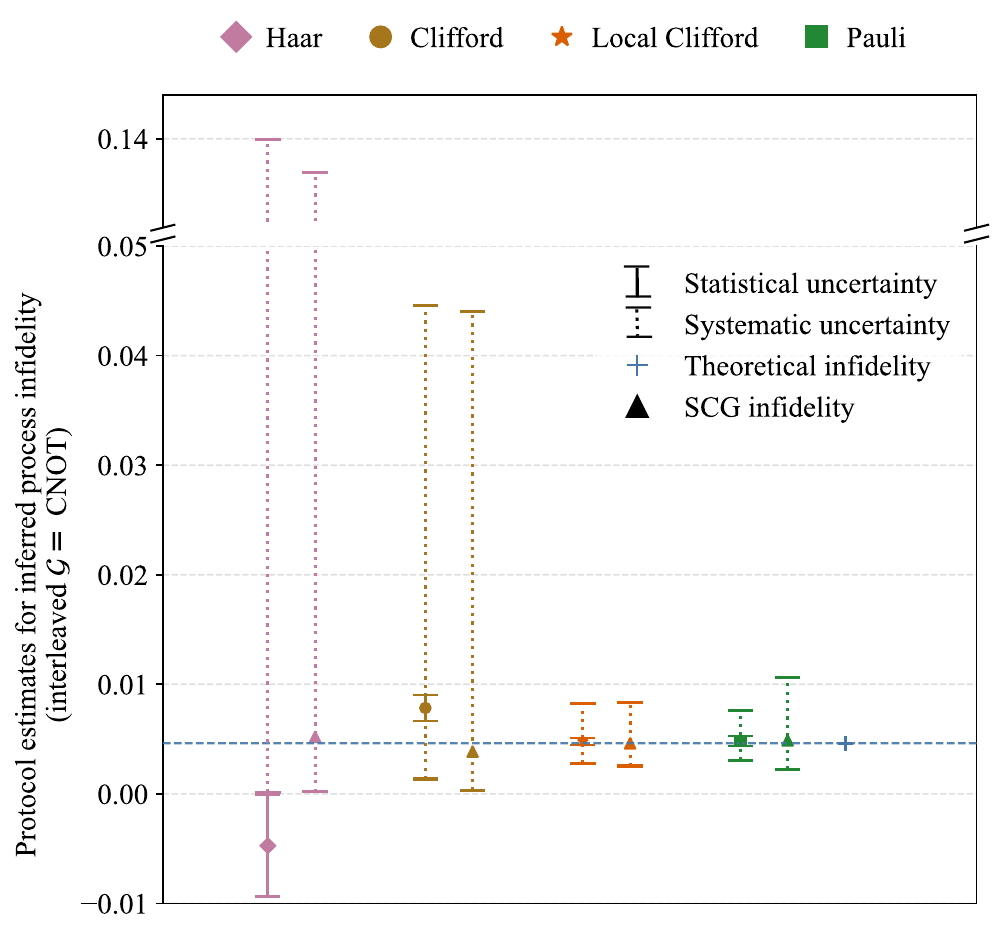} 
    \caption{\footnotesize Numerical plot highlighting the importance of systematic uncertainties for a natural (overrotation) coherent error model, where we see that the theoretical infidelity value is well outside the statistical bar of the estimates of the protocols such as Haar and Clifford, but is within the systematic bound. This figures also highlights how interference of errors can lead to negative (unphysical) estimate of process infidelity, as is observed for the Haar protocol, an effect we also see experimentally as described in the next section.
    The overrotation error model is of the form $U^{1+\delta}$ where $U$ is the target unitary, with $\delta=0.05$ for two-qubit gates  and $\delta=0.01$ for single-qubit gates.  Other simulation parameters are the same as in Fig.~\ref{fig:sys uncertainty as function of twirling group}.
    The triangles beside each estimate are the infidelities computed using the self-consistent gauge (SCG) for each twirling group; we see that the SCG yields an almost negligible difference to the naive ``error on the gate" method for computing the theoretical infidelity. See Appendix \ref{Optimal gauge} for more details on how to construct the self-consistent gauge. 
 }
    \label{fig:true and theoretical}
\end{figure}

\subsection{Numerical results}
\label{subsection:numerical results}
We now present numerical simulations to address the central questions motivating this work, the data for which are provided in Appendix. \ref{Numerical and experimental data}
We first demonstrate in Fig.~\ref{fig:sys uncertainty as function of twirling group} how dramatically the systematic uncertainties from Eqn.~\ref{eq:systematic-bound}  vary under the four different pairs of benchmarking protocols and their associated twirling groups (see Fig.~\ref{fig:Circuits_for_all_protocols} and Table \ref{tab:TwirlingGroups}). The error model comprises coherent error on the two-qubit gates of the form $e^{i\theta_2 ZZ}$ with $\theta_2=10^{\circ} $ and coherent error on the single-qubit gates of the form $e^{i\theta_1 Z}$ with $\theta_1=1^{\circ}$. The figure demonstrates that the magnitude of systematic uncertainty varies substantially across protocols with different twirling groups, where  protocols employing  twirling groups with more total error, such as Haar and Clifford, which require two qubit gates, exhibit significantly larger systematic uncertainties than those based on minimal twirling groups, including Local Clifford and Pauli. 
Notably, despite the Local Clifford being strictly larger than the Pauli group, the two yield nearly identical systematic uncertainties in our simulations. 
This similarity in performance is because we implement all single qubit gates via the decomposition $Z(\theta)X(\pi/2)Z(\omega)X(\pi/2)Z(\phi)$ decomposition  \cite{Knapp1988LieGB,PhysRevA.96.022330}, and assume that only the $X(\pi/2)$ gates have error where the $Z$ rotations are error free as they can be implemented experimentally virtually in software. As a result of this approach, single-qubit Pauli gates and single-qubit Clifford gates have almost the same average error rates. 
The number of shots used is $s=1500$ which for the Pauli twirling group is 100 randomizations distributed equally across 15 input states.

In Fig.~\ref{fig:rel error vs ratio} we explore how the size of the systematic error varies as we decrease the error rate on the single-qubit gates.  The trend confirms that systematic error decreases as the error rate on the twirling gates decreases, and again shows that protocols with larger twirling groups (and more total error) consistently exhibiting significantly larger systematic uncertainties than those employing minimal twirling groups.

We now demonstrate the practical importance of accounting for the systematic error bounds, where in the presence of simple coherent errors, statistical error bars alone are insufficient and systematic effects can dominate the interpretation of benchmarking estimates.
Fig.~\ref{fig:true and theoretical} is a repetition of the numerical simulations from Fig.~\ref{fig:sys uncertainty as function of twirling group} but under an overrotation error model of coherent errors, for which the noisy implementation of any unitary $U$ is $\tilde{U}=U^{1+\delta}$. This figure highlights the critical role of the systematic uncertainty, because we observe clear signatures of coherent interference between the twirling gates and the interleaved gate: the theoretical infidelity lies outside the statistical error bars for both of the Haar and Clifford protocols. Moreover, in the case of the Haar protocol, this interference drives the estimated process infidelity to a negative value, which is a physically impossible estimate. As we will show in the next section, physically impossible estimates of the fidelity (due to coherent error cancellation) is not a contrived occurrence in our numerics; on the contrary this is an effect we also observed experimentally. 

\begin{figure}
    \centering
    \includegraphics[width=0.48\textwidth]{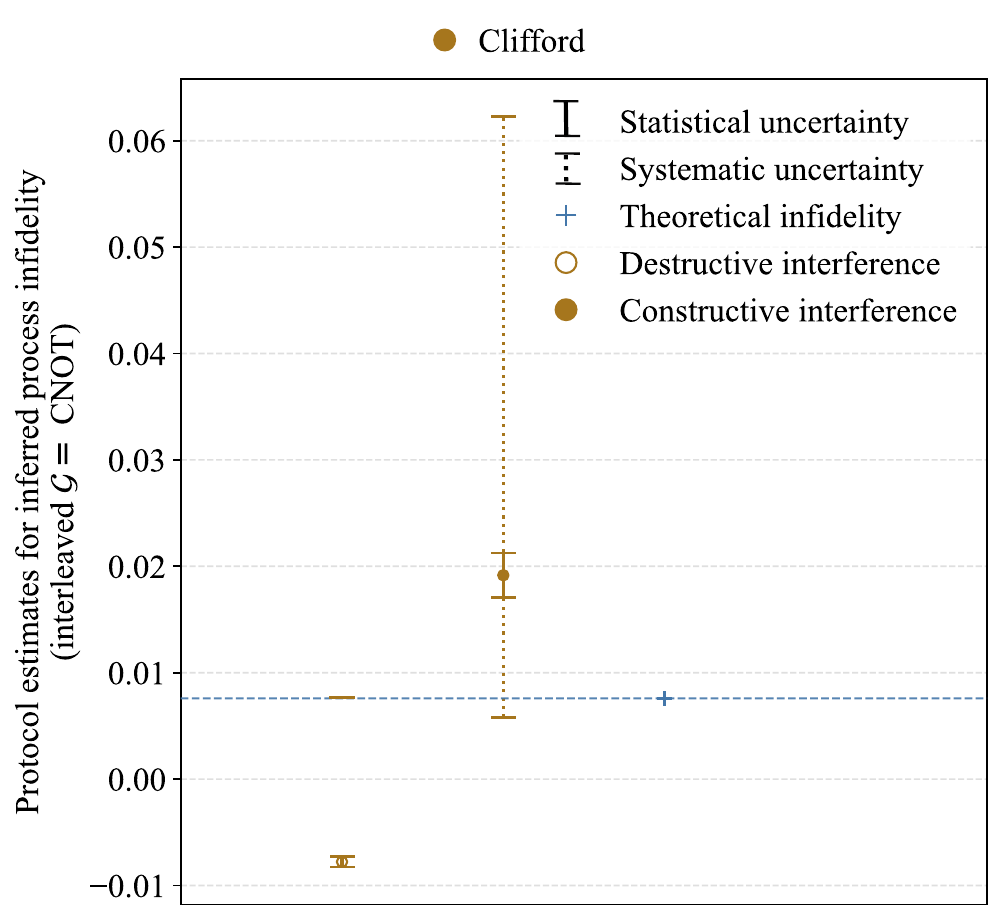} 
    \caption{\footnotesize Numerical plot showing how coherent errors between random twirling gates and fixed hard gate can constructively and destructively interfere to give entirely different infidelity estimates using Clifford protocols. For the destructive interference shown by yellow open circle we use the coherent error model of the form $e^{i\theta YY}$ and for constructive interference  shown by yellow closed circle we use $e^{-i\theta YY}$ on the  random clifford gates and $e^{i \theta XZ }$ on the interleaved CNOT gates with $\theta=10\degree$. The process infidelity of the reference experiments corresponding to both error models of the form $e^{i\theta YY}$ and $e^{-i\theta YY}$ are same as it only depends on the magnitude of $\theta$. Notice that in the destructive interference case the inferred process infidelity of interleaved CNOT is negative (unphysical) with the lower and upper systematic bound collapsing to a single value using Eqn.~\ref{eq:systematic-bound} which is consistent with the theoretical value. We also observe similar unphysical estimate due to interference of coherent errors in experimental data, as shown in Fig.~\ref{fig:expeiments}.} 
    \label{fig:construction_destruction}
\end{figure}

We can however contrive adversarial coherent error models that show the extent of the systematic error that can occur in practice. For the standard IRB protocol, i.e., the Clifford protocol, we fix the error on the interleaved CNOT gate to be $e^{i\theta XZ}$, and consider two different error models for the random two-qubit Clifford gates:  $e^{i\theta YY}$ or $e^{-i\theta YY}$. As shown in Fig.~\ref{fig:construction_destruction}, whereas the reference RB experiment, i.e., Clifford($\mathcal{I}$), has the same  process infidelity in both the cases, the inferred process infidelity for the interleaved gate differs dramatically due to constructive (amplification of error) vs destructive (cancellation of error) coherence. This error model is of course contrived so that under conjugation by the CNOT gate, a $YY$ error transforms into $-XZ$. Consequently, depending on the relative sign, the coherent errors on the interleaved and twirling gates can either add constructively or cancel destructively.  In the case, where the errors cancel, we yet again obtain a negative, and therefore unphysical, estimate of the process infidelity of the interleaved CNOT gate. While similar interference can in principle occur in other benchmarking protocols, the effect is far less pronounced when Local Clifford or Pauli groups are used, since errors on these twirling gates are unlikely to match the structure and magnitude of the error on a two-qubit CNOT. This can be interpreted in a platform-dependent manner. If there is a coherent error within the randomizing layer that is similar to that of the target two-qubit gate error, e.g., an equivalent residual $ZZ$ interaction to the coherent over-rotation on a CPhase type of gate, then it is possible to achieve constructive or destructive interference even when applying local randomization. It can be said that in this case, the local randomizing layer is not ``easy'' in the sense relevant here, because its error contribution $\epsilon^{\mathcal E}$ is no longer small compared with that of the target gate. Hence, in the typical case where minimal twirling groups such as the Local Clifford and Pauli groups exhibit substantially lower systematic uncertainty and exhibit much more limited sensitivity to these extreme effects of coherent interference, making them far more reliable for estimating gate performance. 

All numerical simulations in this work are performed under a fixed resource budget, with the total number of shots $s$ across all the protocols set equal to ensure an even-handed comparison. We are working in the \textit{single-shot-per-randomization} limit, where each shot of an experiment corresponds to a new, distinct random circuit. An experimental demonstration of this limit where each shot is a new randomization has been shown in \cite{fruitwala2024hardware}. Since the Pauli and Local Clifford twirling groups do not form a unitary 2-design \cite{PhysRevA.80.012304}, we use different state preparation, measurement and decay curve fitting strategy as described in \cite{Erhard_2019,mckay2311benchmarking}. For instance in Pauli($\mathcal{G}$) and Pauli($\mathcal{I}$) as described in Table \ref{tab:TwirlingGroups} and Fig.~\ref{fig:Circuits_for_all_protocols}, we need to prepare the state in $K$ different +1 eigenstate of the Paulis where each of the input state is assigned $s'=\frac{s}{K}$ shots such that the total number of shots across each all the experiments remain $s$ and we extract the fidelity decay parameter corresponding to each Pauli of interest. Pauli($\mathcal{G}$) and Pauli($\mathcal{I}$) are equivalent to the extension of Cycle Error Reconstruction (CER) proposed in \cite{CarignanDugas2024estimatingcoherent} where the number of repetitions of the hard cycle is $0$ and $1$ respectively. The general extension employing multiple values of hard gate repetitions provides more statistically stable fidelity estimates, but is more resource-intensive.  For Local Clifford$(\mathcal{I})$, two single-qubit fidelity decay parameters are obtained, one for each qubit.

\subsection{Experimental results}
\begin{figure*}[t]
    \centering
    \includegraphics[width=
    \textwidth]{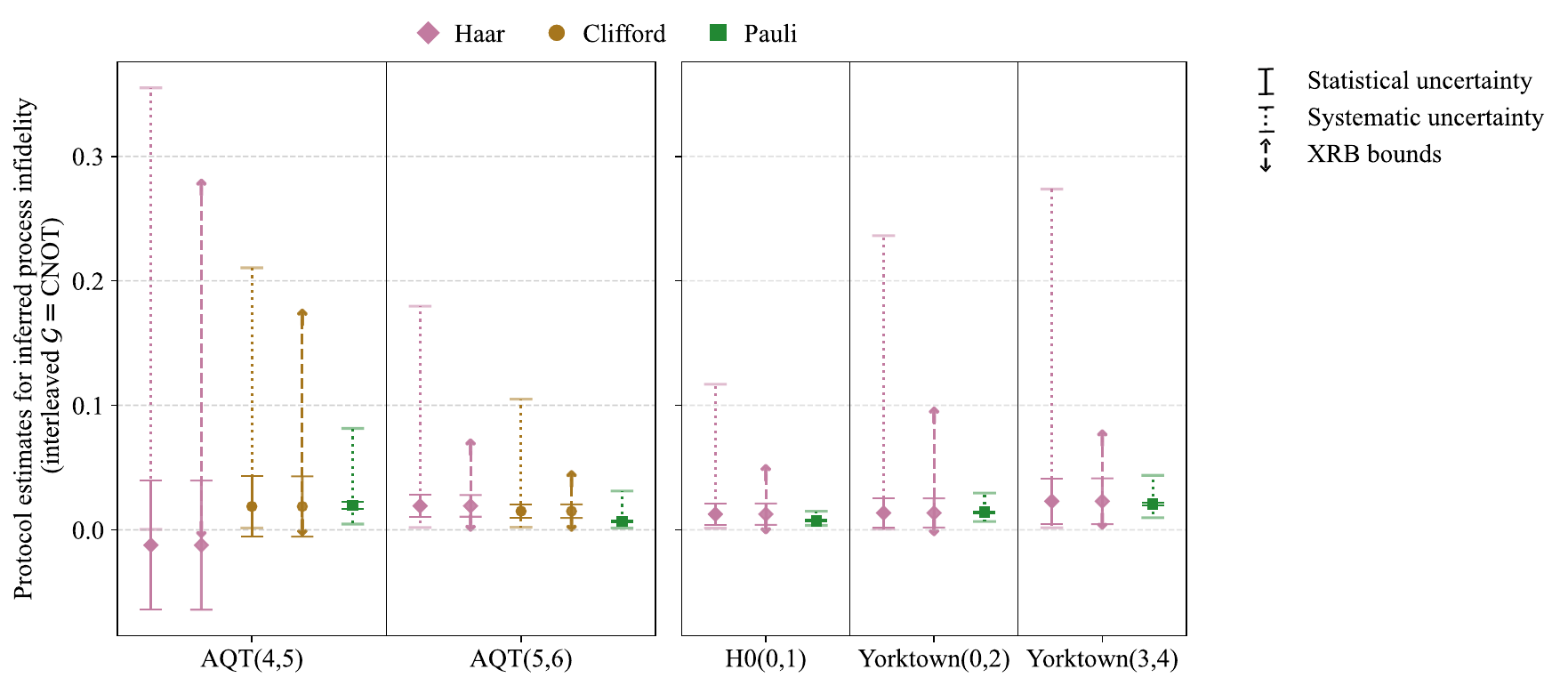} 
    \caption{ \footnotesize \
    Experimental data across multiple platforms illustrating the significant systematic uncertainties in estimated infidelities of interleaved gates, and showing that the accuracy of interleaved benchmarking improves under randomizing gates with lower error rates: the Pauli randomization in cycle benchmarking \cite{Erhard_2019} shows significantly lower uncertainty than standard IRB with either Clifford or Haar randomization.  The experimental data also show how estimating the unitarity can help reduce the systematic uncertainty as detailed in section \ref{XRB bounds} (XRB bounds), however this reduced bound  still remains substantially larger than the systematic uncertainty from Pauli randomization under cycle benchmarking. Also notable in the first column (AQT(4,5)), the IRB protocol with Haar randomization yields a negative valued (unphysical) estimates of  the process infidelity due to destructive interference, as also discussed in the numerical simulations. Details on the experimental platforms are present in section \ref{experimental results} of the main text.}
    \label{fig:expeiments}

\end{figure*}
\label{experimental results}

We now present experimental results for the three of the interleaved benchmarking protocols across three experimental platforms with   $\mathcal{G}=\text{CNOT}$ as the interleaved gate.  The experimental data are plotted in Fig.~\ref{fig:expeiments} and the data are provided in Table. \ref{Experimental data table} in  Appendix. \ref{Numerical and experimental data}. The first two columns of experimental data were performed on the Advanced Quantum Testbed (AQT) from Lawrence Berkeley National Laboratory, an $8$ qubit device with $4$ operational qubits at the time of implementation. The AQT superconducting system used for the experiments is the same as the one used in \cite{PhysRevX.11.041039}.
These results, repeated for two different pairs of qubits,  illustrate that cycle benchmarking with Pauli twirling \cite{Erhard_2019} consistently provides a much more reliable estimate of process infidelity with significantly reduced systematic uncertainty compared to the standard IRB protocol based on Clifford randomization. Indeed, we see that under the actual experimental errors (not engineered errors) the interleaved protocol with the Haar twirling group can actually produce negatively valued estimated for the infidelity of the interleaved gate, as we also observed  in the numerical simulations of the previous section.

The final three columns provide a further set of experiments designed to illustrate these effects  across two other  experimental platforms.  The cross-platform results consistently show large the systematic uncertainties for infidelity estimates under both Haar and Clifford randomization  compared to the more accurate estimates through the Pauli randomization used in cycle benchmarking \cite{Erhard_2019}. 
These results also explore the reduction in uncertainty that is achievable via benchmarking the degree of coherence in the error via the XRB protocol. The results also highlight that even the tighter XRB bounds are still larger than the systematic bounds of CB and fall short of its greater accuracy. For the cross-platform benchmarking we focused on the Haar protocol because it is the unique protocol that can accommodate cross-platform comparisons with inequivalent and non-Clifford interleaved gates.\\
The pattern of agreement and disagreement among the inferred process infidelities across protocols on the same physical qubit pair should not be interpreted solely from the statistical overlap of the fitted estimates. Agreement within statistical uncertainty confirms reproducibility, but it does not by itself rule out coherent systematic error. As discussed above, the systematic uncertainty interval defines the range of theoretical target infidelities that are compatible with the measured estimates. Since in experiment the theoretical infidelity is not directly accessible, a statistically precise point estimate does not guarantee proximity to the theoretical value when the interval is broad.
This is precisely what we observed in simulation where sometimes under coherent error interference, the inferred infidelity can differ substantially from the theoretical target infidelity. 
 The cross-protocol pattern admits several limiting cases.  First, if the errors were effectively stochastic, 
 the XRB bounds interval would collapse toward the statistical uncertainty of the fitted estimates. The observed nonzero XRB bounds interval already rules this out. Second, if coherent errors were present but structurally unaligned with the errors on the randomizing gates, the systematic bounds interval would remain finite but unsaturated. It would produce protocol dependent shifts without unphysical central estimates. Finally, finite sample fluctuations or ratio-estimator variance could produce randomization instance-to-instance scatter about a common value, but would not naturally produce a reproducible negative central estimate of a process infidelity. The negative Haar-randomized estimate observed for AQT(4,5) is therefore most naturally interpreted as a case of near-saturation of the systematic lower bound where coherent errors in the reference and target components compose destructively, so that the ratio estimator assigns an unphysical negative value even though the compatible systematic interval remains physical. For AQT(5,6) the CB estimate lies well outside the statistical uncertainty of Haar and Clifford estimates. Thus, if one reported only the Haar or Clifford estimate with its statistical error bars, one would incorrectly conclude that the true bare CNOT infidelity is known to a precision that excludes the CB-compatible value. 

In addition to the AQT results, the other two platforms were  a superconducting device  from IBM (Yorktown), a $5$ qubit device, and an ion-trap device (H0), a $6$ qubit device from Honeywell Quantum Solutions, now Quantinuum. It is important to note that the results from these devices were from late $2019$ and therefore do not showcase how well their premium devices currently operate.
The H0 device was a 6 qubit trapped-ion quantum processor using the quantum charge coupled device (QCCD) architecture~\cite{Pino_2021}.
On H0, qubits are encoded in hyperfine levels of Yb$^+$ ions,
which are trapped by AC electric fields in a surface ion-trap.
Single-qubit and two-qubit gates are performed in one of two separated gate zones,
containing up to two qubits at a time,
and all-to-all qubit connectivity is achieved by shuttling ions between zones.

\section{Conclusion}
\label{Conclusion}

In this work we compared the accuracy of four interleaved benchmarking protocols defined by distinct choices of randomizing gates, including most notably the widely-adopted interleaved randomized benchmarking (IRB) protocol based on the standard randomized benchmarking with the multi-qubit Clifford operations, and the more recently proposed cycle benchmarking (CB) protocol which leverages local randomizaton with single qubit Pauli gates. We demonstrated  experimentally on multiple platforms, and also numerically under a family of coherent error models and error strengths, that the standard approach of IRB is subject to a very large systematic uncertainty that makes the protocol highly inaccurate for estimating the infidelity of an interleaved gate in the presence of coherent errors. Fortunately, we found that this inaccuracy is largely mitigated when cycle benchmarking is adopted for estimating the infidelity of an interleaved gate and more generally showed that the accuracy of interleaved protocols improves as a function of decreasing error on the choice of randomizing gates. 

     We also considered the protocol for bounding the amount of coherent errors on the Clifford gate set, the so-called unitarity of the errors, but found that even when IRB is supplemented by this technique, it still dramatically underperforms the accuracy of cycle benchmarking. We observed  that the highest accuracy methods for calibration and benchmarking of an interleaved gate are those leveraging randomizing gates with the lowest error rates, which for most platforms comprises single qubit gates, notably the Pauli gates considered in cycle benchmarking or the local Clifford gates that generate a variant of direct RB. A distinct drawback of the local Clifford randomization, when scaled to parallelized gates (via simultaneous direct RB, or SDRB) is the overcounting of  inter-gate crosstalk errors, which overcounting arises because of an assumption that the error model factorizes along the gate blocks \cite{mckay2311benchmarking,merkel2025clifford}. This is a separate and distinct source of bias (even in the case of incoherent error) from those discussed in this work, and one that CB entirely avoids via \textit{crosstalk-robustness}: CB does not require any assumption on the factorization of the error model. In this context of unknown degree of cross-talk, CB offers a separate accuracy advantage. We leave a detailed analysis of this for future work.

Additionally we observe that the Pauli eigenvalue data that is learned in the cycle benchmarking experiment provides the additional benefit of \textit{data-reusability}: this information  delivers also the fine-grained error rates of the stochastic error channel for that cycle (viz gate layer), which is broadly relevant, e.g., to pulse optimization and analyzing  decoder performance for error correction circuits \cite{PhysRevA.94.052325,PhysRevX.11.041039,ishii2025implementation,Kurita_2023,PhysRevResearch.4.033140,winick2022concepts,PhysRevLett.130.250601,urbanek2021mitigating,Faehrmann_2022,Iyer2022,Jain2023,fazio2025characterizing,iyer2025enhancing}.  In particular, this data-reusability was explicitly invoked in \cite{Erhard_2019}, where the the measured set of Pauli eigenvalues measured in the cycle benchmarking experiment were applied to infer the Pauli error rates through an appropriate reconstruction procedure, later reported in the CER manuscript \cite{carignan2023error}. Finally we observe that CB can also be easily implemented with hardware efficient randomization offering significantly reduced latencies \cite{fruitwala2024hardware}. Hence, our takeaway is that CB is both the most accurate and most resource efficient scheme for fast, high-fidelity gate calibration and benchmarking under minimal error model assumptions.

\hspace{0.1in}

\section*{Acknowledgments}
We thank  Joel Wallman, Nicholas Fazio, Matthew A. Graydon, Ian Hincks, Egor Ospadov, Dar Dahlen for helpful discussions. We would also like to thank the anonymous referee for valuable comments helping us emphasize the difference between statistical and systematic uncertainty . Numerical simulations were performed using True-Q software \cite{beale_2020_3945250}. We acknowledge financial support from the US Army Research Office through Grant Number W911NF-21-1-0007.

\section*{Author's Contribution}
D.S.~and K.B.~led the numerical, theoretical and experimental work and co-wrote the manuscript. A.C.D led the analysis of self-consistent gauge (Appendix A).  A.H., I.S., and K.M.~provided experimental results. J.E.~conceived the project and contributed to all theoretical aspects.

\section*{Data Availability Statement}
Data for the figures are available in Appendix \ref{Numerical and experimental data}. The codes used for the analysis are publicly available in \cite{my_repo}.

\bibliography{ref}

\begin{thebibliography}{74}%
\makeatletter
\providecommand \@ifxundefined [1]{%
 \@ifx{#1\undefined}
}%
\providecommand \@ifnum [1]{%
 \ifnum #1\expandafter \@firstoftwo
 \else \expandafter \@secondoftwo
 \fi
}%
\providecommand \@ifx [1]{%
 \ifx #1\expandafter \@firstoftwo
 \else \expandafter \@secondoftwo
 \fi
}%
\providecommand \natexlab [1]{#1}%
\providecommand \enquote  [1]{``#1''}%
\providecommand \bibnamefont  [1]{#1}%
\providecommand \bibfnamefont [1]{#1}%
\providecommand \citenamefont [1]{#1}%
\providecommand \href@noop [0]{\@secondoftwo}%
\providecommand \href [0]{\begingroup \@sanitize@url \@href}%
\providecommand \@href[1]{\@@startlink{#1}\@@href}%
\providecommand \@@href[1]{\endgroup#1\@@endlink}%
\providecommand \@sanitize@url [0]{\catcode `\\12\catcode `\$12\catcode `\&12\catcode `\#12\catcode `\^12\catcode `\_12\catcode `\%12\relax}%
\providecommand \@@startlink[1]{}%
\providecommand \@@endlink[0]{}%
\providecommand \url  [0]{\begingroup\@sanitize@url \@url }%
\providecommand \@url [1]{\endgroup\@href {#1}{\urlprefix }}%
\providecommand \urlprefix  [0]{URL }%
\providecommand \Eprint [0]{\href }%
\providecommand \doibase [0]{https://doi.org/}%
\providecommand \selectlanguage [0]{\@gobble}%
\providecommand \bibinfo  [0]{\@secondoftwo}%
\providecommand \bibfield  [0]{\@secondoftwo}%
\providecommand \translation [1]{[#1]}%
\providecommand \BibitemOpen [0]{}%
\providecommand \bibitemStop [0]{}%
\providecommand \bibitemNoStop [0]{.\EOS\space}%
\providecommand \EOS [0]{\spacefactor3000\relax}%
\providecommand \BibitemShut  [1]{\csname bibitem#1\endcsname}%
\let\auto@bib@innerbib\@empty
\bibitem [{\citenamefont {Emerson}\ \emph {et~al.}(2005)\citenamefont {Emerson}, \citenamefont {Alicki},\ and\ \citenamefont {Życzkowski}}]{Emerson_2005}%
  \BibitemOpen
  \bibfield  {author} {\bibinfo {author} {\bibfnamefont {J.}~\bibnamefont {Emerson}}, \bibinfo {author} {\bibfnamefont {R.}~\bibnamefont {Alicki}},\ and\ \bibinfo {author} {\bibfnamefont {K.}~\bibnamefont {Życzkowski}},\ }\bibfield  {title} {\bibinfo {title} {Scalable noise estimation with random unitary operators},\ }\href {https://doi.org/10.1088/1464-4266/7/10/021} {\bibfield  {journal} {\bibinfo  {journal} {Journal of Optics B: Quantum and Semiclassical Optics}\ }\textbf {\bibinfo {volume} {7}},\ \bibinfo {pages} {S347–S352} (\bibinfo {year} {2005})}\BibitemShut {NoStop}%
\bibitem [{\citenamefont {Dankert}\ \emph {et~al.}(2009)\citenamefont {Dankert}, \citenamefont {Cleve}, \citenamefont {Emerson},\ and\ \citenamefont {Livine}}]{PhysRevA.80.012304}%
  \BibitemOpen
  \bibfield  {author} {\bibinfo {author} {\bibfnamefont {C.}~\bibnamefont {Dankert}}, \bibinfo {author} {\bibfnamefont {R.}~\bibnamefont {Cleve}}, \bibinfo {author} {\bibfnamefont {J.}~\bibnamefont {Emerson}},\ and\ \bibinfo {author} {\bibfnamefont {E.}~\bibnamefont {Livine}},\ }\bibfield  {title} {\bibinfo {title} {Exact and approximate unitary 2-designs and their application to fidelity estimation},\ }\href {https://doi.org/10.1103/PhysRevA.80.012304} {\bibfield  {journal} {\bibinfo  {journal} {Phys. Rev. A}\ }\textbf {\bibinfo {volume} {80}},\ \bibinfo {pages} {012304} (\bibinfo {year} {2009})}\BibitemShut {NoStop}%
\bibitem [{\citenamefont {Magesan}\ \emph {et~al.}(2011)\citenamefont {Magesan}, \citenamefont {Gambetta},\ and\ \citenamefont {Emerson}}]{PhysRevLett.106.180504}%
  \BibitemOpen
  \bibfield  {author} {\bibinfo {author} {\bibfnamefont {E.}~\bibnamefont {Magesan}}, \bibinfo {author} {\bibfnamefont {J.~M.}\ \bibnamefont {Gambetta}},\ and\ \bibinfo {author} {\bibfnamefont {J.}~\bibnamefont {Emerson}},\ }\bibfield  {title} {\bibinfo {title} {Scalable and robust randomized benchmarking of quantum processes},\ }\href {https://doi.org/10.1103/PhysRevLett.106.180504} {\bibfield  {journal} {\bibinfo  {journal} {Phys. Rev. Lett.}\ }\textbf {\bibinfo {volume} {106}},\ \bibinfo {pages} {180504} (\bibinfo {year} {2011})}\BibitemShut {NoStop}%
\bibitem [{\citenamefont {Magesan}\ \emph {et~al.}(2012{\natexlab{a}})\citenamefont {Magesan}, \citenamefont {Gambetta},\ and\ \citenamefont {Emerson}}]{PhysRevA.85.042311}%
  \BibitemOpen
  \bibfield  {author} {\bibinfo {author} {\bibfnamefont {E.}~\bibnamefont {Magesan}}, \bibinfo {author} {\bibfnamefont {J.~M.}\ \bibnamefont {Gambetta}},\ and\ \bibinfo {author} {\bibfnamefont {J.}~\bibnamefont {Emerson}},\ }\bibfield  {title} {\bibinfo {title} {Characterizing quantum gates via randomized benchmarking},\ }\href {https://doi.org/10.1103/PhysRevA.85.042311} {\bibfield  {journal} {\bibinfo  {journal} {Phys. Rev. A}\ }\textbf {\bibinfo {volume} {85}},\ \bibinfo {pages} {042311} (\bibinfo {year} {2012}{\natexlab{a}})}\BibitemShut {NoStop}%
\bibitem [{\citenamefont {L\'evi}\ \emph {et~al.}(2007)\citenamefont {L\'evi}, \citenamefont {L\'opez}, \citenamefont {Emerson},\ and\ \citenamefont {Cory}}]{PhysRevA.75.022314}%
  \BibitemOpen
  \bibfield  {author} {\bibinfo {author} {\bibfnamefont {B.}~\bibnamefont {L\'evi}}, \bibinfo {author} {\bibfnamefont {C.~C.}\ \bibnamefont {L\'opez}}, \bibinfo {author} {\bibfnamefont {J.}~\bibnamefont {Emerson}},\ and\ \bibinfo {author} {\bibfnamefont {D.~G.}\ \bibnamefont {Cory}},\ }\bibfield  {title} {\bibinfo {title} {Efficient error characterization in quantum information processing},\ }\href {https://doi.org/10.1103/PhysRevA.75.022314} {\bibfield  {journal} {\bibinfo  {journal} {Phys. Rev. A}\ }\textbf {\bibinfo {volume} {75}},\ \bibinfo {pages} {022314} (\bibinfo {year} {2007})}\BibitemShut {NoStop}%
\bibitem [{\citenamefont {Helsen}\ \emph {et~al.}(2022)\citenamefont {Helsen}, \citenamefont {Roth}, \citenamefont {Onorati}, \citenamefont {Werner},\ and\ \citenamefont {Eisert}}]{PRXQuantum.3.020357}%
  \BibitemOpen
  \bibfield  {author} {\bibinfo {author} {\bibfnamefont {J.}~\bibnamefont {Helsen}}, \bibinfo {author} {\bibfnamefont {I.}~\bibnamefont {Roth}}, \bibinfo {author} {\bibfnamefont {E.}~\bibnamefont {Onorati}}, \bibinfo {author} {\bibfnamefont {A.}~\bibnamefont {Werner}},\ and\ \bibinfo {author} {\bibfnamefont {J.}~\bibnamefont {Eisert}},\ }\bibfield  {title} {\bibinfo {title} {General framework for randomized benchmarking},\ }\href {https://doi.org/10.1103/PRXQuantum.3.020357} {\bibfield  {journal} {\bibinfo  {journal} {PRX Quantum}\ }\textbf {\bibinfo {volume} {3}},\ \bibinfo {pages} {020357} (\bibinfo {year} {2022})}\BibitemShut {NoStop}%
\bibitem [{\citenamefont {Hashim}\ \emph {et~al.}(2025)\citenamefont {Hashim}, \citenamefont {Nguyen}, \citenamefont {Goss}, \citenamefont {Marinelli}, \citenamefont {Naik}, \citenamefont {Chistolini}, \citenamefont {Hines}, \citenamefont {Marceaux}, \citenamefont {Kim}, \citenamefont {Gokhale}, \citenamefont {Tomesh}, \citenamefont {Chen}, \citenamefont {Jiang}, \citenamefont {Ferracin}, \citenamefont {Rudinger}, \citenamefont {Proctor}, \citenamefont {Young}, \citenamefont {Siddiqi},\ and\ \citenamefont {Blume-Kohout}}]{PRXQuantum.6.030202}%
  \BibitemOpen
  \bibfield  {author} {\bibinfo {author} {\bibfnamefont {A.}~\bibnamefont {Hashim}}, \bibinfo {author} {\bibfnamefont {L.~B.}\ \bibnamefont {Nguyen}}, \bibinfo {author} {\bibfnamefont {N.}~\bibnamefont {Goss}}, \bibinfo {author} {\bibfnamefont {B.}~\bibnamefont {Marinelli}}, \bibinfo {author} {\bibfnamefont {R.~K.}\ \bibnamefont {Naik}}, \bibinfo {author} {\bibfnamefont {T.}~\bibnamefont {Chistolini}}, \bibinfo {author} {\bibfnamefont {J.}~\bibnamefont {Hines}}, \bibinfo {author} {\bibfnamefont {J.}~\bibnamefont {Marceaux}}, \bibinfo {author} {\bibfnamefont {Y.}~\bibnamefont {Kim}}, \bibinfo {author} {\bibfnamefont {P.}~\bibnamefont {Gokhale}}, \bibinfo {author} {\bibfnamefont {T.}~\bibnamefont {Tomesh}}, \bibinfo {author} {\bibfnamefont {S.}~\bibnamefont {Chen}}, \bibinfo {author} {\bibfnamefont {L.}~\bibnamefont {Jiang}}, \bibinfo {author} {\bibfnamefont {S.}~\bibnamefont {Ferracin}}, \bibinfo {author} {\bibfnamefont {K.}~\bibnamefont {Rudinger}}, \bibinfo {author} {\bibfnamefont {T.}~\bibnamefont
  {Proctor}}, \bibinfo {author} {\bibfnamefont {K.~C.}\ \bibnamefont {Young}}, \bibinfo {author} {\bibfnamefont {I.}~\bibnamefont {Siddiqi}},\ and\ \bibinfo {author} {\bibfnamefont {R.}~\bibnamefont {Blume-Kohout}},\ }\bibfield  {title} {\bibinfo {title} {Practical introduction to benchmarking and characterization of quantum computers},\ }\href {https://doi.org/10.1103/PRXQuantum.6.030202} {\bibfield  {journal} {\bibinfo  {journal} {PRX Quantum}\ }\textbf {\bibinfo {volume} {6}},\ \bibinfo {pages} {030202} (\bibinfo {year} {2025})}\BibitemShut {NoStop}%
\bibitem [{\citenamefont {Magesan}\ \emph {et~al.}(2012{\natexlab{b}})\citenamefont {Magesan}, \citenamefont {Gambetta}, \citenamefont {Johnson}, \citenamefont {Ryan}, \citenamefont {Chow}, \citenamefont {Merkel}, \citenamefont {da~Silva}, \citenamefont {Keefe}, \citenamefont {Rothwell}, \citenamefont {Ohki}, \citenamefont {Ketchen},\ and\ \citenamefont {Steffen}}]{PhysRevLett.109.080505}%
  \BibitemOpen
  \bibfield  {author} {\bibinfo {author} {\bibfnamefont {E.}~\bibnamefont {Magesan}}, \bibinfo {author} {\bibfnamefont {J.~M.}\ \bibnamefont {Gambetta}}, \bibinfo {author} {\bibfnamefont {B.~R.}\ \bibnamefont {Johnson}}, \bibinfo {author} {\bibfnamefont {C.~A.}\ \bibnamefont {Ryan}}, \bibinfo {author} {\bibfnamefont {J.~M.}\ \bibnamefont {Chow}}, \bibinfo {author} {\bibfnamefont {S.~T.}\ \bibnamefont {Merkel}}, \bibinfo {author} {\bibfnamefont {M.~P.}\ \bibnamefont {da~Silva}}, \bibinfo {author} {\bibfnamefont {G.~A.}\ \bibnamefont {Keefe}}, \bibinfo {author} {\bibfnamefont {M.~B.}\ \bibnamefont {Rothwell}}, \bibinfo {author} {\bibfnamefont {T.~A.}\ \bibnamefont {Ohki}}, \bibinfo {author} {\bibfnamefont {M.~B.}\ \bibnamefont {Ketchen}},\ and\ \bibinfo {author} {\bibfnamefont {M.}~\bibnamefont {Steffen}},\ }\bibfield  {title} {\bibinfo {title} {Efficient measurement of quantum gate error by interleaved randomized benchmarking},\ }\href {https://doi.org/10.1103/PhysRevLett.109.080505} {\bibfield  {journal}
  {\bibinfo  {journal} {Phys. Rev. Lett.}\ }\textbf {\bibinfo {volume} {109}},\ \bibinfo {pages} {080505} (\bibinfo {year} {2012}{\natexlab{b}})}\BibitemShut {NoStop}%
\bibitem [{\citenamefont {C\'orcoles}\ \emph {et~al.}(2013)\citenamefont {C\'orcoles}, \citenamefont {Gambetta}, \citenamefont {Chow}, \citenamefont {Smolin}, \citenamefont {Ware}, \citenamefont {Strand}, \citenamefont {Plourde},\ and\ \citenamefont {Steffen}}]{PhysRevA.87.030301}%
  \BibitemOpen
  \bibfield  {author} {\bibinfo {author} {\bibfnamefont {A.~D.}\ \bibnamefont {C\'orcoles}}, \bibinfo {author} {\bibfnamefont {J.~M.}\ \bibnamefont {Gambetta}}, \bibinfo {author} {\bibfnamefont {J.~M.}\ \bibnamefont {Chow}}, \bibinfo {author} {\bibfnamefont {J.~A.}\ \bibnamefont {Smolin}}, \bibinfo {author} {\bibfnamefont {M.}~\bibnamefont {Ware}}, \bibinfo {author} {\bibfnamefont {J.}~\bibnamefont {Strand}}, \bibinfo {author} {\bibfnamefont {B.~L.~T.}\ \bibnamefont {Plourde}},\ and\ \bibinfo {author} {\bibfnamefont {M.}~\bibnamefont {Steffen}},\ }\bibfield  {title} {\bibinfo {title} {Process verification of two-qubit quantum gates by randomized benchmarking},\ }\href {https://doi.org/10.1103/PhysRevA.87.030301} {\bibfield  {journal} {\bibinfo  {journal} {Phys. Rev. A}\ }\textbf {\bibinfo {volume} {87}},\ \bibinfo {pages} {030301} (\bibinfo {year} {2013})}\BibitemShut {NoStop}%
\bibitem [{\citenamefont {Barends}\ \emph {et~al.}(2014)\citenamefont {Barends}, \citenamefont {Kelly}, \citenamefont {Megrant}, \citenamefont {Veitia}, \citenamefont {Sank}, \citenamefont {Jeffrey}, \citenamefont {White}, \citenamefont {Mutus}, \citenamefont {Fowler}, \citenamefont {Campbell}, \citenamefont {Chen}, \citenamefont {Chen}, \citenamefont {Chiaro}, \citenamefont {Dunsworth}, \citenamefont {Neill}, \citenamefont {O’Malley}, \citenamefont {Roushan}, \citenamefont {Vainsencher}, \citenamefont {Wenner}, \citenamefont {Korotkov}, \citenamefont {Cleland},\ and\ \citenamefont {Martinis}}]{Barends_2014}%
  \BibitemOpen
  \bibfield  {author} {\bibinfo {author} {\bibfnamefont {R.}~\bibnamefont {Barends}}, \bibinfo {author} {\bibfnamefont {J.}~\bibnamefont {Kelly}}, \bibinfo {author} {\bibfnamefont {A.}~\bibnamefont {Megrant}}, \bibinfo {author} {\bibfnamefont {A.}~\bibnamefont {Veitia}}, \bibinfo {author} {\bibfnamefont {D.}~\bibnamefont {Sank}}, \bibinfo {author} {\bibfnamefont {E.}~\bibnamefont {Jeffrey}}, \bibinfo {author} {\bibfnamefont {T.~C.}\ \bibnamefont {White}}, \bibinfo {author} {\bibfnamefont {J.}~\bibnamefont {Mutus}}, \bibinfo {author} {\bibfnamefont {A.~G.}\ \bibnamefont {Fowler}}, \bibinfo {author} {\bibfnamefont {B.}~\bibnamefont {Campbell}}, \bibinfo {author} {\bibfnamefont {Y.}~\bibnamefont {Chen}}, \bibinfo {author} {\bibfnamefont {Z.}~\bibnamefont {Chen}}, \bibinfo {author} {\bibfnamefont {B.}~\bibnamefont {Chiaro}}, \bibinfo {author} {\bibfnamefont {A.}~\bibnamefont {Dunsworth}}, \bibinfo {author} {\bibfnamefont {C.}~\bibnamefont {Neill}}, \bibinfo {author} {\bibfnamefont {P.}~\bibnamefont {O’Malley}},
  \bibinfo {author} {\bibfnamefont {P.}~\bibnamefont {Roushan}}, \bibinfo {author} {\bibfnamefont {A.}~\bibnamefont {Vainsencher}}, \bibinfo {author} {\bibfnamefont {J.}~\bibnamefont {Wenner}}, \bibinfo {author} {\bibfnamefont {A.~N.}\ \bibnamefont {Korotkov}}, \bibinfo {author} {\bibfnamefont {A.~N.}\ \bibnamefont {Cleland}},\ and\ \bibinfo {author} {\bibfnamefont {J.~M.}\ \bibnamefont {Martinis}},\ }\bibfield  {title} {\bibinfo {title} {Superconducting quantum circuits at the surface code threshold for fault tolerance},\ }\href {https://doi.org/10.1038/nature13171} {\bibfield  {journal} {\bibinfo  {journal} {Nature}\ }\textbf {\bibinfo {volume} {508}},\ \bibinfo {pages} {500–503} (\bibinfo {year} {2014})}\BibitemShut {NoStop}%
\bibitem [{\citenamefont {Kelly}\ \emph {et~al.}(2014)\citenamefont {Kelly}, \citenamefont {Barends}, \citenamefont {Campbell}, \citenamefont {Chen}, \citenamefont {Chen}, \citenamefont {Chiaro}, \citenamefont {Dunsworth}, \citenamefont {Fowler}, \citenamefont {Hoi}, \citenamefont {Jeffrey}, \citenamefont {Megrant}, \citenamefont {Mutus}, \citenamefont {Neill}, \citenamefont {O'Malley}, \citenamefont {Quintana}, \citenamefont {Roushan}, \citenamefont {Sank}, \citenamefont {Vainsencher}, \citenamefont {Wenner}, \citenamefont {White}, \citenamefont {Cleland},\ and\ \citenamefont {Martinis}}]{PhysRevLett.112.240504}%
  \BibitemOpen
  \bibfield  {author} {\bibinfo {author} {\bibfnamefont {J.}~\bibnamefont {Kelly}}, \bibinfo {author} {\bibfnamefont {R.}~\bibnamefont {Barends}}, \bibinfo {author} {\bibfnamefont {B.}~\bibnamefont {Campbell}}, \bibinfo {author} {\bibfnamefont {Y.}~\bibnamefont {Chen}}, \bibinfo {author} {\bibfnamefont {Z.}~\bibnamefont {Chen}}, \bibinfo {author} {\bibfnamefont {B.}~\bibnamefont {Chiaro}}, \bibinfo {author} {\bibfnamefont {A.}~\bibnamefont {Dunsworth}}, \bibinfo {author} {\bibfnamefont {A.~G.}\ \bibnamefont {Fowler}}, \bibinfo {author} {\bibfnamefont {I.-C.}\ \bibnamefont {Hoi}}, \bibinfo {author} {\bibfnamefont {E.}~\bibnamefont {Jeffrey}}, \bibinfo {author} {\bibfnamefont {A.}~\bibnamefont {Megrant}}, \bibinfo {author} {\bibfnamefont {J.}~\bibnamefont {Mutus}}, \bibinfo {author} {\bibfnamefont {C.}~\bibnamefont {Neill}}, \bibinfo {author} {\bibfnamefont {P.~J.~J.}\ \bibnamefont {O'Malley}}, \bibinfo {author} {\bibfnamefont {C.}~\bibnamefont {Quintana}}, \bibinfo {author} {\bibfnamefont {P.}~\bibnamefont
  {Roushan}}, \bibinfo {author} {\bibfnamefont {D.}~\bibnamefont {Sank}}, \bibinfo {author} {\bibfnamefont {A.}~\bibnamefont {Vainsencher}}, \bibinfo {author} {\bibfnamefont {J.}~\bibnamefont {Wenner}}, \bibinfo {author} {\bibfnamefont {T.~C.}\ \bibnamefont {White}}, \bibinfo {author} {\bibfnamefont {A.~N.}\ \bibnamefont {Cleland}},\ and\ \bibinfo {author} {\bibfnamefont {J.~M.}\ \bibnamefont {Martinis}},\ }\bibfield  {title} {\bibinfo {title} {Optimal quantum control using randomized benchmarking},\ }\href {https://doi.org/10.1103/PhysRevLett.112.240504} {\bibfield  {journal} {\bibinfo  {journal} {Phys. Rev. Lett.}\ }\textbf {\bibinfo {volume} {112}},\ \bibinfo {pages} {240504} (\bibinfo {year} {2014})}\BibitemShut {NoStop}%
\bibitem [{\citenamefont {McKay}\ \emph {et~al.}(2019)\citenamefont {McKay}, \citenamefont {Sheldon}, \citenamefont {Smolin}, \citenamefont {Chow},\ and\ \citenamefont {Gambetta}}]{PhysRevLett.122.200502}%
  \BibitemOpen
  \bibfield  {author} {\bibinfo {author} {\bibfnamefont {D.~C.}\ \bibnamefont {McKay}}, \bibinfo {author} {\bibfnamefont {S.}~\bibnamefont {Sheldon}}, \bibinfo {author} {\bibfnamefont {J.~A.}\ \bibnamefont {Smolin}}, \bibinfo {author} {\bibfnamefont {J.~M.}\ \bibnamefont {Chow}},\ and\ \bibinfo {author} {\bibfnamefont {J.~M.}\ \bibnamefont {Gambetta}},\ }\bibfield  {title} {\bibinfo {title} {Three-qubit randomized benchmarking},\ }\href {https://doi.org/10.1103/PhysRevLett.122.200502} {\bibfield  {journal} {\bibinfo  {journal} {Phys. Rev. Lett.}\ }\textbf {\bibinfo {volume} {122}},\ \bibinfo {pages} {200502} (\bibinfo {year} {2019})}\BibitemShut {NoStop}%
\bibitem [{\citenamefont {Muhonen}\ \emph {et~al.}(2015)\citenamefont {Muhonen}, \citenamefont {Laucht}, \citenamefont {Simmons}, \citenamefont {Dehollain}, \citenamefont {Kalra}, \citenamefont {Hudson}, \citenamefont {Freer}, \citenamefont {Itoh}, \citenamefont {Jamieson}, \citenamefont {McCallum}, \citenamefont {Dzurak},\ and\ \citenamefont {Morello}}]{Muhonen_2015}%
  \BibitemOpen
  \bibfield  {author} {\bibinfo {author} {\bibfnamefont {J.~T.}\ \bibnamefont {Muhonen}}, \bibinfo {author} {\bibfnamefont {A.}~\bibnamefont {Laucht}}, \bibinfo {author} {\bibfnamefont {S.}~\bibnamefont {Simmons}}, \bibinfo {author} {\bibfnamefont {J.~P.}\ \bibnamefont {Dehollain}}, \bibinfo {author} {\bibfnamefont {R.}~\bibnamefont {Kalra}}, \bibinfo {author} {\bibfnamefont {F.~E.}\ \bibnamefont {Hudson}}, \bibinfo {author} {\bibfnamefont {S.}~\bibnamefont {Freer}}, \bibinfo {author} {\bibfnamefont {K.~M.}\ \bibnamefont {Itoh}}, \bibinfo {author} {\bibfnamefont {D.~N.}\ \bibnamefont {Jamieson}}, \bibinfo {author} {\bibfnamefont {J.~C.}\ \bibnamefont {McCallum}}, \bibinfo {author} {\bibfnamefont {A.~S.}\ \bibnamefont {Dzurak}},\ and\ \bibinfo {author} {\bibfnamefont {A.}~\bibnamefont {Morello}},\ }\bibfield  {title} {\bibinfo {title} {Quantifying the quantum gate fidelity of single-atom spin qubits in silicon by randomized benchmarking},\ }\href {https://doi.org/10.1088/0953-8984/27/15/154205} {\bibfield
  {journal} {\bibinfo  {journal} {Journal of Physics: Condensed Matter}\ }\textbf {\bibinfo {volume} {27}},\ \bibinfo {pages} {154205} (\bibinfo {year} {2015})}\BibitemShut {NoStop}%
\bibitem [{\citenamefont {Kawakami}\ \emph {et~al.}(2016)\citenamefont {Kawakami}, \citenamefont {Jullien}, \citenamefont {Scarlino}, \citenamefont {Ward}, \citenamefont {Savage}, \citenamefont {Lagally}, \citenamefont {Dobrovitski}, \citenamefont {Friesen}, \citenamefont {Coppersmith}, \citenamefont {Eriksson},\ and\ \citenamefont {Vandersypen}}]{Kawakami_2016}%
  \BibitemOpen
  \bibfield  {author} {\bibinfo {author} {\bibfnamefont {E.}~\bibnamefont {Kawakami}}, \bibinfo {author} {\bibfnamefont {T.}~\bibnamefont {Jullien}}, \bibinfo {author} {\bibfnamefont {P.}~\bibnamefont {Scarlino}}, \bibinfo {author} {\bibfnamefont {D.~R.}\ \bibnamefont {Ward}}, \bibinfo {author} {\bibfnamefont {D.~E.}\ \bibnamefont {Savage}}, \bibinfo {author} {\bibfnamefont {M.~G.}\ \bibnamefont {Lagally}}, \bibinfo {author} {\bibfnamefont {V.~V.}\ \bibnamefont {Dobrovitski}}, \bibinfo {author} {\bibfnamefont {M.}~\bibnamefont {Friesen}}, \bibinfo {author} {\bibfnamefont {S.~N.}\ \bibnamefont {Coppersmith}}, \bibinfo {author} {\bibfnamefont {M.~A.}\ \bibnamefont {Eriksson}},\ and\ \bibinfo {author} {\bibfnamefont {L.~M.~K.}\ \bibnamefont {Vandersypen}},\ }\bibfield  {title} {\bibinfo {title} {Gate fidelity and coherence of an electron spin in an si/sige quantum dot with micromagnet},\ }\href {https://doi.org/10.1073/pnas.1603251113} {\bibfield  {journal} {\bibinfo  {journal} {Proceedings of the National
  Academy of Sciences}\ }\textbf {\bibinfo {volume} {113}},\ \bibinfo {pages} {11738–11743} (\bibinfo {year} {2016})}\BibitemShut {NoStop}%
\bibitem [{\citenamefont {Gaebler}\ \emph {et~al.}(2012)\citenamefont {Gaebler}, \citenamefont {Meier}, \citenamefont {Tan}, \citenamefont {Bowler}, \citenamefont {Lin}, \citenamefont {Hanneke}, \citenamefont {Jost}, \citenamefont {Home}, \citenamefont {Knill}, \citenamefont {Leibfried},\ and\ \citenamefont {Wineland}}]{PhysRevLett.108.260503}%
  \BibitemOpen
  \bibfield  {author} {\bibinfo {author} {\bibfnamefont {J.~P.}\ \bibnamefont {Gaebler}}, \bibinfo {author} {\bibfnamefont {A.~M.}\ \bibnamefont {Meier}}, \bibinfo {author} {\bibfnamefont {T.~R.}\ \bibnamefont {Tan}}, \bibinfo {author} {\bibfnamefont {R.}~\bibnamefont {Bowler}}, \bibinfo {author} {\bibfnamefont {Y.}~\bibnamefont {Lin}}, \bibinfo {author} {\bibfnamefont {D.}~\bibnamefont {Hanneke}}, \bibinfo {author} {\bibfnamefont {J.~D.}\ \bibnamefont {Jost}}, \bibinfo {author} {\bibfnamefont {J.~P.}\ \bibnamefont {Home}}, \bibinfo {author} {\bibfnamefont {E.}~\bibnamefont {Knill}}, \bibinfo {author} {\bibfnamefont {D.}~\bibnamefont {Leibfried}},\ and\ \bibinfo {author} {\bibfnamefont {D.~J.}\ \bibnamefont {Wineland}},\ }\bibfield  {title} {\bibinfo {title} {Randomized benchmarking of multiqubit gates},\ }\href {https://doi.org/10.1103/PhysRevLett.108.260503} {\bibfield  {journal} {\bibinfo  {journal} {Phys. Rev. Lett.}\ }\textbf {\bibinfo {volume} {108}},\ \bibinfo {pages} {260503} (\bibinfo {year}
  {2012})}\BibitemShut {NoStop}%
\bibitem [{\citenamefont {Manetsch}\ \emph {et~al.}(2025)\citenamefont {Manetsch}, \citenamefont {Nomura}, \citenamefont {Bataille}, \citenamefont {Lv}, \citenamefont {Leung},\ and\ \citenamefont {Endres}}]{Manetsch_2025}%
  \BibitemOpen
  \bibfield  {author} {\bibinfo {author} {\bibfnamefont {H.~J.}\ \bibnamefont {Manetsch}}, \bibinfo {author} {\bibfnamefont {G.}~\bibnamefont {Nomura}}, \bibinfo {author} {\bibfnamefont {E.}~\bibnamefont {Bataille}}, \bibinfo {author} {\bibfnamefont {X.}~\bibnamefont {Lv}}, \bibinfo {author} {\bibfnamefont {K.~H.}\ \bibnamefont {Leung}},\ and\ \bibinfo {author} {\bibfnamefont {M.}~\bibnamefont {Endres}},\ }\bibfield  {title} {\bibinfo {title} {A tweezer array with 6,100 highly coherent atomic qubits},\ }\href {https://doi.org/10.1038/s41586-025-09641-4} {\bibfield  {journal} {\bibinfo  {journal} {Nature}\ }\textbf {\bibinfo {volume} {647}},\ \bibinfo {pages} {60–67} (\bibinfo {year} {2025})}\BibitemShut {NoStop}%
\bibitem [{\citenamefont {Weinstein}\ \emph {et~al.}(2023)\citenamefont {Weinstein}, \citenamefont {Reed}, \citenamefont {Jones}, \citenamefont {Andrews}, \citenamefont {Barnes}, \citenamefont {Blumoff}, \citenamefont {Euliss}, \citenamefont {Eng}, \citenamefont {Fong}, \citenamefont {Ha}, \citenamefont {Hulbert}, \citenamefont {Jackson}, \citenamefont {Jura}, \citenamefont {Keating}, \citenamefont {Kerckhoff}, \citenamefont {Kiselev}, \citenamefont {Matten}, \citenamefont {Sabbir}, \citenamefont {Smith}, \citenamefont {Wright}, \citenamefont {Rakher}, \citenamefont {Ladd},\ and\ \citenamefont {Borselli}}]{Weinstein_2023}%
  \BibitemOpen
  \bibfield  {author} {\bibinfo {author} {\bibfnamefont {A.~J.}\ \bibnamefont {Weinstein}}, \bibinfo {author} {\bibfnamefont {M.~D.}\ \bibnamefont {Reed}}, \bibinfo {author} {\bibfnamefont {A.~M.}\ \bibnamefont {Jones}}, \bibinfo {author} {\bibfnamefont {R.~W.}\ \bibnamefont {Andrews}}, \bibinfo {author} {\bibfnamefont {D.}~\bibnamefont {Barnes}}, \bibinfo {author} {\bibfnamefont {J.~Z.}\ \bibnamefont {Blumoff}}, \bibinfo {author} {\bibfnamefont {L.~E.}\ \bibnamefont {Euliss}}, \bibinfo {author} {\bibfnamefont {K.}~\bibnamefont {Eng}}, \bibinfo {author} {\bibfnamefont {B.~H.}\ \bibnamefont {Fong}}, \bibinfo {author} {\bibfnamefont {S.~D.}\ \bibnamefont {Ha}}, \bibinfo {author} {\bibfnamefont {D.~R.}\ \bibnamefont {Hulbert}}, \bibinfo {author} {\bibfnamefont {C.~A.~C.}\ \bibnamefont {Jackson}}, \bibinfo {author} {\bibfnamefont {M.}~\bibnamefont {Jura}}, \bibinfo {author} {\bibfnamefont {T.~E.}\ \bibnamefont {Keating}}, \bibinfo {author} {\bibfnamefont {J.}~\bibnamefont {Kerckhoff}}, \bibinfo {author}
  {\bibfnamefont {A.~A.}\ \bibnamefont {Kiselev}}, \bibinfo {author} {\bibfnamefont {J.}~\bibnamefont {Matten}}, \bibinfo {author} {\bibfnamefont {G.}~\bibnamefont {Sabbir}}, \bibinfo {author} {\bibfnamefont {A.}~\bibnamefont {Smith}}, \bibinfo {author} {\bibfnamefont {J.}~\bibnamefont {Wright}}, \bibinfo {author} {\bibfnamefont {M.~T.}\ \bibnamefont {Rakher}}, \bibinfo {author} {\bibfnamefont {T.~D.}\ \bibnamefont {Ladd}},\ and\ \bibinfo {author} {\bibfnamefont {M.~G.}\ \bibnamefont {Borselli}},\ }\bibfield  {title} {\bibinfo {title} {Universal logic with encoded spin qubits in silicon},\ }\href {https://doi.org/10.1038/s41586-023-05777-3} {\bibfield  {journal} {\bibinfo  {journal} {Nature}\ }\textbf {\bibinfo {volume} {615}},\ \bibinfo {pages} {817–822} (\bibinfo {year} {2023})}\BibitemShut {NoStop}%
\bibitem [{\citenamefont {Moses}\ \emph {et~al.}(2023)\citenamefont {Moses}, \citenamefont {Baldwin}, \citenamefont {Allman}, \citenamefont {Ancona}, \citenamefont {Ascarrunz}, \citenamefont {Barnes}, \citenamefont {Bartolotta}, \citenamefont {Bjork}, \citenamefont {Blanchard}, \citenamefont {Bohn}, \citenamefont {Bohnet}, \citenamefont {Brown}, \citenamefont {Burdick}, \citenamefont {Burton}, \citenamefont {Campbell}, \citenamefont {Campora}, \citenamefont {Carron}, \citenamefont {Chambers}, \citenamefont {Chan}, \citenamefont {Chen}, \citenamefont {Chernoguzov}, \citenamefont {Chertkov}, \citenamefont {Colina}, \citenamefont {Curtis}, \citenamefont {Daniel}, \citenamefont {DeCross}, \citenamefont {Deen}, \citenamefont {Delaney}, \citenamefont {Dreiling}, \citenamefont {Ertsgaard}, \citenamefont {Esposito}, \citenamefont {Estey}, \citenamefont {Fabrikant}, \citenamefont {Figgatt}, \citenamefont {Foltz}, \citenamefont {Foss-Feig}, \citenamefont {Francois}, \citenamefont {Gaebler}, \citenamefont {Gatterman},
  \citenamefont {Gilbreth}, \citenamefont {Giles}, \citenamefont {Glynn}, \citenamefont {Hall}, \citenamefont {Hankin}, \citenamefont {Hansen}, \citenamefont {Hayes}, \citenamefont {Higashi}, \citenamefont {Hoffman}, \citenamefont {Horning}, \citenamefont {Hout}, \citenamefont {Jacobs}, \citenamefont {Johansen}, \citenamefont {Jones}, \citenamefont {Karcz}, \citenamefont {Klein}, \citenamefont {Lauria}, \citenamefont {Lee}, \citenamefont {Liefer}, \citenamefont {Lu}, \citenamefont {Lucchetti}, \citenamefont {Lytle}, \citenamefont {Malm}, \citenamefont {Matheny}, \citenamefont {Mathewson}, \citenamefont {Mayer}, \citenamefont {Miller}, \citenamefont {Mills}, \citenamefont {Neyenhuis}, \citenamefont {Nugent}, \citenamefont {Olson}, \citenamefont {Parks}, \citenamefont {Price}, \citenamefont {Price}, \citenamefont {Pugh}, \citenamefont {Ransford}, \citenamefont {Reed}, \citenamefont {Roman}, \citenamefont {Rowe}, \citenamefont {Ryan-Anderson}, \citenamefont {Sanders}, \citenamefont {Sedlacek}, \citenamefont
  {Shevchuk}, \citenamefont {Siegfried}, \citenamefont {Skripka}, \citenamefont {Spaun}, \citenamefont {Sprenkle}, \citenamefont {Stutz}, \citenamefont {Swallows}, \citenamefont {Tobey}, \citenamefont {Tran}, \citenamefont {Tran}, \citenamefont {Vogt}, \citenamefont {Volin}, \citenamefont {Walker}, \citenamefont {Zolot},\ and\ \citenamefont {Pino}}]{PhysRevX.13.041052}%
  \BibitemOpen
  \bibfield  {author} {\bibinfo {author} {\bibfnamefont {S.~A.}\ \bibnamefont {Moses}}, \bibinfo {author} {\bibfnamefont {C.~H.}\ \bibnamefont {Baldwin}}, \bibinfo {author} {\bibfnamefont {M.~S.}\ \bibnamefont {Allman}}, \bibinfo {author} {\bibfnamefont {R.}~\bibnamefont {Ancona}}, \bibinfo {author} {\bibfnamefont {L.}~\bibnamefont {Ascarrunz}}, \bibinfo {author} {\bibfnamefont {C.}~\bibnamefont {Barnes}}, \bibinfo {author} {\bibfnamefont {J.}~\bibnamefont {Bartolotta}}, \bibinfo {author} {\bibfnamefont {B.}~\bibnamefont {Bjork}}, \bibinfo {author} {\bibfnamefont {P.}~\bibnamefont {Blanchard}}, \bibinfo {author} {\bibfnamefont {M.}~\bibnamefont {Bohn}}, \bibinfo {author} {\bibfnamefont {J.~G.}\ \bibnamefont {Bohnet}}, \bibinfo {author} {\bibfnamefont {N.~C.}\ \bibnamefont {Brown}}, \bibinfo {author} {\bibfnamefont {N.~Q.}\ \bibnamefont {Burdick}}, \bibinfo {author} {\bibfnamefont {W.~C.}\ \bibnamefont {Burton}}, \bibinfo {author} {\bibfnamefont {S.~L.}\ \bibnamefont {Campbell}}, \bibinfo {author}
  {\bibfnamefont {J.~P.}\ \bibnamefont {Campora}}, \bibinfo {author} {\bibfnamefont {C.}~\bibnamefont {Carron}}, \bibinfo {author} {\bibfnamefont {J.}~\bibnamefont {Chambers}}, \bibinfo {author} {\bibfnamefont {J.~W.}\ \bibnamefont {Chan}}, \bibinfo {author} {\bibfnamefont {Y.~H.}\ \bibnamefont {Chen}}, \bibinfo {author} {\bibfnamefont {A.}~\bibnamefont {Chernoguzov}}, \bibinfo {author} {\bibfnamefont {E.}~\bibnamefont {Chertkov}}, \bibinfo {author} {\bibfnamefont {J.}~\bibnamefont {Colina}}, \bibinfo {author} {\bibfnamefont {J.~P.}\ \bibnamefont {Curtis}}, \bibinfo {author} {\bibfnamefont {R.}~\bibnamefont {Daniel}}, \bibinfo {author} {\bibfnamefont {M.}~\bibnamefont {DeCross}}, \bibinfo {author} {\bibfnamefont {D.}~\bibnamefont {Deen}}, \bibinfo {author} {\bibfnamefont {C.}~\bibnamefont {Delaney}}, \bibinfo {author} {\bibfnamefont {J.~M.}\ \bibnamefont {Dreiling}}, \bibinfo {author} {\bibfnamefont {C.~T.}\ \bibnamefont {Ertsgaard}}, \bibinfo {author} {\bibfnamefont {J.}~\bibnamefont {Esposito}}, \bibinfo
  {author} {\bibfnamefont {B.}~\bibnamefont {Estey}}, \bibinfo {author} {\bibfnamefont {M.}~\bibnamefont {Fabrikant}}, \bibinfo {author} {\bibfnamefont {C.}~\bibnamefont {Figgatt}}, \bibinfo {author} {\bibfnamefont {C.}~\bibnamefont {Foltz}}, \bibinfo {author} {\bibfnamefont {M.}~\bibnamefont {Foss-Feig}}, \bibinfo {author} {\bibfnamefont {D.}~\bibnamefont {Francois}}, \bibinfo {author} {\bibfnamefont {J.~P.}\ \bibnamefont {Gaebler}}, \bibinfo {author} {\bibfnamefont {T.~M.}\ \bibnamefont {Gatterman}}, \bibinfo {author} {\bibfnamefont {C.~N.}\ \bibnamefont {Gilbreth}}, \bibinfo {author} {\bibfnamefont {J.}~\bibnamefont {Giles}}, \bibinfo {author} {\bibfnamefont {E.}~\bibnamefont {Glynn}}, \bibinfo {author} {\bibfnamefont {A.}~\bibnamefont {Hall}}, \bibinfo {author} {\bibfnamefont {A.~M.}\ \bibnamefont {Hankin}}, \bibinfo {author} {\bibfnamefont {A.}~\bibnamefont {Hansen}}, \bibinfo {author} {\bibfnamefont {D.}~\bibnamefont {Hayes}}, \bibinfo {author} {\bibfnamefont {B.}~\bibnamefont {Higashi}}, \bibinfo
  {author} {\bibfnamefont {I.~M.}\ \bibnamefont {Hoffman}}, \bibinfo {author} {\bibfnamefont {B.}~\bibnamefont {Horning}}, \bibinfo {author} {\bibfnamefont {J.~J.}\ \bibnamefont {Hout}}, \bibinfo {author} {\bibfnamefont {R.}~\bibnamefont {Jacobs}}, \bibinfo {author} {\bibfnamefont {J.}~\bibnamefont {Johansen}}, \bibinfo {author} {\bibfnamefont {L.}~\bibnamefont {Jones}}, \bibinfo {author} {\bibfnamefont {J.}~\bibnamefont {Karcz}}, \bibinfo {author} {\bibfnamefont {T.}~\bibnamefont {Klein}}, \bibinfo {author} {\bibfnamefont {P.}~\bibnamefont {Lauria}}, \bibinfo {author} {\bibfnamefont {P.}~\bibnamefont {Lee}}, \bibinfo {author} {\bibfnamefont {D.}~\bibnamefont {Liefer}}, \bibinfo {author} {\bibfnamefont {S.~T.}\ \bibnamefont {Lu}}, \bibinfo {author} {\bibfnamefont {D.}~\bibnamefont {Lucchetti}}, \bibinfo {author} {\bibfnamefont {C.}~\bibnamefont {Lytle}}, \bibinfo {author} {\bibfnamefont {A.}~\bibnamefont {Malm}}, \bibinfo {author} {\bibfnamefont {M.}~\bibnamefont {Matheny}}, \bibinfo {author} {\bibfnamefont
  {B.}~\bibnamefont {Mathewson}}, \bibinfo {author} {\bibfnamefont {K.}~\bibnamefont {Mayer}}, \bibinfo {author} {\bibfnamefont {D.~B.}\ \bibnamefont {Miller}}, \bibinfo {author} {\bibfnamefont {M.}~\bibnamefont {Mills}}, \bibinfo {author} {\bibfnamefont {B.}~\bibnamefont {Neyenhuis}}, \bibinfo {author} {\bibfnamefont {L.}~\bibnamefont {Nugent}}, \bibinfo {author} {\bibfnamefont {S.}~\bibnamefont {Olson}}, \bibinfo {author} {\bibfnamefont {J.}~\bibnamefont {Parks}}, \bibinfo {author} {\bibfnamefont {G.~N.}\ \bibnamefont {Price}}, \bibinfo {author} {\bibfnamefont {Z.}~\bibnamefont {Price}}, \bibinfo {author} {\bibfnamefont {M.}~\bibnamefont {Pugh}}, \bibinfo {author} {\bibfnamefont {A.}~\bibnamefont {Ransford}}, \bibinfo {author} {\bibfnamefont {A.~P.}\ \bibnamefont {Reed}}, \bibinfo {author} {\bibfnamefont {C.}~\bibnamefont {Roman}}, \bibinfo {author} {\bibfnamefont {M.}~\bibnamefont {Rowe}}, \bibinfo {author} {\bibfnamefont {C.}~\bibnamefont {Ryan-Anderson}}, \bibinfo {author} {\bibfnamefont
  {S.}~\bibnamefont {Sanders}}, \bibinfo {author} {\bibfnamefont {J.}~\bibnamefont {Sedlacek}}, \bibinfo {author} {\bibfnamefont {P.}~\bibnamefont {Shevchuk}}, \bibinfo {author} {\bibfnamefont {P.}~\bibnamefont {Siegfried}}, \bibinfo {author} {\bibfnamefont {T.}~\bibnamefont {Skripka}}, \bibinfo {author} {\bibfnamefont {B.}~\bibnamefont {Spaun}}, \bibinfo {author} {\bibfnamefont {R.~T.}\ \bibnamefont {Sprenkle}}, \bibinfo {author} {\bibfnamefont {R.~P.}\ \bibnamefont {Stutz}}, \bibinfo {author} {\bibfnamefont {M.}~\bibnamefont {Swallows}}, \bibinfo {author} {\bibfnamefont {R.~I.}\ \bibnamefont {Tobey}}, \bibinfo {author} {\bibfnamefont {A.}~\bibnamefont {Tran}}, \bibinfo {author} {\bibfnamefont {T.}~\bibnamefont {Tran}}, \bibinfo {author} {\bibfnamefont {E.}~\bibnamefont {Vogt}}, \bibinfo {author} {\bibfnamefont {C.}~\bibnamefont {Volin}}, \bibinfo {author} {\bibfnamefont {J.}~\bibnamefont {Walker}}, \bibinfo {author} {\bibfnamefont {A.~M.}\ \bibnamefont {Zolot}},\ and\ \bibinfo {author} {\bibfnamefont
  {J.~M.}\ \bibnamefont {Pino}},\ }\bibfield  {title} {\bibinfo {title} {A race-track trapped-ion quantum processor},\ }\href {https://doi.org/10.1103/PhysRevX.13.041052} {\bibfield  {journal} {\bibinfo  {journal} {Phys. Rev. X}\ }\textbf {\bibinfo {volume} {13}},\ \bibinfo {pages} {041052} (\bibinfo {year} {2023})}\BibitemShut {NoStop}%
\bibitem [{\citenamefont {Li}\ \emph {et~al.}(2023)\citenamefont {Li}, \citenamefont {Liu}, \citenamefont {Zhao}, \citenamefont {Mi}, \citenamefont {Xu}, \citenamefont {Liang}, \citenamefont {Su}, \citenamefont {Sun}, \citenamefont {Xue}, \citenamefont {Zhang}, \citenamefont {Liu}, \citenamefont {Jin},\ and\ \citenamefont {Yu}}]{Li_2023}%
  \BibitemOpen
  \bibfield  {author} {\bibinfo {author} {\bibfnamefont {Z.}~\bibnamefont {Li}}, \bibinfo {author} {\bibfnamefont {P.}~\bibnamefont {Liu}}, \bibinfo {author} {\bibfnamefont {P.}~\bibnamefont {Zhao}}, \bibinfo {author} {\bibfnamefont {Z.}~\bibnamefont {Mi}}, \bibinfo {author} {\bibfnamefont {H.}~\bibnamefont {Xu}}, \bibinfo {author} {\bibfnamefont {X.}~\bibnamefont {Liang}}, \bibinfo {author} {\bibfnamefont {T.}~\bibnamefont {Su}}, \bibinfo {author} {\bibfnamefont {W.}~\bibnamefont {Sun}}, \bibinfo {author} {\bibfnamefont {G.}~\bibnamefont {Xue}}, \bibinfo {author} {\bibfnamefont {J.-N.}\ \bibnamefont {Zhang}}, \bibinfo {author} {\bibfnamefont {W.}~\bibnamefont {Liu}}, \bibinfo {author} {\bibfnamefont {Y.}~\bibnamefont {Jin}},\ and\ \bibinfo {author} {\bibfnamefont {H.}~\bibnamefont {Yu}},\ }\bibfield  {title} {\bibinfo {title} {Error per single-qubit gate below 10−4 in a superconducting qubit},\ }\bibfield  {journal} {\bibinfo  {journal} {npj Quantum Information}\ }\textbf {\bibinfo {volume} {9}},\ \href
  {https://doi.org/10.1038/s41534-023-00781-x} {10.1038/s41534-023-00781-x} (\bibinfo {year} {2023})\BibitemShut {NoStop}%
\bibitem [{\citenamefont {Wu}\ \emph {et~al.}()\citenamefont {Wu}, \citenamefont {Camenzind}, \citenamefont {B{\"u}tler}, \citenamefont {Jin}, \citenamefont {Noiri}, \citenamefont {Takeda}, \citenamefont {Nakajima}, \citenamefont {Kobayashi}, \citenamefont {Scappucci}, \citenamefont {Goan} \emph {et~al.}}]{wu2507simultaneous}%
  \BibitemOpen
  \bibfield  {author} {\bibinfo {author} {\bibfnamefont {Y.}~\bibnamefont {Wu}}, \bibinfo {author} {\bibfnamefont {L.}~\bibnamefont {Camenzind}}, \bibinfo {author} {\bibfnamefont {P.}~\bibnamefont {B{\"u}tler}}, \bibinfo {author} {\bibfnamefont {I.}~\bibnamefont {Jin}}, \bibinfo {author} {\bibfnamefont {A.}~\bibnamefont {Noiri}}, \bibinfo {author} {\bibfnamefont {K.}~\bibnamefont {Takeda}}, \bibinfo {author} {\bibfnamefont {T.}~\bibnamefont {Nakajima}}, \bibinfo {author} {\bibfnamefont {T.}~\bibnamefont {Kobayashi}}, \bibinfo {author} {\bibfnamefont {G.}~\bibnamefont {Scappucci}}, \bibinfo {author} {\bibfnamefont {H.}~\bibnamefont {Goan}}, \emph {et~al.},\ }\bibfield  {title} {\bibinfo {title} {Simultaneous high-fidelity single-qubit gates in a spin qubit array (2025)},\ }\href {https://arxiv.org/abs/2507.11918} {\bibinfo  {journal} {arXiv preprint arXiv:2507.11918}\ }\BibitemShut {NoStop}%
\bibitem [{\citenamefont {Song}\ \emph {et~al.}(2025)\citenamefont {Song}, \citenamefont {Beltr\'an}, \citenamefont {Besedin}, \citenamefont {Kerschbaum}, \citenamefont {Pechal}, \citenamefont {Swiadek}, \citenamefont {Hellings}, \citenamefont {Colao~Zanuz}, \citenamefont {Flasby}, \citenamefont {Besse},\ and\ \citenamefont {Wallraff}}]{q418-pydy}%
  \BibitemOpen
\bibfield  {journal} {  }\bibfield  {author} {\bibinfo {author} {\bibfnamefont {Y.}~\bibnamefont {Song}}, \bibinfo {author} {\bibfnamefont {L.}~\bibnamefont {Beltr\'an}}, \bibinfo {author} {\bibfnamefont {I.}~\bibnamefont {Besedin}}, \bibinfo {author} {\bibfnamefont {M.}~\bibnamefont {Kerschbaum}}, \bibinfo {author} {\bibfnamefont {M.}~\bibnamefont {Pechal}}, \bibinfo {author} {\bibfnamefont {F.~m.~c.}\ \bibnamefont {Swiadek}}, \bibinfo {author} {\bibfnamefont {C.}~\bibnamefont {Hellings}}, \bibinfo {author} {\bibfnamefont {D.}~\bibnamefont {Colao~Zanuz}}, \bibinfo {author} {\bibfnamefont {A.}~\bibnamefont {Flasby}}, \bibinfo {author} {\bibfnamefont {J.-C.}\ \bibnamefont {Besse}},\ and\ \bibinfo {author} {\bibfnamefont {A.}~\bibnamefont {Wallraff}},\ }\bibfield  {title} {\bibinfo {title} {Constant-depth fan-out with real-time feedforward on a superconducting quantum processor},\ }\href {https://doi.org/10.1103/q418-pydy} {\bibfield  {journal} {\bibinfo  {journal} {Phys. Rev. Appl.}\ }\textbf {\bibinfo
  {volume} {24}},\ \bibinfo {pages} {024068} (\bibinfo {year} {2025})}\BibitemShut {NoStop}%
\bibitem [{\citenamefont {Champion}\ \emph {et~al.}(2025)\citenamefont {Champion}, \citenamefont {Wang}, \citenamefont {Parker},\ and\ \citenamefont {Blok}}]{vbh4-lysv}%
  \BibitemOpen
  \bibfield  {author} {\bibinfo {author} {\bibfnamefont {E.}~\bibnamefont {Champion}}, \bibinfo {author} {\bibfnamefont {Z.}~\bibnamefont {Wang}}, \bibinfo {author} {\bibfnamefont {R.~W.}\ \bibnamefont {Parker}},\ and\ \bibinfo {author} {\bibfnamefont {M.~S.}\ \bibnamefont {Blok}},\ }\bibfield  {title} {\bibinfo {title} {Efficient control of a transmon qudit using effective spin-$7/2$ rotations},\ }\href {https://doi.org/10.1103/vbh4-lysv} {\bibfield  {journal} {\bibinfo  {journal} {Phys. Rev. X}\ }\textbf {\bibinfo {volume} {15}},\ \bibinfo {pages} {021096} (\bibinfo {year} {2025})}\BibitemShut {NoStop}%
\bibitem [{\citenamefont {Norris}\ \emph {et~al.}(2025)\citenamefont {Norris}, \citenamefont {Dalton}, \citenamefont {Zanuz}, \citenamefont {Rommens}, \citenamefont {Flasby}, \citenamefont {Panah}, \citenamefont {Swiadek}, \citenamefont {Scarato}, \citenamefont {Hellings}, \citenamefont {Besse} \emph {et~al.}}]{norris2025performance}%
  \BibitemOpen
  \bibfield  {author} {\bibinfo {author} {\bibfnamefont {G.~J.}\ \bibnamefont {Norris}}, \bibinfo {author} {\bibfnamefont {K.}~\bibnamefont {Dalton}}, \bibinfo {author} {\bibfnamefont {D.~C.}\ \bibnamefont {Zanuz}}, \bibinfo {author} {\bibfnamefont {A.}~\bibnamefont {Rommens}}, \bibinfo {author} {\bibfnamefont {A.}~\bibnamefont {Flasby}}, \bibinfo {author} {\bibfnamefont {M.~B.}\ \bibnamefont {Panah}}, \bibinfo {author} {\bibfnamefont {F.}~\bibnamefont {Swiadek}}, \bibinfo {author} {\bibfnamefont {C.}~\bibnamefont {Scarato}}, \bibinfo {author} {\bibfnamefont {C.}~\bibnamefont {Hellings}}, \bibinfo {author} {\bibfnamefont {J.-C.}\ \bibnamefont {Besse}}, \emph {et~al.},\ }\bibfield  {title} {\bibinfo {title} {Performance characterization of a multi-module quantum processor with static inter-chip couplers},\ }\href {https://arxiv.org/abs/2503.12603} {\bibfield  {journal} {\bibinfo  {journal} {arXiv preprint arXiv:2503.12603}\ } (\bibinfo {year} {2025})}\BibitemShut {NoStop}%
\bibitem [{\citenamefont {Scarato}\ \emph {et~al.}(2025)\citenamefont {Scarato}, \citenamefont {Hanke}, \citenamefont {Remm}, \citenamefont {Laz\ifmmode~\u{a}\else \u{a}\fi{}r}, \citenamefont {Lacroix}, \citenamefont {Colao~Zanuz}, \citenamefont {Flasby}, \citenamefont {Wallraff},\ and\ \citenamefont {Hellings}}]{h7cv-xgw2}%
  \BibitemOpen
  \bibfield  {author} {\bibinfo {author} {\bibfnamefont {C.}~\bibnamefont {Scarato}}, \bibinfo {author} {\bibfnamefont {K.}~\bibnamefont {Hanke}}, \bibinfo {author} {\bibfnamefont {A.}~\bibnamefont {Remm}}, \bibinfo {author} {\bibfnamefont {S.}~\bibnamefont {Laz\ifmmode~\u{a}\else \u{a}\fi{}r}}, \bibinfo {author} {\bibfnamefont {N.}~\bibnamefont {Lacroix}}, \bibinfo {author} {\bibfnamefont {D.}~\bibnamefont {Colao~Zanuz}}, \bibinfo {author} {\bibfnamefont {A.}~\bibnamefont {Flasby}}, \bibinfo {author} {\bibfnamefont {A.}~\bibnamefont {Wallraff}},\ and\ \bibinfo {author} {\bibfnamefont {C.}~\bibnamefont {Hellings}},\ }\bibfield  {title} {\bibinfo {title} {Realizing a continuous set of two-qubit gates parameterized by an idle time},\ }\href {https://doi.org/10.1103/h7cv-xgw2} {\bibfield  {journal} {\bibinfo  {journal} {PRX Quantum}\ }\textbf {\bibinfo {volume} {6}},\ \bibinfo {pages} {040317} (\bibinfo {year} {2025})}\BibitemShut {NoStop}%
\bibitem [{\citenamefont {Sung}\ \emph {et~al.}(2021)\citenamefont {Sung}, \citenamefont {Ding}, \citenamefont {Braum\"uller}, \citenamefont {Veps\"al\"ainen}, \citenamefont {Kannan}, \citenamefont {Kjaergaard}, \citenamefont {Greene}, \citenamefont {Samach}, \citenamefont {McNally}, \citenamefont {Kim}, \citenamefont {Melville}, \citenamefont {Niedzielski}, \citenamefont {Schwartz}, \citenamefont {Yoder}, \citenamefont {Orlando}, \citenamefont {Gustavsson},\ and\ \citenamefont {Oliver}}]{PhysRevX.11.021058}%
  \BibitemOpen
  \bibfield  {author} {\bibinfo {author} {\bibfnamefont {Y.}~\bibnamefont {Sung}}, \bibinfo {author} {\bibfnamefont {L.}~\bibnamefont {Ding}}, \bibinfo {author} {\bibfnamefont {J.}~\bibnamefont {Braum\"uller}}, \bibinfo {author} {\bibfnamefont {A.}~\bibnamefont {Veps\"al\"ainen}}, \bibinfo {author} {\bibfnamefont {B.}~\bibnamefont {Kannan}}, \bibinfo {author} {\bibfnamefont {M.}~\bibnamefont {Kjaergaard}}, \bibinfo {author} {\bibfnamefont {A.}~\bibnamefont {Greene}}, \bibinfo {author} {\bibfnamefont {G.~O.}\ \bibnamefont {Samach}}, \bibinfo {author} {\bibfnamefont {C.}~\bibnamefont {McNally}}, \bibinfo {author} {\bibfnamefont {D.}~\bibnamefont {Kim}}, \bibinfo {author} {\bibfnamefont {A.}~\bibnamefont {Melville}}, \bibinfo {author} {\bibfnamefont {B.~M.}\ \bibnamefont {Niedzielski}}, \bibinfo {author} {\bibfnamefont {M.~E.}\ \bibnamefont {Schwartz}}, \bibinfo {author} {\bibfnamefont {J.~L.}\ \bibnamefont {Yoder}}, \bibinfo {author} {\bibfnamefont {T.~P.}\ \bibnamefont {Orlando}}, \bibinfo {author}
  {\bibfnamefont {S.}~\bibnamefont {Gustavsson}},\ and\ \bibinfo {author} {\bibfnamefont {W.~D.}\ \bibnamefont {Oliver}},\ }\bibfield  {title} {\bibinfo {title} {Realization of high-fidelity cz and $zz$-free iswap gates with a tunable coupler},\ }\href {https://doi.org/10.1103/PhysRevX.11.021058} {\bibfield  {journal} {\bibinfo  {journal} {Phys. Rev. X}\ }\textbf {\bibinfo {volume} {11}},\ \bibinfo {pages} {021058} (\bibinfo {year} {2021})}\BibitemShut {NoStop}%
\bibitem [{\citenamefont {Erhard}\ \emph {et~al.}(2019)\citenamefont {Erhard}, \citenamefont {Wallman}, \citenamefont {Postler}, \citenamefont {Meth}, \citenamefont {Stricker}, \citenamefont {Martinez}, \citenamefont {Schindler}, \citenamefont {Monz}, \citenamefont {Emerson},\ and\ \citenamefont {Blatt}}]{Erhard_2019}%
  \BibitemOpen
  \bibfield  {author} {\bibinfo {author} {\bibfnamefont {A.}~\bibnamefont {Erhard}}, \bibinfo {author} {\bibfnamefont {J.~J.}\ \bibnamefont {Wallman}}, \bibinfo {author} {\bibfnamefont {L.}~\bibnamefont {Postler}}, \bibinfo {author} {\bibfnamefont {M.}~\bibnamefont {Meth}}, \bibinfo {author} {\bibfnamefont {R.}~\bibnamefont {Stricker}}, \bibinfo {author} {\bibfnamefont {E.~A.}\ \bibnamefont {Martinez}}, \bibinfo {author} {\bibfnamefont {P.}~\bibnamefont {Schindler}}, \bibinfo {author} {\bibfnamefont {T.}~\bibnamefont {Monz}}, \bibinfo {author} {\bibfnamefont {J.}~\bibnamefont {Emerson}},\ and\ \bibinfo {author} {\bibfnamefont {R.}~\bibnamefont {Blatt}},\ }\bibfield  {title} {\bibinfo {title} {Characterizing large-scale quantum computers via cycle benchmarking},\ }\bibfield  {journal} {\bibinfo  {journal} {Nature Communications}\ }\textbf {\bibinfo {volume} {10}},\ \href {https://doi.org/10.1038/s41467-019-13068-7} {10.1038/s41467-019-13068-7} (\bibinfo {year} {2019})\BibitemShut {NoStop}%
\bibitem [{\citenamefont {Emerson}\ \emph {et~al.}(2007)\citenamefont {Emerson}, \citenamefont {Silva}, \citenamefont {Moussa}, \citenamefont {Ryan}, \citenamefont {Laforest}, \citenamefont {Baugh}, \citenamefont {Cory},\ and\ \citenamefont {Laflamme}}]{Emerson_2007}%
  \BibitemOpen
  \bibfield  {author} {\bibinfo {author} {\bibfnamefont {J.}~\bibnamefont {Emerson}}, \bibinfo {author} {\bibfnamefont {M.}~\bibnamefont {Silva}}, \bibinfo {author} {\bibfnamefont {O.}~\bibnamefont {Moussa}}, \bibinfo {author} {\bibfnamefont {C.}~\bibnamefont {Ryan}}, \bibinfo {author} {\bibfnamefont {M.}~\bibnamefont {Laforest}}, \bibinfo {author} {\bibfnamefont {J.}~\bibnamefont {Baugh}}, \bibinfo {author} {\bibfnamefont {D.~G.}\ \bibnamefont {Cory}},\ and\ \bibinfo {author} {\bibfnamefont {R.}~\bibnamefont {Laflamme}},\ }\bibfield  {title} {\bibinfo {title} {Symmetrized characterization of noisy quantum processes},\ }\href {https://doi.org/10.1126/science.1145699} {\bibfield  {journal} {\bibinfo  {journal} {Science}\ }\textbf {\bibinfo {volume} {317}},\ \bibinfo {pages} {1893–1896} (\bibinfo {year} {2007})}\BibitemShut {NoStop}%
\bibitem [{\citenamefont {McKay}\ \emph {et~al.}()\citenamefont {McKay}, \citenamefont {Hincks}, \citenamefont {Pritchett}, \citenamefont {Carroll}, \citenamefont {Govia},\ and\ \citenamefont {Merkel}}]{mckay2311benchmarking}%
  \BibitemOpen
  \bibfield  {author} {\bibinfo {author} {\bibfnamefont {D.~C.}\ \bibnamefont {McKay}}, \bibinfo {author} {\bibfnamefont {I.}~\bibnamefont {Hincks}}, \bibinfo {author} {\bibfnamefont {E.~J.}\ \bibnamefont {Pritchett}}, \bibinfo {author} {\bibfnamefont {M.}~\bibnamefont {Carroll}}, \bibinfo {author} {\bibfnamefont {L.~C.}\ \bibnamefont {Govia}},\ and\ \bibinfo {author} {\bibfnamefont {S.~T.}\ \bibnamefont {Merkel}},\ }\bibfield  {title} {\bibinfo {title} {Benchmarking quantum processor performance at scale, november 2023},\ }\href {https://arxiv.org/abs/2311.05933} {\bibinfo  {journal} {arxiv:2311.05933}\ }\BibitemShut {NoStop}%
\bibitem [{\citenamefont {Carignan-Dugas}\ \emph {et~al.}(2023)\citenamefont {Carignan-Dugas}, \citenamefont {Dahlen}, \citenamefont {Hincks}, \citenamefont {Ospadov}, \citenamefont {Beale}, \citenamefont {Ferracin}, \citenamefont {Skanes-Norman}, \citenamefont {Emerson},\ and\ \citenamefont {Wallman}}]{carignan2023error}%
  \BibitemOpen
\bibfield  {journal} {  }\bibfield  {author} {\bibinfo {author} {\bibfnamefont {A.}~\bibnamefont {Carignan-Dugas}}, \bibinfo {author} {\bibfnamefont {D.}~\bibnamefont {Dahlen}}, \bibinfo {author} {\bibfnamefont {I.}~\bibnamefont {Hincks}}, \bibinfo {author} {\bibfnamefont {E.}~\bibnamefont {Ospadov}}, \bibinfo {author} {\bibfnamefont {S.~J.}\ \bibnamefont {Beale}}, \bibinfo {author} {\bibfnamefont {S.}~\bibnamefont {Ferracin}}, \bibinfo {author} {\bibfnamefont {J.}~\bibnamefont {Skanes-Norman}}, \bibinfo {author} {\bibfnamefont {J.}~\bibnamefont {Emerson}},\ and\ \bibinfo {author} {\bibfnamefont {J.~J.}\ \bibnamefont {Wallman}},\ }\bibfield  {title} {\bibinfo {title} {The error reconstruction and compiled calibration of quantum computing cycles},\ }\href {https://arxiv.org/abs/2303.17714} {\bibfield  {journal} {\bibinfo  {journal} {arXiv preprint arXiv:2303.17714}\ } (\bibinfo {year} {2023})}\BibitemShut {NoStop}%
\bibitem [{\citenamefont {Beale}\ \emph {et~al.}(2020)\citenamefont {Beale}, \citenamefont {Boone}, \citenamefont {Carignan-Dugas}, \citenamefont {Chytros}, \citenamefont {Dahlen}, \citenamefont {Dawkins}, \citenamefont {Emerson}, \citenamefont {Ferracin}, \citenamefont {Frey}, \citenamefont {Hincks}, \citenamefont {Hufnagel}, \citenamefont {Iyer}, \citenamefont {Jain}, \citenamefont {Kolbush}, \citenamefont {Ospadov}, \citenamefont {Pino}, \citenamefont {Qassim}, \citenamefont {Saunders}, \citenamefont {Skanes-Norman}, \citenamefont {Stasiuk}, \citenamefont {Wallman}, \citenamefont {Winick},\ and\ \citenamefont {Wright}}]{beale_2020_3945250}%
  \BibitemOpen
  \bibfield  {author} {\bibinfo {author} {\bibfnamefont {S.~J.}\ \bibnamefont {Beale}}, \bibinfo {author} {\bibfnamefont {K.}~\bibnamefont {Boone}}, \bibinfo {author} {\bibfnamefont {A.}~\bibnamefont {Carignan-Dugas}}, \bibinfo {author} {\bibfnamefont {A.}~\bibnamefont {Chytros}}, \bibinfo {author} {\bibfnamefont {D.}~\bibnamefont {Dahlen}}, \bibinfo {author} {\bibfnamefont {H.}~\bibnamefont {Dawkins}}, \bibinfo {author} {\bibfnamefont {J.}~\bibnamefont {Emerson}}, \bibinfo {author} {\bibfnamefont {S.}~\bibnamefont {Ferracin}}, \bibinfo {author} {\bibfnamefont {V.}~\bibnamefont {Frey}}, \bibinfo {author} {\bibfnamefont {I.}~\bibnamefont {Hincks}}, \bibinfo {author} {\bibfnamefont {D.}~\bibnamefont {Hufnagel}}, \bibinfo {author} {\bibfnamefont {P.}~\bibnamefont {Iyer}}, \bibinfo {author} {\bibfnamefont {A.}~\bibnamefont {Jain}}, \bibinfo {author} {\bibfnamefont {J.}~\bibnamefont {Kolbush}}, \bibinfo {author} {\bibfnamefont {E.}~\bibnamefont {Ospadov}}, \bibinfo {author} {\bibfnamefont {J.~L.}\ \bibnamefont
  {Pino}}, \bibinfo {author} {\bibfnamefont {H.}~\bibnamefont {Qassim}}, \bibinfo {author} {\bibfnamefont {J.}~\bibnamefont {Saunders}}, \bibinfo {author} {\bibfnamefont {J.}~\bibnamefont {Skanes-Norman}}, \bibinfo {author} {\bibfnamefont {A.}~\bibnamefont {Stasiuk}}, \bibinfo {author} {\bibfnamefont {J.~J.}\ \bibnamefont {Wallman}}, \bibinfo {author} {\bibfnamefont {A.}~\bibnamefont {Winick}},\ and\ \bibinfo {author} {\bibfnamefont {E.}~\bibnamefont {Wright}},\ }\href {https://doi.org/10.5281/zenodo.3945250} {\bibinfo {title} {True-q}} (\bibinfo {year} {2020})\BibitemShut {NoStop}%
\bibitem [{\citenamefont {Carignan-Dugas}\ \emph {et~al.}(2024)\citenamefont {Carignan-Dugas}, \citenamefont {Ranu},\ and\ \citenamefont {Dreher}}]{CarignanDugas2024estimatingcoherent}%
  \BibitemOpen
  \bibfield  {author} {\bibinfo {author} {\bibfnamefont {A.}~\bibnamefont {Carignan-Dugas}}, \bibinfo {author} {\bibfnamefont {S.~K.}\ \bibnamefont {Ranu}},\ and\ \bibinfo {author} {\bibfnamefont {P.}~\bibnamefont {Dreher}},\ }\bibfield  {title} {\bibinfo {title} {Estimating {C}oherent {C}ontributions to the {E}rror {P}rofile {U}sing {C}ycle {E}rror {R}econstruction},\ }\href {https://doi.org/10.22331/q-2024-06-13-1367} {\bibfield  {journal} {\bibinfo  {journal} {{Quantum}}\ }\textbf {\bibinfo {volume} {8}},\ \bibinfo {pages} {1367} (\bibinfo {year} {2024})}\BibitemShut {NoStop}%
\bibitem [{\citenamefont {Chen}\ \emph {et~al.}(2023)\citenamefont {Chen}, \citenamefont {Liu}, \citenamefont {Otten}, \citenamefont {Seif}, \citenamefont {Fefferman},\ and\ \citenamefont {Jiang}}]{Chen_2023}%
  \BibitemOpen
  \bibfield  {author} {\bibinfo {author} {\bibfnamefont {S.}~\bibnamefont {Chen}}, \bibinfo {author} {\bibfnamefont {Y.}~\bibnamefont {Liu}}, \bibinfo {author} {\bibfnamefont {M.}~\bibnamefont {Otten}}, \bibinfo {author} {\bibfnamefont {A.}~\bibnamefont {Seif}}, \bibinfo {author} {\bibfnamefont {B.}~\bibnamefont {Fefferman}},\ and\ \bibinfo {author} {\bibfnamefont {L.}~\bibnamefont {Jiang}},\ }\bibfield  {title} {\bibinfo {title} {The learnability of pauli noise},\ }\bibfield  {journal} {\bibinfo  {journal} {Nature Communications}\ }\textbf {\bibinfo {volume} {14}},\ \href {https://doi.org/10.1038/s41467-022-35759-4} {10.1038/s41467-022-35759-4} (\bibinfo {year} {2023})\BibitemShut {NoStop}%
\bibitem [{\citenamefont {Calzona}\ \emph {et~al.}(2024)\citenamefont {Calzona}, \citenamefont {Papi{\v{c}}}, \citenamefont {Figueroa-Romero},\ and\ \citenamefont {Auer}}]{calzona2024multi}%
  \BibitemOpen
  \bibfield  {author} {\bibinfo {author} {\bibfnamefont {A.}~\bibnamefont {Calzona}}, \bibinfo {author} {\bibfnamefont {M.}~\bibnamefont {Papi{\v{c}}}}, \bibinfo {author} {\bibfnamefont {P.}~\bibnamefont {Figueroa-Romero}},\ and\ \bibinfo {author} {\bibfnamefont {A.}~\bibnamefont {Auer}},\ }\bibfield  {title} {\bibinfo {title} {Multi-layer cycle benchmarking for high-accuracy error characterization},\ }\href {http://dx.doi.org/10.1103/4hbl-hpl3} {\bibfield  {journal} {\bibinfo  {journal} {arXiv preprint arXiv:2412.09332}\ } (\bibinfo {year} {2024})}\BibitemShut {NoStop}%
\bibitem [{\citenamefont {Zhang}\ \emph {et~al.}(2025)\citenamefont {Zhang}, \citenamefont {Chen}, \citenamefont {Liu},\ and\ \citenamefont {Jiang}}]{PRXQuantum.6.010310}%
  \BibitemOpen
  \bibfield  {author} {\bibinfo {author} {\bibfnamefont {Z.}~\bibnamefont {Zhang}}, \bibinfo {author} {\bibfnamefont {S.}~\bibnamefont {Chen}}, \bibinfo {author} {\bibfnamefont {Y.}~\bibnamefont {Liu}},\ and\ \bibinfo {author} {\bibfnamefont {L.}~\bibnamefont {Jiang}},\ }\bibfield  {title} {\bibinfo {title} {Generalized cycle benchmarking algorithm for characterizing midcircuit measurements},\ }\href {https://doi.org/10.1103/PRXQuantum.6.010310} {\bibfield  {journal} {\bibinfo  {journal} {PRX Quantum}\ }\textbf {\bibinfo {volume} {6}},\ \bibinfo {pages} {010310} (\bibinfo {year} {2025})}\BibitemShut {NoStop}%
\bibitem [{\citenamefont {Flammia}\ and\ \citenamefont {Wallman}(2020)}]{Flammia_2020}%
  \BibitemOpen
  \bibfield  {author} {\bibinfo {author} {\bibfnamefont {S.~T.}\ \bibnamefont {Flammia}}\ and\ \bibinfo {author} {\bibfnamefont {J.~J.}\ \bibnamefont {Wallman}},\ }\bibfield  {title} {\bibinfo {title} {Efficient estimation of pauli channels},\ }\href {https://doi.org/10.1145/3408039} {\bibfield  {journal} {\bibinfo  {journal} {ACM Transactions on Quantum Computing}\ }\textbf {\bibinfo {volume} {1}},\ \bibinfo {pages} {1–32} (\bibinfo {year} {2020})}\BibitemShut {NoStop}%
\bibitem [{\citenamefont {Hashim}\ \emph {et~al.}(2021)\citenamefont {Hashim}, \citenamefont {Naik}, \citenamefont {Morvan}, \citenamefont {Ville}, \citenamefont {Mitchell}, \citenamefont {Kreikebaum}, \citenamefont {Davis}, \citenamefont {Smith}, \citenamefont {Iancu}, \citenamefont {O'Brien}, \citenamefont {Hincks}, \citenamefont {Wallman}, \citenamefont {Emerson},\ and\ \citenamefont {Siddiqi}}]{PhysRevX.11.041039}%
  \BibitemOpen
  \bibfield  {author} {\bibinfo {author} {\bibfnamefont {A.}~\bibnamefont {Hashim}}, \bibinfo {author} {\bibfnamefont {R.~K.}\ \bibnamefont {Naik}}, \bibinfo {author} {\bibfnamefont {A.}~\bibnamefont {Morvan}}, \bibinfo {author} {\bibfnamefont {J.-L.}\ \bibnamefont {Ville}}, \bibinfo {author} {\bibfnamefont {B.}~\bibnamefont {Mitchell}}, \bibinfo {author} {\bibfnamefont {J.~M.}\ \bibnamefont {Kreikebaum}}, \bibinfo {author} {\bibfnamefont {M.}~\bibnamefont {Davis}}, \bibinfo {author} {\bibfnamefont {E.}~\bibnamefont {Smith}}, \bibinfo {author} {\bibfnamefont {C.}~\bibnamefont {Iancu}}, \bibinfo {author} {\bibfnamefont {K.~P.}\ \bibnamefont {O'Brien}}, \bibinfo {author} {\bibfnamefont {I.}~\bibnamefont {Hincks}}, \bibinfo {author} {\bibfnamefont {J.~J.}\ \bibnamefont {Wallman}}, \bibinfo {author} {\bibfnamefont {J.}~\bibnamefont {Emerson}},\ and\ \bibinfo {author} {\bibfnamefont {I.}~\bibnamefont {Siddiqi}},\ }\bibfield  {title} {\bibinfo {title} {Randomized compiling for scalable quantum computing on a noisy
  superconducting quantum processor},\ }\href {https://doi.org/10.1103/PhysRevX.11.041039} {\bibfield  {journal} {\bibinfo  {journal} {Phys. Rev. X}\ }\textbf {\bibinfo {volume} {11}},\ \bibinfo {pages} {041039} (\bibinfo {year} {2021})}\BibitemShut {NoStop}%
\bibitem [{\citenamefont {Fazio}\ \emph {et~al.}(2025)\citenamefont {Fazio}, \citenamefont {Freund}, \citenamefont {Sannamoth}, \citenamefont {Steiner}, \citenamefont {Marciniak}, \citenamefont {Rispler}, \citenamefont {Harper}, \citenamefont {Monz}, \citenamefont {Emerson},\ and\ \citenamefont {Bartlett}}]{fazio2025characterizing}%
  \BibitemOpen
  \bibfield  {author} {\bibinfo {author} {\bibfnamefont {N.}~\bibnamefont {Fazio}}, \bibinfo {author} {\bibfnamefont {R.}~\bibnamefont {Freund}}, \bibinfo {author} {\bibfnamefont {D.}~\bibnamefont {Sannamoth}}, \bibinfo {author} {\bibfnamefont {A.}~\bibnamefont {Steiner}}, \bibinfo {author} {\bibfnamefont {C.~D.}\ \bibnamefont {Marciniak}}, \bibinfo {author} {\bibfnamefont {M.}~\bibnamefont {Rispler}}, \bibinfo {author} {\bibfnamefont {R.}~\bibnamefont {Harper}}, \bibinfo {author} {\bibfnamefont {T.}~\bibnamefont {Monz}}, \bibinfo {author} {\bibfnamefont {J.}~\bibnamefont {Emerson}},\ and\ \bibinfo {author} {\bibfnamefont {S.~D.}\ \bibnamefont {Bartlett}},\ }\bibfield  {title} {\bibinfo {title} {Characterizing physical and logical errors in a transversal cnot via cycle error reconstruction},\ }\href {https://arxiv.org/abs/2504.11980} {\bibfield  {journal} {\bibinfo  {journal} {arXiv preprint arXiv:2504.11980}\ } (\bibinfo {year} {2025})}\BibitemShut {NoStop}%
\bibitem [{\citenamefont {Pulido-Mateo}\ \emph {et~al.}(2024)\citenamefont {Pulido-Mateo}, \citenamefont {Mendpara}, \citenamefont {Duwe}, \citenamefont {Dubielzig}, \citenamefont {Zarantonello}, \citenamefont {Krinner},\ and\ \citenamefont {Ospelkaus}}]{PhysRevResearch.6.L022067}%
  \BibitemOpen
  \bibfield  {author} {\bibinfo {author} {\bibfnamefont {N.}~\bibnamefont {Pulido-Mateo}}, \bibinfo {author} {\bibfnamefont {H.}~\bibnamefont {Mendpara}}, \bibinfo {author} {\bibfnamefont {M.}~\bibnamefont {Duwe}}, \bibinfo {author} {\bibfnamefont {T.}~\bibnamefont {Dubielzig}}, \bibinfo {author} {\bibfnamefont {G.}~\bibnamefont {Zarantonello}}, \bibinfo {author} {\bibfnamefont {L.}~\bibnamefont {Krinner}},\ and\ \bibinfo {author} {\bibfnamefont {C.}~\bibnamefont {Ospelkaus}},\ }\bibfield  {title} {\bibinfo {title} {Arbitrary quantum circuits on a fully integrated two-qubit computation register for a trapped-ion quantum processor},\ }\href {https://doi.org/10.1103/PhysRevResearch.6.L022067} {\bibfield  {journal} {\bibinfo  {journal} {Phys. Rev. Res.}\ }\textbf {\bibinfo {volume} {6}},\ \bibinfo {pages} {L022067} (\bibinfo {year} {2024})}\BibitemShut {NoStop}%
\bibitem [{\citenamefont {Haghshenas}\ \emph {et~al.}(2025)\citenamefont {Haghshenas}, \citenamefont {Chertkov}, \citenamefont {Mills}, \citenamefont {Kadow}, \citenamefont {Lin}, \citenamefont {Chen}, \citenamefont {Cade}, \citenamefont {Niesen}, \citenamefont {Begu{\v{s}}i{\'c}}, \citenamefont {Rudolph} \emph {et~al.}}]{haghshenas2025digital}%
  \BibitemOpen
  \bibfield  {author} {\bibinfo {author} {\bibfnamefont {R.}~\bibnamefont {Haghshenas}}, \bibinfo {author} {\bibfnamefont {E.}~\bibnamefont {Chertkov}}, \bibinfo {author} {\bibfnamefont {M.}~\bibnamefont {Mills}}, \bibinfo {author} {\bibfnamefont {W.}~\bibnamefont {Kadow}}, \bibinfo {author} {\bibfnamefont {S.-H.}\ \bibnamefont {Lin}}, \bibinfo {author} {\bibfnamefont {Y.-H.}\ \bibnamefont {Chen}}, \bibinfo {author} {\bibfnamefont {C.}~\bibnamefont {Cade}}, \bibinfo {author} {\bibfnamefont {I.}~\bibnamefont {Niesen}}, \bibinfo {author} {\bibfnamefont {T.}~\bibnamefont {Begu{\v{s}}i{\'c}}}, \bibinfo {author} {\bibfnamefont {M.~S.}\ \bibnamefont {Rudolph}}, \emph {et~al.},\ }\bibfield  {title} {\bibinfo {title} {Digital quantum magnetism at the frontier of classical simulations},\ }\href {https://arxiv.org/abs/2503.20870} {\bibfield  {journal} {\bibinfo  {journal} {arXiv preprint arXiv:2503.20870}\ } (\bibinfo {year} {2025})}\BibitemShut {NoStop}%
\bibitem [{\citenamefont {Carignan-Dugas}\ \emph {et~al.}(2019{\natexlab{a}})\citenamefont {Carignan-Dugas}, \citenamefont {Wallman},\ and\ \citenamefont {Emerson}}]{Carignan_Dugas_2019}%
  \BibitemOpen
  \bibfield  {author} {\bibinfo {author} {\bibfnamefont {A.}~\bibnamefont {Carignan-Dugas}}, \bibinfo {author} {\bibfnamefont {J.~J.}\ \bibnamefont {Wallman}},\ and\ \bibinfo {author} {\bibfnamefont {J.}~\bibnamefont {Emerson}},\ }\bibfield  {title} {\bibinfo {title} {Bounding the average gate fidelity of composite channels using the unitarity},\ }\href {https://doi.org/10.1088/1367-2630/ab1800} {\bibfield  {journal} {\bibinfo  {journal} {New Journal of Physics}\ }\textbf {\bibinfo {volume} {21}},\ \bibinfo {pages} {053016} (\bibinfo {year} {2019}{\natexlab{a}})}\BibitemShut {NoStop}%
\bibitem [{\citenamefont {Proctor}\ \emph {et~al.}(2019)\citenamefont {Proctor}, \citenamefont {Carignan-Dugas}, \citenamefont {Rudinger}, \citenamefont {Nielsen}, \citenamefont {Blume-Kohout},\ and\ \citenamefont {Young}}]{PhysRevLett.123.030503}%
  \BibitemOpen
  \bibfield  {author} {\bibinfo {author} {\bibfnamefont {T.~J.}\ \bibnamefont {Proctor}}, \bibinfo {author} {\bibfnamefont {A.}~\bibnamefont {Carignan-Dugas}}, \bibinfo {author} {\bibfnamefont {K.}~\bibnamefont {Rudinger}}, \bibinfo {author} {\bibfnamefont {E.}~\bibnamefont {Nielsen}}, \bibinfo {author} {\bibfnamefont {R.}~\bibnamefont {Blume-Kohout}},\ and\ \bibinfo {author} {\bibfnamefont {K.}~\bibnamefont {Young}},\ }\bibfield  {title} {\bibinfo {title} {Direct randomized benchmarking for multiqubit devices},\ }\href {https://doi.org/10.1103/PhysRevLett.123.030503} {\bibfield  {journal} {\bibinfo  {journal} {Phys. Rev. Lett.}\ }\textbf {\bibinfo {volume} {123}},\ \bibinfo {pages} {030503} (\bibinfo {year} {2019})}\BibitemShut {NoStop}%
\bibitem [{\citenamefont {Polloreno}\ \emph {et~al.}(2025)\citenamefont {Polloreno}, \citenamefont {Carignan-Dugas}, \citenamefont {Hines}, \citenamefont {Blume-Kohout}, \citenamefont {Young},\ and\ \citenamefont {Proctor}}]{Polloreno_2025}%
  \BibitemOpen
  \bibfield  {author} {\bibinfo {author} {\bibfnamefont {A.~M.}\ \bibnamefont {Polloreno}}, \bibinfo {author} {\bibfnamefont {A.}~\bibnamefont {Carignan-Dugas}}, \bibinfo {author} {\bibfnamefont {J.}~\bibnamefont {Hines}}, \bibinfo {author} {\bibfnamefont {R.}~\bibnamefont {Blume-Kohout}}, \bibinfo {author} {\bibfnamefont {K.}~\bibnamefont {Young}},\ and\ \bibinfo {author} {\bibfnamefont {T.}~\bibnamefont {Proctor}},\ }\bibfield  {title} {\bibinfo {title} {A theory of direct randomized benchmarking},\ }\href {https://doi.org/10.22331/q-2025-09-05-1848} {\bibfield  {journal} {\bibinfo  {journal} {Quantum}\ }\textbf {\bibinfo {volume} {9}},\ \bibinfo {pages} {1848} (\bibinfo {year} {2025})}\BibitemShut {NoStop}%
\bibitem [{\citenamefont {Knill}\ \emph {et~al.}(2008)\citenamefont {Knill}, \citenamefont {Leibfried}, \citenamefont {Reichle}, \citenamefont {Britton}, \citenamefont {Blakestad}, \citenamefont {Jost}, \citenamefont {Langer}, \citenamefont {Ozeri}, \citenamefont {Seidelin},\ and\ \citenamefont {Wineland}}]{PhysRevA.77.012307}%
  \BibitemOpen
  \bibfield  {author} {\bibinfo {author} {\bibfnamefont {E.}~\bibnamefont {Knill}}, \bibinfo {author} {\bibfnamefont {D.}~\bibnamefont {Leibfried}}, \bibinfo {author} {\bibfnamefont {R.}~\bibnamefont {Reichle}}, \bibinfo {author} {\bibfnamefont {J.}~\bibnamefont {Britton}}, \bibinfo {author} {\bibfnamefont {R.~B.}\ \bibnamefont {Blakestad}}, \bibinfo {author} {\bibfnamefont {J.~D.}\ \bibnamefont {Jost}}, \bibinfo {author} {\bibfnamefont {C.}~\bibnamefont {Langer}}, \bibinfo {author} {\bibfnamefont {R.}~\bibnamefont {Ozeri}}, \bibinfo {author} {\bibfnamefont {S.}~\bibnamefont {Seidelin}},\ and\ \bibinfo {author} {\bibfnamefont {D.~J.}\ \bibnamefont {Wineland}},\ }\bibfield  {title} {\bibinfo {title} {Randomized benchmarking of quantum gates},\ }\href {https://doi.org/10.1103/PhysRevA.77.012307} {\bibfield  {journal} {\bibinfo  {journal} {Phys. Rev. A}\ }\textbf {\bibinfo {volume} {77}},\ \bibinfo {pages} {012307} (\bibinfo {year} {2008})}\BibitemShut {NoStop}%
\bibitem [{\citenamefont {Boone}\ \emph {et~al.}(2019)\citenamefont {Boone}, \citenamefont {Carignan-Dugas}, \citenamefont {Wallman},\ and\ \citenamefont {Emerson}}]{PhysRevA.99.032329}%
  \BibitemOpen
  \bibfield  {author} {\bibinfo {author} {\bibfnamefont {K.}~\bibnamefont {Boone}}, \bibinfo {author} {\bibfnamefont {A.}~\bibnamefont {Carignan-Dugas}}, \bibinfo {author} {\bibfnamefont {J.~J.}\ \bibnamefont {Wallman}},\ and\ \bibinfo {author} {\bibfnamefont {J.}~\bibnamefont {Emerson}},\ }\bibfield  {title} {\bibinfo {title} {Randomized benchmarking under different gate sets},\ }\href {https://doi.org/10.1103/PhysRevA.99.032329} {\bibfield  {journal} {\bibinfo  {journal} {Phys. Rev. A}\ }\textbf {\bibinfo {volume} {99}},\ \bibinfo {pages} {032329} (\bibinfo {year} {2019})}\BibitemShut {NoStop}%
\bibitem [{\citenamefont {Wallman}(2018)}]{Wallman_2018}%
  \BibitemOpen
  \bibfield  {author} {\bibinfo {author} {\bibfnamefont {J.~J.}\ \bibnamefont {Wallman}},\ }\bibfield  {title} {\bibinfo {title} {Randomized benchmarking with gate-dependent noise},\ }\href {https://doi.org/10.22331/q-2018-01-29-47} {\bibfield  {journal} {\bibinfo  {journal} {Quantum}\ }\textbf {\bibinfo {volume} {2}},\ \bibinfo {pages} {47} (\bibinfo {year} {2018})}\BibitemShut {NoStop}%
\bibitem [{\citenamefont {Proctor}\ \emph {et~al.}(2017)\citenamefont {Proctor}, \citenamefont {Rudinger}, \citenamefont {Young}, \citenamefont {Sarovar},\ and\ \citenamefont {Blume-Kohout}}]{PhysRevLett.119.130502}%
  \BibitemOpen
  \bibfield  {author} {\bibinfo {author} {\bibfnamefont {T.}~\bibnamefont {Proctor}}, \bibinfo {author} {\bibfnamefont {K.}~\bibnamefont {Rudinger}}, \bibinfo {author} {\bibfnamefont {K.}~\bibnamefont {Young}}, \bibinfo {author} {\bibfnamefont {M.}~\bibnamefont {Sarovar}},\ and\ \bibinfo {author} {\bibfnamefont {R.}~\bibnamefont {Blume-Kohout}},\ }\bibfield  {title} {\bibinfo {title} {What randomized benchmarking actually measures},\ }\href {https://doi.org/10.1103/PhysRevLett.119.130502} {\bibfield  {journal} {\bibinfo  {journal} {Phys. Rev. Lett.}\ }\textbf {\bibinfo {volume} {119}},\ \bibinfo {pages} {130502} (\bibinfo {year} {2017})}\BibitemShut {NoStop}%
\bibitem [{\citenamefont {Proctor}\ \emph {et~al.}(2022)\citenamefont {Proctor}, \citenamefont {Seritan}, \citenamefont {Nielsen}, \citenamefont {Rudinger}, \citenamefont {Young}, \citenamefont {Blume-Kohout},\ and\ \citenamefont {Sarovar}}]{proctor2022establishing}%
  \BibitemOpen
  \bibfield  {author} {\bibinfo {author} {\bibfnamefont {T.}~\bibnamefont {Proctor}}, \bibinfo {author} {\bibfnamefont {S.}~\bibnamefont {Seritan}}, \bibinfo {author} {\bibfnamefont {E.}~\bibnamefont {Nielsen}}, \bibinfo {author} {\bibfnamefont {K.}~\bibnamefont {Rudinger}}, \bibinfo {author} {\bibfnamefont {K.}~\bibnamefont {Young}}, \bibinfo {author} {\bibfnamefont {R.}~\bibnamefont {Blume-Kohout}},\ and\ \bibinfo {author} {\bibfnamefont {M.}~\bibnamefont {Sarovar}},\ }\bibfield  {title} {\bibinfo {title} {Establishing trust in quantum computations},\ }\href {https://arxiv.org/abs/2204.07568} {\bibfield  {journal} {\bibinfo  {journal} {arXiv preprint arXiv:2204.07568}\ } (\bibinfo {year} {2022})}\BibitemShut {NoStop}%
\bibitem [{\citenamefont {Foxen}\ \emph {et~al.}(2020)\citenamefont {Foxen}, \citenamefont {Neill}, \citenamefont {Dunsworth}, \citenamefont {Roushan}, \citenamefont {Chiaro}, \citenamefont {Megrant}, \citenamefont {Kelly}, \citenamefont {Chen}, \citenamefont {Satzinger}, \citenamefont {Barends}, \citenamefont {Arute}, \citenamefont {Arya}, \citenamefont {Babbush}, \citenamefont {Bacon}, \citenamefont {Bardin}, \citenamefont {Boixo}, \citenamefont {Buell}, \citenamefont {Burkett}, \citenamefont {Chen}, \citenamefont {Collins}, \citenamefont {Farhi}, \citenamefont {Fowler}, \citenamefont {Gidney}, \citenamefont {Giustina}, \citenamefont {Graff}, \citenamefont {Harrigan}, \citenamefont {Huang}, \citenamefont {Isakov}, \citenamefont {Jeffrey}, \citenamefont {Jiang}, \citenamefont {Kafri}, \citenamefont {Kechedzhi}, \citenamefont {Klimov}, \citenamefont {Korotkov}, \citenamefont {Kostritsa}, \citenamefont {Landhuis}, \citenamefont {Lucero}, \citenamefont {McClean}, \citenamefont {McEwen}, \citenamefont {Mi},
  \citenamefont {Mohseni}, \citenamefont {Mutus}, \citenamefont {Naaman}, \citenamefont {Neeley}, \citenamefont {Niu}, \citenamefont {Petukhov}, \citenamefont {Quintana}, \citenamefont {Rubin}, \citenamefont {Sank}, \citenamefont {Smelyanskiy}, \citenamefont {Vainsencher}, \citenamefont {White}, \citenamefont {Yao}, \citenamefont {Yeh}, \citenamefont {Zalcman}, \citenamefont {Neven},\ and\ \citenamefont {Martinis}}]{PhysRevLett.125.120504}%
  \BibitemOpen
  \bibfield  {author} {\bibinfo {author} {\bibfnamefont {B.}~\bibnamefont {Foxen}}, \bibinfo {author} {\bibfnamefont {C.}~\bibnamefont {Neill}}, \bibinfo {author} {\bibfnamefont {A.}~\bibnamefont {Dunsworth}}, \bibinfo {author} {\bibfnamefont {P.}~\bibnamefont {Roushan}}, \bibinfo {author} {\bibfnamefont {B.}~\bibnamefont {Chiaro}}, \bibinfo {author} {\bibfnamefont {A.}~\bibnamefont {Megrant}}, \bibinfo {author} {\bibfnamefont {J.}~\bibnamefont {Kelly}}, \bibinfo {author} {\bibfnamefont {Z.}~\bibnamefont {Chen}}, \bibinfo {author} {\bibfnamefont {K.}~\bibnamefont {Satzinger}}, \bibinfo {author} {\bibfnamefont {R.}~\bibnamefont {Barends}}, \bibinfo {author} {\bibfnamefont {F.}~\bibnamefont {Arute}}, \bibinfo {author} {\bibfnamefont {K.}~\bibnamefont {Arya}}, \bibinfo {author} {\bibfnamefont {R.}~\bibnamefont {Babbush}}, \bibinfo {author} {\bibfnamefont {D.}~\bibnamefont {Bacon}}, \bibinfo {author} {\bibfnamefont {J.~C.}\ \bibnamefont {Bardin}}, \bibinfo {author} {\bibfnamefont {S.}~\bibnamefont {Boixo}},
  \bibinfo {author} {\bibfnamefont {D.}~\bibnamefont {Buell}}, \bibinfo {author} {\bibfnamefont {B.}~\bibnamefont {Burkett}}, \bibinfo {author} {\bibfnamefont {Y.}~\bibnamefont {Chen}}, \bibinfo {author} {\bibfnamefont {R.}~\bibnamefont {Collins}}, \bibinfo {author} {\bibfnamefont {E.}~\bibnamefont {Farhi}}, \bibinfo {author} {\bibfnamefont {A.}~\bibnamefont {Fowler}}, \bibinfo {author} {\bibfnamefont {C.}~\bibnamefont {Gidney}}, \bibinfo {author} {\bibfnamefont {M.}~\bibnamefont {Giustina}}, \bibinfo {author} {\bibfnamefont {R.}~\bibnamefont {Graff}}, \bibinfo {author} {\bibfnamefont {M.}~\bibnamefont {Harrigan}}, \bibinfo {author} {\bibfnamefont {T.}~\bibnamefont {Huang}}, \bibinfo {author} {\bibfnamefont {S.~V.}\ \bibnamefont {Isakov}}, \bibinfo {author} {\bibfnamefont {E.}~\bibnamefont {Jeffrey}}, \bibinfo {author} {\bibfnamefont {Z.}~\bibnamefont {Jiang}}, \bibinfo {author} {\bibfnamefont {D.}~\bibnamefont {Kafri}}, \bibinfo {author} {\bibfnamefont {K.}~\bibnamefont {Kechedzhi}}, \bibinfo {author}
  {\bibfnamefont {P.}~\bibnamefont {Klimov}}, \bibinfo {author} {\bibfnamefont {A.}~\bibnamefont {Korotkov}}, \bibinfo {author} {\bibfnamefont {F.}~\bibnamefont {Kostritsa}}, \bibinfo {author} {\bibfnamefont {D.}~\bibnamefont {Landhuis}}, \bibinfo {author} {\bibfnamefont {E.}~\bibnamefont {Lucero}}, \bibinfo {author} {\bibfnamefont {J.}~\bibnamefont {McClean}}, \bibinfo {author} {\bibfnamefont {M.}~\bibnamefont {McEwen}}, \bibinfo {author} {\bibfnamefont {X.}~\bibnamefont {Mi}}, \bibinfo {author} {\bibfnamefont {M.}~\bibnamefont {Mohseni}}, \bibinfo {author} {\bibfnamefont {J.~Y.}\ \bibnamefont {Mutus}}, \bibinfo {author} {\bibfnamefont {O.}~\bibnamefont {Naaman}}, \bibinfo {author} {\bibfnamefont {M.}~\bibnamefont {Neeley}}, \bibinfo {author} {\bibfnamefont {M.}~\bibnamefont {Niu}}, \bibinfo {author} {\bibfnamefont {A.}~\bibnamefont {Petukhov}}, \bibinfo {author} {\bibfnamefont {C.}~\bibnamefont {Quintana}}, \bibinfo {author} {\bibfnamefont {N.}~\bibnamefont {Rubin}}, \bibinfo {author} {\bibfnamefont
  {D.}~\bibnamefont {Sank}}, \bibinfo {author} {\bibfnamefont {V.}~\bibnamefont {Smelyanskiy}}, \bibinfo {author} {\bibfnamefont {A.}~\bibnamefont {Vainsencher}}, \bibinfo {author} {\bibfnamefont {T.~C.}\ \bibnamefont {White}}, \bibinfo {author} {\bibfnamefont {Z.}~\bibnamefont {Yao}}, \bibinfo {author} {\bibfnamefont {P.}~\bibnamefont {Yeh}}, \bibinfo {author} {\bibfnamefont {A.}~\bibnamefont {Zalcman}}, \bibinfo {author} {\bibfnamefont {H.}~\bibnamefont {Neven}},\ and\ \bibinfo {author} {\bibfnamefont {J.~M.}\ \bibnamefont {Martinis}} (\bibinfo {collaboration} {Google AI Quantum}),\ }\bibfield  {title} {\bibinfo {title} {Demonstrating a continuous set of two-qubit gates for near-term quantum algorithms},\ }\href {https://doi.org/10.1103/PhysRevLett.125.120504} {\bibfield  {journal} {\bibinfo  {journal} {Phys. Rev. Lett.}\ }\textbf {\bibinfo {volume} {125}},\ \bibinfo {pages} {120504} (\bibinfo {year} {2020})}\BibitemShut {NoStop}%
\bibitem [{\citenamefont {Wu}\ \emph {et~al.}(2021)\citenamefont {Wu}, \citenamefont {Bao}, \citenamefont {Cao}, \citenamefont {Chen}, \citenamefont {Chen}, \citenamefont {Chen}, \citenamefont {Chung}, \citenamefont {Deng}, \citenamefont {Du}, \citenamefont {Fan}, \citenamefont {Gong}, \citenamefont {Guo}, \citenamefont {Guo}, \citenamefont {Guo}, \citenamefont {Han}, \citenamefont {Hong}, \citenamefont {Huang}, \citenamefont {Huo}, \citenamefont {Li}, \citenamefont {Li}, \citenamefont {Li}, \citenamefont {Li}, \citenamefont {Liang}, \citenamefont {Lin}, \citenamefont {Lin}, \citenamefont {Qian}, \citenamefont {Qiao}, \citenamefont {Rong}, \citenamefont {Su}, \citenamefont {Sun}, \citenamefont {Wang}, \citenamefont {Wang}, \citenamefont {Wu}, \citenamefont {Xu}, \citenamefont {Yan}, \citenamefont {Yang}, \citenamefont {Yang}, \citenamefont {Ye}, \citenamefont {Yin}, \citenamefont {Ying}, \citenamefont {Yu}, \citenamefont {Zha}, \citenamefont {Zhang}, \citenamefont {Zhang}, \citenamefont {Zhang}, \citenamefont
  {Zhang}, \citenamefont {Zhao}, \citenamefont {Zhao}, \citenamefont {Zhou}, \citenamefont {Zhu}, \citenamefont {Lu}, \citenamefont {Peng}, \citenamefont {Zhu},\ and\ \citenamefont {Pan}}]{Wu_2021}%
  \BibitemOpen
  \bibfield  {author} {\bibinfo {author} {\bibfnamefont {Y.}~\bibnamefont {Wu}}, \bibinfo {author} {\bibfnamefont {W.-S.}\ \bibnamefont {Bao}}, \bibinfo {author} {\bibfnamefont {S.}~\bibnamefont {Cao}}, \bibinfo {author} {\bibfnamefont {F.}~\bibnamefont {Chen}}, \bibinfo {author} {\bibfnamefont {M.-C.}\ \bibnamefont {Chen}}, \bibinfo {author} {\bibfnamefont {X.}~\bibnamefont {Chen}}, \bibinfo {author} {\bibfnamefont {T.-H.}\ \bibnamefont {Chung}}, \bibinfo {author} {\bibfnamefont {H.}~\bibnamefont {Deng}}, \bibinfo {author} {\bibfnamefont {Y.}~\bibnamefont {Du}}, \bibinfo {author} {\bibfnamefont {D.}~\bibnamefont {Fan}}, \bibinfo {author} {\bibfnamefont {M.}~\bibnamefont {Gong}}, \bibinfo {author} {\bibfnamefont {C.}~\bibnamefont {Guo}}, \bibinfo {author} {\bibfnamefont {C.}~\bibnamefont {Guo}}, \bibinfo {author} {\bibfnamefont {S.}~\bibnamefont {Guo}}, \bibinfo {author} {\bibfnamefont {L.}~\bibnamefont {Han}}, \bibinfo {author} {\bibfnamefont {L.}~\bibnamefont {Hong}}, \bibinfo {author} {\bibfnamefont
  {H.-L.}\ \bibnamefont {Huang}}, \bibinfo {author} {\bibfnamefont {Y.-H.}\ \bibnamefont {Huo}}, \bibinfo {author} {\bibfnamefont {L.}~\bibnamefont {Li}}, \bibinfo {author} {\bibfnamefont {N.}~\bibnamefont {Li}}, \bibinfo {author} {\bibfnamefont {S.}~\bibnamefont {Li}}, \bibinfo {author} {\bibfnamefont {Y.}~\bibnamefont {Li}}, \bibinfo {author} {\bibfnamefont {F.}~\bibnamefont {Liang}}, \bibinfo {author} {\bibfnamefont {C.}~\bibnamefont {Lin}}, \bibinfo {author} {\bibfnamefont {J.}~\bibnamefont {Lin}}, \bibinfo {author} {\bibfnamefont {H.}~\bibnamefont {Qian}}, \bibinfo {author} {\bibfnamefont {D.}~\bibnamefont {Qiao}}, \bibinfo {author} {\bibfnamefont {H.}~\bibnamefont {Rong}}, \bibinfo {author} {\bibfnamefont {H.}~\bibnamefont {Su}}, \bibinfo {author} {\bibfnamefont {L.}~\bibnamefont {Sun}}, \bibinfo {author} {\bibfnamefont {L.}~\bibnamefont {Wang}}, \bibinfo {author} {\bibfnamefont {S.}~\bibnamefont {Wang}}, \bibinfo {author} {\bibfnamefont {D.}~\bibnamefont {Wu}}, \bibinfo {author} {\bibfnamefont
  {Y.}~\bibnamefont {Xu}}, \bibinfo {author} {\bibfnamefont {K.}~\bibnamefont {Yan}}, \bibinfo {author} {\bibfnamefont {W.}~\bibnamefont {Yang}}, \bibinfo {author} {\bibfnamefont {Y.}~\bibnamefont {Yang}}, \bibinfo {author} {\bibfnamefont {Y.}~\bibnamefont {Ye}}, \bibinfo {author} {\bibfnamefont {J.}~\bibnamefont {Yin}}, \bibinfo {author} {\bibfnamefont {C.}~\bibnamefont {Ying}}, \bibinfo {author} {\bibfnamefont {J.}~\bibnamefont {Yu}}, \bibinfo {author} {\bibfnamefont {C.}~\bibnamefont {Zha}}, \bibinfo {author} {\bibfnamefont {C.}~\bibnamefont {Zhang}}, \bibinfo {author} {\bibfnamefont {H.}~\bibnamefont {Zhang}}, \bibinfo {author} {\bibfnamefont {K.}~\bibnamefont {Zhang}}, \bibinfo {author} {\bibfnamefont {Y.}~\bibnamefont {Zhang}}, \bibinfo {author} {\bibfnamefont {H.}~\bibnamefont {Zhao}}, \bibinfo {author} {\bibfnamefont {Y.}~\bibnamefont {Zhao}}, \bibinfo {author} {\bibfnamefont {L.}~\bibnamefont {Zhou}}, \bibinfo {author} {\bibfnamefont {Q.}~\bibnamefont {Zhu}}, \bibinfo {author} {\bibfnamefont {C.-Y.}\
  \bibnamefont {Lu}}, \bibinfo {author} {\bibfnamefont {C.-Z.}\ \bibnamefont {Peng}}, \bibinfo {author} {\bibfnamefont {X.}~\bibnamefont {Zhu}},\ and\ \bibinfo {author} {\bibfnamefont {J.-W.}\ \bibnamefont {Pan}},\ }\bibfield  {title} {\bibinfo {title} {Strong quantum computational advantage using a superconducting quantum processor},\ }\bibfield  {journal} {\bibinfo  {journal} {Physical Review Letters}\ }\textbf {\bibinfo {volume} {127}},\ \href {https://doi.org/10.1103/physrevlett.127.180501} {10.1103/physrevlett.127.180501} (\bibinfo {year} {2021})\BibitemShut {NoStop}%
\bibitem [{\citenamefont {Arute}\ \emph {et~al.}(2019)\citenamefont {Arute}, \citenamefont {Arya}, \citenamefont {Babbush}, \citenamefont {Bacon}, \citenamefont {Bardin}, \citenamefont {Barends}, \citenamefont {Biswas}, \citenamefont {Boixo}, \citenamefont {Brandao}, \citenamefont {Buell}, \citenamefont {Burkett}, \citenamefont {Chen}, \citenamefont {Chen}, \citenamefont {Chiaro}, \citenamefont {Collins}, \citenamefont {Courtney}, \citenamefont {Dunsworth}, \citenamefont {Farhi}, \citenamefont {Foxen}, \citenamefont {Fowler}, \citenamefont {Gidney}, \citenamefont {Giustina}, \citenamefont {Graff}, \citenamefont {Guerin}, \citenamefont {Habegger}, \citenamefont {Harrigan}, \citenamefont {Hartmann}, \citenamefont {Ho}, \citenamefont {Hoffmann}, \citenamefont {Huang}, \citenamefont {Humble}, \citenamefont {Isakov}, \citenamefont {Jeffrey}, \citenamefont {Jiang}, \citenamefont {Kafri}, \citenamefont {Kechedzhi}, \citenamefont {Kelly}, \citenamefont {Klimov}, \citenamefont {Knysh}, \citenamefont {Korotkov},
  \citenamefont {Kostritsa}, \citenamefont {Landhuis}, \citenamefont {Lindmark}, \citenamefont {Lucero}, \citenamefont {Lyakh}, \citenamefont {Mandrà}, \citenamefont {McClean}, \citenamefont {McEwen}, \citenamefont {Megrant}, \citenamefont {Mi}, \citenamefont {Michielsen}, \citenamefont {Mohseni}, \citenamefont {Mutus}, \citenamefont {Naaman}, \citenamefont {Neeley}, \citenamefont {Neill}, \citenamefont {Niu}, \citenamefont {Ostby}, \citenamefont {Petukhov}, \citenamefont {Platt}, \citenamefont {Quintana}, \citenamefont {Rieffel}, \citenamefont {Roushan}, \citenamefont {Rubin}, \citenamefont {Sank}, \citenamefont {Satzinger}, \citenamefont {Smelyanskiy}, \citenamefont {Sung}, \citenamefont {Trevithick}, \citenamefont {Vainsencher}, \citenamefont {Villalonga}, \citenamefont {White}, \citenamefont {Yao}, \citenamefont {Yeh}, \citenamefont {Zalcman}, \citenamefont {Neven},\ and\ \citenamefont {Martinis}}]{arute_quantum_2019}%
  \BibitemOpen
  \bibfield  {author} {\bibinfo {author} {\bibfnamefont {F.}~\bibnamefont {Arute}}, \bibinfo {author} {\bibfnamefont {K.}~\bibnamefont {Arya}}, \bibinfo {author} {\bibfnamefont {R.}~\bibnamefont {Babbush}}, \bibinfo {author} {\bibfnamefont {D.}~\bibnamefont {Bacon}}, \bibinfo {author} {\bibfnamefont {J.~C.}\ \bibnamefont {Bardin}}, \bibinfo {author} {\bibfnamefont {R.}~\bibnamefont {Barends}}, \bibinfo {author} {\bibfnamefont {R.}~\bibnamefont {Biswas}}, \bibinfo {author} {\bibfnamefont {S.}~\bibnamefont {Boixo}}, \bibinfo {author} {\bibfnamefont {F.~G. S.~L.}\ \bibnamefont {Brandao}}, \bibinfo {author} {\bibfnamefont {D.~A.}\ \bibnamefont {Buell}}, \bibinfo {author} {\bibfnamefont {B.}~\bibnamefont {Burkett}}, \bibinfo {author} {\bibfnamefont {Y.}~\bibnamefont {Chen}}, \bibinfo {author} {\bibfnamefont {Z.}~\bibnamefont {Chen}}, \bibinfo {author} {\bibfnamefont {B.}~\bibnamefont {Chiaro}}, \bibinfo {author} {\bibfnamefont {R.}~\bibnamefont {Collins}}, \bibinfo {author} {\bibfnamefont {W.}~\bibnamefont
  {Courtney}}, \bibinfo {author} {\bibfnamefont {A.}~\bibnamefont {Dunsworth}}, \bibinfo {author} {\bibfnamefont {E.}~\bibnamefont {Farhi}}, \bibinfo {author} {\bibfnamefont {B.}~\bibnamefont {Foxen}}, \bibinfo {author} {\bibfnamefont {A.}~\bibnamefont {Fowler}}, \bibinfo {author} {\bibfnamefont {C.}~\bibnamefont {Gidney}}, \bibinfo {author} {\bibfnamefont {M.}~\bibnamefont {Giustina}}, \bibinfo {author} {\bibfnamefont {R.}~\bibnamefont {Graff}}, \bibinfo {author} {\bibfnamefont {K.}~\bibnamefont {Guerin}}, \bibinfo {author} {\bibfnamefont {S.}~\bibnamefont {Habegger}}, \bibinfo {author} {\bibfnamefont {M.~P.}\ \bibnamefont {Harrigan}}, \bibinfo {author} {\bibfnamefont {M.~J.}\ \bibnamefont {Hartmann}}, \bibinfo {author} {\bibfnamefont {A.}~\bibnamefont {Ho}}, \bibinfo {author} {\bibfnamefont {M.}~\bibnamefont {Hoffmann}}, \bibinfo {author} {\bibfnamefont {T.}~\bibnamefont {Huang}}, \bibinfo {author} {\bibfnamefont {T.~S.}\ \bibnamefont {Humble}}, \bibinfo {author} {\bibfnamefont {S.~V.}\ \bibnamefont
  {Isakov}}, \bibinfo {author} {\bibfnamefont {E.}~\bibnamefont {Jeffrey}}, \bibinfo {author} {\bibfnamefont {Z.}~\bibnamefont {Jiang}}, \bibinfo {author} {\bibfnamefont {D.}~\bibnamefont {Kafri}}, \bibinfo {author} {\bibfnamefont {K.}~\bibnamefont {Kechedzhi}}, \bibinfo {author} {\bibfnamefont {J.}~\bibnamefont {Kelly}}, \bibinfo {author} {\bibfnamefont {P.~V.}\ \bibnamefont {Klimov}}, \bibinfo {author} {\bibfnamefont {S.}~\bibnamefont {Knysh}}, \bibinfo {author} {\bibfnamefont {A.}~\bibnamefont {Korotkov}}, \bibinfo {author} {\bibfnamefont {F.}~\bibnamefont {Kostritsa}}, \bibinfo {author} {\bibfnamefont {D.}~\bibnamefont {Landhuis}}, \bibinfo {author} {\bibfnamefont {M.}~\bibnamefont {Lindmark}}, \bibinfo {author} {\bibfnamefont {E.}~\bibnamefont {Lucero}}, \bibinfo {author} {\bibfnamefont {D.}~\bibnamefont {Lyakh}}, \bibinfo {author} {\bibfnamefont {S.}~\bibnamefont {Mandrà}}, \bibinfo {author} {\bibfnamefont {J.~R.}\ \bibnamefont {McClean}}, \bibinfo {author} {\bibfnamefont {M.}~\bibnamefont {McEwen}},
  \bibinfo {author} {\bibfnamefont {A.}~\bibnamefont {Megrant}}, \bibinfo {author} {\bibfnamefont {X.}~\bibnamefont {Mi}}, \bibinfo {author} {\bibfnamefont {K.}~\bibnamefont {Michielsen}}, \bibinfo {author} {\bibfnamefont {M.}~\bibnamefont {Mohseni}}, \bibinfo {author} {\bibfnamefont {J.}~\bibnamefont {Mutus}}, \bibinfo {author} {\bibfnamefont {O.}~\bibnamefont {Naaman}}, \bibinfo {author} {\bibfnamefont {M.}~\bibnamefont {Neeley}}, \bibinfo {author} {\bibfnamefont {C.}~\bibnamefont {Neill}}, \bibinfo {author} {\bibfnamefont {M.~Y.}\ \bibnamefont {Niu}}, \bibinfo {author} {\bibfnamefont {E.}~\bibnamefont {Ostby}}, \bibinfo {author} {\bibfnamefont {A.}~\bibnamefont {Petukhov}}, \bibinfo {author} {\bibfnamefont {J.~C.}\ \bibnamefont {Platt}}, \bibinfo {author} {\bibfnamefont {C.}~\bibnamefont {Quintana}}, \bibinfo {author} {\bibfnamefont {E.~G.}\ \bibnamefont {Rieffel}}, \bibinfo {author} {\bibfnamefont {P.}~\bibnamefont {Roushan}}, \bibinfo {author} {\bibfnamefont {N.~C.}\ \bibnamefont {Rubin}}, \bibinfo
  {author} {\bibfnamefont {D.}~\bibnamefont {Sank}}, \bibinfo {author} {\bibfnamefont {K.~J.}\ \bibnamefont {Satzinger}}, \bibinfo {author} {\bibfnamefont {V.}~\bibnamefont {Smelyanskiy}}, \bibinfo {author} {\bibfnamefont {K.~J.}\ \bibnamefont {Sung}}, \bibinfo {author} {\bibfnamefont {M.~D.}\ \bibnamefont {Trevithick}}, \bibinfo {author} {\bibfnamefont {A.}~\bibnamefont {Vainsencher}}, \bibinfo {author} {\bibfnamefont {B.}~\bibnamefont {Villalonga}}, \bibinfo {author} {\bibfnamefont {T.}~\bibnamefont {White}}, \bibinfo {author} {\bibfnamefont {Z.~J.}\ \bibnamefont {Yao}}, \bibinfo {author} {\bibfnamefont {P.}~\bibnamefont {Yeh}}, \bibinfo {author} {\bibfnamefont {A.}~\bibnamefont {Zalcman}}, \bibinfo {author} {\bibfnamefont {H.}~\bibnamefont {Neven}},\ and\ \bibinfo {author} {\bibfnamefont {J.~M.}\ \bibnamefont {Martinis}},\ }\bibfield  {title} {\bibinfo {title} {Quantum supremacy using a programmable superconducting processor},\ }\href {https://doi.org/10.1038/s41586-019-1666-5} {\bibfield  {journal}
  {\bibinfo  {journal} {Nature}\ }\textbf {\bibinfo {volume} {574}},\ \bibinfo {pages} {505} (\bibinfo {year} {2019})}\BibitemShut {NoStop}%
\bibitem [{\citenamefont {Simakov}\ \emph {et~al.}(2026)\citenamefont {Simakov}, \citenamefont {Zotova}, \citenamefont {Chudakova}, \citenamefont {Kazmina}, \citenamefont {Polyanskiy}, \citenamefont {Abramov}, \citenamefont {Tarkhov}, \citenamefont {Mumlyakov}, \citenamefont {Trofimov}, \citenamefont {Rudenko}, \citenamefont {Chichkov}, \citenamefont {Chichkov},\ and\ \citenamefont {Mazhorin}}]{simakov2026decayratesinterleavedbenchmarking}%
  \BibitemOpen
  \bibfield  {author} {\bibinfo {author} {\bibfnamefont {I.~A.}\ \bibnamefont {Simakov}}, \bibinfo {author} {\bibfnamefont {A.~V.}\ \bibnamefont {Zotova}}, \bibinfo {author} {\bibfnamefont {T.~A.}\ \bibnamefont {Chudakova}}, \bibinfo {author} {\bibfnamefont {A.~S.}\ \bibnamefont {Kazmina}}, \bibinfo {author} {\bibfnamefont {A.~M.}\ \bibnamefont {Polyanskiy}}, \bibinfo {author} {\bibfnamefont {N.~N.}\ \bibnamefont {Abramov}}, \bibinfo {author} {\bibfnamefont {M.~A.}\ \bibnamefont {Tarkhov}}, \bibinfo {author} {\bibfnamefont {A.~M.}\ \bibnamefont {Mumlyakov}}, \bibinfo {author} {\bibfnamefont {I.~V.}\ \bibnamefont {Trofimov}}, \bibinfo {author} {\bibfnamefont {N.~Y.}\ \bibnamefont {Rudenko}}, \bibinfo {author} {\bibfnamefont {M.~V.}\ \bibnamefont {Chichkov}}, \bibinfo {author} {\bibfnamefont {V.~I.}\ \bibnamefont {Chichkov}},\ and\ \bibinfo {author} {\bibfnamefont {G.~S.}\ \bibnamefont {Mazhorin}},\ }\href {https://arxiv.org/abs/2603.05422} {\bibinfo {title} {Decay rates in interleaved benchmarking with
  single-qubit references}} (\bibinfo {year} {2026}),\ \Eprint {https://arxiv.org/abs/2603.05422} {arXiv:2603.05422 [quant-ph]} \BibitemShut {NoStop}%
\bibitem [{\citenamefont {Merkel}\ \emph {et~al.}(2025)\citenamefont {Merkel}, \citenamefont {Proctor}, \citenamefont {Ferracin}, \citenamefont {Hines}, \citenamefont {Barron}, \citenamefont {Govia},\ and\ \citenamefont {McKay}}]{merkel2025clifford}%
  \BibitemOpen
  \bibfield  {author} {\bibinfo {author} {\bibfnamefont {S.}~\bibnamefont {Merkel}}, \bibinfo {author} {\bibfnamefont {T.}~\bibnamefont {Proctor}}, \bibinfo {author} {\bibfnamefont {S.}~\bibnamefont {Ferracin}}, \bibinfo {author} {\bibfnamefont {J.}~\bibnamefont {Hines}}, \bibinfo {author} {\bibfnamefont {S.}~\bibnamefont {Barron}}, \bibinfo {author} {\bibfnamefont {L.~C.}\ \bibnamefont {Govia}},\ and\ \bibinfo {author} {\bibfnamefont {D.}~\bibnamefont {McKay}},\ }\bibfield  {title} {\bibinfo {title} {When clifford benchmarks are sufficient; estimating application performance with scalable proxy circuits},\ }\href {https://arxiv.org/abs/2503.05943} {\bibfield  {journal} {\bibinfo  {journal} {arXiv preprint arXiv:2503.05943}\ } (\bibinfo {year} {2025})}\BibitemShut {NoStop}%
\bibitem [{\citenamefont {Gambetta}\ \emph {et~al.}(2012)\citenamefont {Gambetta}, \citenamefont {C\'orcoles}, \citenamefont {Merkel}, \citenamefont {Johnson}, \citenamefont {Smolin}, \citenamefont {Chow}, \citenamefont {Ryan}, \citenamefont {Rigetti}, \citenamefont {Poletto}, \citenamefont {Ohki}, \citenamefont {Ketchen},\ and\ \citenamefont {Steffen}}]{PhysRevLett.109.240504}%
  \BibitemOpen
  \bibfield  {author} {\bibinfo {author} {\bibfnamefont {J.~M.}\ \bibnamefont {Gambetta}}, \bibinfo {author} {\bibfnamefont {A.~D.}\ \bibnamefont {C\'orcoles}}, \bibinfo {author} {\bibfnamefont {S.~T.}\ \bibnamefont {Merkel}}, \bibinfo {author} {\bibfnamefont {B.~R.}\ \bibnamefont {Johnson}}, \bibinfo {author} {\bibfnamefont {J.~A.}\ \bibnamefont {Smolin}}, \bibinfo {author} {\bibfnamefont {J.~M.}\ \bibnamefont {Chow}}, \bibinfo {author} {\bibfnamefont {C.~A.}\ \bibnamefont {Ryan}}, \bibinfo {author} {\bibfnamefont {C.}~\bibnamefont {Rigetti}}, \bibinfo {author} {\bibfnamefont {S.}~\bibnamefont {Poletto}}, \bibinfo {author} {\bibfnamefont {T.~A.}\ \bibnamefont {Ohki}}, \bibinfo {author} {\bibfnamefont {M.~B.}\ \bibnamefont {Ketchen}},\ and\ \bibinfo {author} {\bibfnamefont {M.}~\bibnamefont {Steffen}},\ }\bibfield  {title} {\bibinfo {title} {Characterization of addressability by simultaneous randomized benchmarking},\ }\href {https://doi.org/10.1103/PhysRevLett.109.240504} {\bibfield  {journal} {\bibinfo
  {journal} {Phys. Rev. Lett.}\ }\textbf {\bibinfo {volume} {109}},\ \bibinfo {pages} {240504} (\bibinfo {year} {2012})}\BibitemShut {NoStop}%
\bibitem [{\citenamefont {Nielsen}(2002)}]{Nielsen_2002}%
  \BibitemOpen
  \bibfield  {author} {\bibinfo {author} {\bibfnamefont {M.~A.}\ \bibnamefont {Nielsen}},\ }\bibfield  {title} {\bibinfo {title} {A simple formula for the average gate fidelity of a quantum dynamical operation},\ }\href {https://doi.org/10.1016/s0375-9601(02)01272-0} {\bibfield  {journal} {\bibinfo  {journal} {Physics Letters A}\ }\textbf {\bibinfo {volume} {303}},\ \bibinfo {pages} {249–252} (\bibinfo {year} {2002})}\BibitemShut {NoStop}%
\bibitem [{\citenamefont {Carignan-Dugas}\ \emph {et~al.}(2019{\natexlab{b}})\citenamefont {Carignan-Dugas}, \citenamefont {Alexander},\ and\ \citenamefont {Emerson}}]{Polar}%
  \BibitemOpen
  \bibfield  {author} {\bibinfo {author} {\bibfnamefont {A.}~\bibnamefont {Carignan-Dugas}}, \bibinfo {author} {\bibfnamefont {M.}~\bibnamefont {Alexander}},\ and\ \bibinfo {author} {\bibfnamefont {J.}~\bibnamefont {Emerson}},\ }\bibfield  {title} {\bibinfo {title} {A polar decomposition for quantum channels (with applications to bounding error propagation in quantum circuits)},\ }\href {https://doi.org/10.22331/q-2019-08-12-173} {\bibfield  {journal} {\bibinfo  {journal} {Quantum}\ }\textbf {\bibinfo {volume} {3}},\ \bibinfo {pages} {173} (\bibinfo {year} {2019}{\natexlab{b}})}\BibitemShut {NoStop}%
\bibitem [{\citenamefont {Wallman}\ and\ \citenamefont {Flammia}(2016)}]{Wallman_2016}%
  \BibitemOpen
  \bibfield  {author} {\bibinfo {author} {\bibfnamefont {J.~J.}\ \bibnamefont {Wallman}}\ and\ \bibinfo {author} {\bibfnamefont {S.~T.}\ \bibnamefont {Flammia}},\ }\bibfield  {title} {\bibinfo {title} {Corrigendum: Randomized benchmarking with confidence (2014 new j. phys. 16 103032)},\ }\href {https://doi.org/10.1088/1367-2630/18/7/079501} {\bibfield  {journal} {\bibinfo  {journal} {New Journal of Physics}\ }\textbf {\bibinfo {volume} {18}},\ \bibinfo {pages} {079501} (\bibinfo {year} {2016})}\BibitemShut {NoStop}%
\bibitem [{\citenamefont {Wallman}\ \emph {et~al.}(2015)\citenamefont {Wallman}, \citenamefont {Granade}, \citenamefont {Harper},\ and\ \citenamefont {Flammia}}]{Wallman_2015}%
  \BibitemOpen
  \bibfield  {author} {\bibinfo {author} {\bibfnamefont {J.}~\bibnamefont {Wallman}}, \bibinfo {author} {\bibfnamefont {C.}~\bibnamefont {Granade}}, \bibinfo {author} {\bibfnamefont {R.}~\bibnamefont {Harper}},\ and\ \bibinfo {author} {\bibfnamefont {S.~T.}\ \bibnamefont {Flammia}},\ }\bibfield  {title} {\bibinfo {title} {Estimating the coherence of noise},\ }\href {https://doi.org/10.1088/1367-2630/17/11/113020} {\bibfield  {journal} {\bibinfo  {journal} {New Journal of Physics}\ }\textbf {\bibinfo {volume} {17}},\ \bibinfo {pages} {113020} (\bibinfo {year} {2015})}\BibitemShut {NoStop}%
\bibitem [{\citenamefont {Carignan-Dugas}\ \emph {et~al.}(2018)\citenamefont {Carignan-Dugas}, \citenamefont {Boone}, \citenamefont {Wallman},\ and\ \citenamefont {Emerson}}]{Carignan_Dugas_2018}%
  \BibitemOpen
  \bibfield  {author} {\bibinfo {author} {\bibfnamefont {A.}~\bibnamefont {Carignan-Dugas}}, \bibinfo {author} {\bibfnamefont {K.}~\bibnamefont {Boone}}, \bibinfo {author} {\bibfnamefont {J.~J.}\ \bibnamefont {Wallman}},\ and\ \bibinfo {author} {\bibfnamefont {J.}~\bibnamefont {Emerson}},\ }\bibfield  {title} {\bibinfo {title} {From randomized benchmarking experiments to gate-set circuit fidelity: how to interpret randomized benchmarking decay parameters},\ }\href {https://doi.org/10.1088/1367-2630/aadcc7} {\bibfield  {journal} {\bibinfo  {journal} {New Journal of Physics}\ }\textbf {\bibinfo {volume} {20}},\ \bibinfo {pages} {092001} (\bibinfo {year} {2018})}\BibitemShut {NoStop}%
\bibitem [{\citenamefont {Knapp}(1988)}]{Knapp1988LieGB}%
  \BibitemOpen
  \bibfield  {author} {\bibinfo {author} {\bibfnamefont {A.~W.}\ \bibnamefont {Knapp}},\ }\bibfield  {title} {\bibinfo {title} {Lie groups beyond an introduction}\ }(\bibinfo {year} {1988})\BibitemShut {NoStop}%
\bibitem [{\citenamefont {McKay}\ \emph {et~al.}(2017)\citenamefont {McKay}, \citenamefont {Wood}, \citenamefont {Sheldon}, \citenamefont {Chow},\ and\ \citenamefont {Gambetta}}]{PhysRevA.96.022330}%
  \BibitemOpen
  \bibfield  {author} {\bibinfo {author} {\bibfnamefont {D.~C.}\ \bibnamefont {McKay}}, \bibinfo {author} {\bibfnamefont {C.~J.}\ \bibnamefont {Wood}}, \bibinfo {author} {\bibfnamefont {S.}~\bibnamefont {Sheldon}}, \bibinfo {author} {\bibfnamefont {J.~M.}\ \bibnamefont {Chow}},\ and\ \bibinfo {author} {\bibfnamefont {J.~M.}\ \bibnamefont {Gambetta}},\ }\bibfield  {title} {\bibinfo {title} {Efficient $z$ gates for quantum computing},\ }\href {https://doi.org/10.1103/PhysRevA.96.022330} {\bibfield  {journal} {\bibinfo  {journal} {Phys. Rev. A}\ }\textbf {\bibinfo {volume} {96}},\ \bibinfo {pages} {022330} (\bibinfo {year} {2017})}\BibitemShut {NoStop}%
\bibitem [{\citenamefont {Fruitwala}\ \emph {et~al.}(2024)\citenamefont {Fruitwala}, \citenamefont {Hashim}, \citenamefont {Rajagopala}, \citenamefont {Xu}, \citenamefont {Hines}, \citenamefont {Naik}, \citenamefont {Siddiqi}, \citenamefont {Klymko}, \citenamefont {Huang},\ and\ \citenamefont {Nowrouzi}}]{fruitwala2024hardware}%
  \BibitemOpen
  \bibfield  {author} {\bibinfo {author} {\bibfnamefont {N.}~\bibnamefont {Fruitwala}}, \bibinfo {author} {\bibfnamefont {A.}~\bibnamefont {Hashim}}, \bibinfo {author} {\bibfnamefont {A.~D.}\ \bibnamefont {Rajagopala}}, \bibinfo {author} {\bibfnamefont {Y.}~\bibnamefont {Xu}}, \bibinfo {author} {\bibfnamefont {J.}~\bibnamefont {Hines}}, \bibinfo {author} {\bibfnamefont {R.~K.}\ \bibnamefont {Naik}}, \bibinfo {author} {\bibfnamefont {I.}~\bibnamefont {Siddiqi}}, \bibinfo {author} {\bibfnamefont {K.}~\bibnamefont {Klymko}}, \bibinfo {author} {\bibfnamefont {G.}~\bibnamefont {Huang}},\ and\ \bibinfo {author} {\bibfnamefont {K.}~\bibnamefont {Nowrouzi}},\ }\bibfield  {title} {\bibinfo {title} {Hardware-efficient randomized compiling},\ }\href {https://arxiv.org/abs/2406.13967} {\bibfield  {journal} {\bibinfo  {journal} {arXiv preprint arXiv:2406.13967}\ } (\bibinfo {year} {2024})}\BibitemShut {NoStop}%
\bibitem [{\citenamefont {Pino}\ \emph {et~al.}(2021)\citenamefont {Pino}, \citenamefont {Dreiling}, \citenamefont {Figgatt}, \citenamefont {Gaebler}, \citenamefont {Moses}, \citenamefont {Allman}, \citenamefont {Baldwin}, \citenamefont {Foss-Feig}, \citenamefont {Hayes}, \citenamefont {Mayer}, \citenamefont {Ryan-Anderson},\ and\ \citenamefont {Neyenhuis}}]{Pino_2021}%
  \BibitemOpen
  \bibfield  {author} {\bibinfo {author} {\bibfnamefont {J.~M.}\ \bibnamefont {Pino}}, \bibinfo {author} {\bibfnamefont {J.~M.}\ \bibnamefont {Dreiling}}, \bibinfo {author} {\bibfnamefont {C.}~\bibnamefont {Figgatt}}, \bibinfo {author} {\bibfnamefont {J.~P.}\ \bibnamefont {Gaebler}}, \bibinfo {author} {\bibfnamefont {S.~A.}\ \bibnamefont {Moses}}, \bibinfo {author} {\bibfnamefont {M.~S.}\ \bibnamefont {Allman}}, \bibinfo {author} {\bibfnamefont {C.~H.}\ \bibnamefont {Baldwin}}, \bibinfo {author} {\bibfnamefont {M.}~\bibnamefont {Foss-Feig}}, \bibinfo {author} {\bibfnamefont {D.}~\bibnamefont {Hayes}}, \bibinfo {author} {\bibfnamefont {K.}~\bibnamefont {Mayer}}, \bibinfo {author} {\bibfnamefont {C.}~\bibnamefont {Ryan-Anderson}},\ and\ \bibinfo {author} {\bibfnamefont {B.}~\bibnamefont {Neyenhuis}},\ }\bibfield  {title} {\bibinfo {title} {Demonstration of the trapped-ion quantum ccd computer architecture},\ }\href {https://doi.org/10.1038/s41586-021-03318-4} {\bibfield  {journal} {\bibinfo  {journal} {Nature}\
  }\textbf {\bibinfo {volume} {592}},\ \bibinfo {pages} {209–213} (\bibinfo {year} {2021})}\BibitemShut {NoStop}%
\bibitem [{\citenamefont {Wallman}\ and\ \citenamefont {Emerson}(2016)}]{PhysRevA.94.052325}%
  \BibitemOpen
  \bibfield  {author} {\bibinfo {author} {\bibfnamefont {J.~J.}\ \bibnamefont {Wallman}}\ and\ \bibinfo {author} {\bibfnamefont {J.}~\bibnamefont {Emerson}},\ }\bibfield  {title} {\bibinfo {title} {Noise tailoring for scalable quantum computation via randomized compiling},\ }\href {https://doi.org/10.1103/PhysRevA.94.052325} {\bibfield  {journal} {\bibinfo  {journal} {Phys. Rev. A}\ }\textbf {\bibinfo {volume} {94}},\ \bibinfo {pages} {052325} (\bibinfo {year} {2016})}\BibitemShut {NoStop}%
\bibitem [{\citenamefont {Ishii}\ \emph {et~al.}(2025)\citenamefont {Ishii}, \citenamefont {Qassim}, \citenamefont {Kurita}, \citenamefont {Emerson}, \citenamefont {Maruyama}, \citenamefont {Oshima},\ and\ \citenamefont {Sato}}]{ishii2025implementation}%
  \BibitemOpen
  \bibfield  {author} {\bibinfo {author} {\bibfnamefont {M.}~\bibnamefont {Ishii}}, \bibinfo {author} {\bibfnamefont {H.}~\bibnamefont {Qassim}}, \bibinfo {author} {\bibfnamefont {T.}~\bibnamefont {Kurita}}, \bibinfo {author} {\bibfnamefont {J.}~\bibnamefont {Emerson}}, \bibinfo {author} {\bibfnamefont {K.}~\bibnamefont {Maruyama}}, \bibinfo {author} {\bibfnamefont {H.}~\bibnamefont {Oshima}},\ and\ \bibinfo {author} {\bibfnamefont {S.}~\bibnamefont {Sato}},\ }\bibfield  {title} {\bibinfo {title} {Implementation and verification of coherent error suppression using randomized compiling for grover's algorithm on a trapped-ion device},\ }\href {https://arxiv.org/abs/2503.05344} {\bibfield  {journal} {\bibinfo  {journal} {arXiv preprint arXiv:2503.05344}\ } (\bibinfo {year} {2025})}\BibitemShut {NoStop}%
\bibitem [{\citenamefont {Kurita}\ \emph {et~al.}(2023)\citenamefont {Kurita}, \citenamefont {Qassim}, \citenamefont {Ishii}, \citenamefont {Oshima}, \citenamefont {Sato},\ and\ \citenamefont {Emerson}}]{Kurita_2023}%
  \BibitemOpen
  \bibfield  {author} {\bibinfo {author} {\bibfnamefont {T.}~\bibnamefont {Kurita}}, \bibinfo {author} {\bibfnamefont {H.}~\bibnamefont {Qassim}}, \bibinfo {author} {\bibfnamefont {M.}~\bibnamefont {Ishii}}, \bibinfo {author} {\bibfnamefont {H.}~\bibnamefont {Oshima}}, \bibinfo {author} {\bibfnamefont {S.}~\bibnamefont {Sato}},\ and\ \bibinfo {author} {\bibfnamefont {J.}~\bibnamefont {Emerson}},\ }\bibfield  {title} {\bibinfo {title} {Synergetic quantum error mitigation by randomized compiling and zero-noise extrapolation for the variational quantum eigensolver},\ }\href {https://doi.org/10.22331/q-2023-11-20-1184} {\bibfield  {journal} {\bibinfo  {journal} {Quantum}\ }\textbf {\bibinfo {volume} {7}},\ \bibinfo {pages} {1184} (\bibinfo {year} {2023})}\BibitemShut {NoStop}%
\bibitem [{\citenamefont {Ville}\ \emph {et~al.}(2022)\citenamefont {Ville}, \citenamefont {Morvan}, \citenamefont {Hashim}, \citenamefont {Naik}, \citenamefont {Lu}, \citenamefont {Mitchell}, \citenamefont {Kreikebaum}, \citenamefont {O'Brien}, \citenamefont {Wallman}, \citenamefont {Hincks}, \citenamefont {Emerson}, \citenamefont {Smith}, \citenamefont {Younis}, \citenamefont {Iancu}, \citenamefont {Santiago},\ and\ \citenamefont {Siddiqi}}]{PhysRevResearch.4.033140}%
  \BibitemOpen
  \bibfield  {author} {\bibinfo {author} {\bibfnamefont {J.-L.}\ \bibnamefont {Ville}}, \bibinfo {author} {\bibfnamefont {A.}~\bibnamefont {Morvan}}, \bibinfo {author} {\bibfnamefont {A.}~\bibnamefont {Hashim}}, \bibinfo {author} {\bibfnamefont {R.~K.}\ \bibnamefont {Naik}}, \bibinfo {author} {\bibfnamefont {M.}~\bibnamefont {Lu}}, \bibinfo {author} {\bibfnamefont {B.}~\bibnamefont {Mitchell}}, \bibinfo {author} {\bibfnamefont {J.-M.}\ \bibnamefont {Kreikebaum}}, \bibinfo {author} {\bibfnamefont {K.~P.}\ \bibnamefont {O'Brien}}, \bibinfo {author} {\bibfnamefont {J.~J.}\ \bibnamefont {Wallman}}, \bibinfo {author} {\bibfnamefont {I.}~\bibnamefont {Hincks}}, \bibinfo {author} {\bibfnamefont {J.}~\bibnamefont {Emerson}}, \bibinfo {author} {\bibfnamefont {E.}~\bibnamefont {Smith}}, \bibinfo {author} {\bibfnamefont {E.}~\bibnamefont {Younis}}, \bibinfo {author} {\bibfnamefont {C.}~\bibnamefont {Iancu}}, \bibinfo {author} {\bibfnamefont {D.~I.}\ \bibnamefont {Santiago}},\ and\ \bibinfo {author} {\bibfnamefont
  {I.}~\bibnamefont {Siddiqi}},\ }\bibfield  {title} {\bibinfo {title} {Leveraging randomized compiling for the quantum imaginary-time-evolution algorithm},\ }\href {https://doi.org/10.1103/PhysRevResearch.4.033140} {\bibfield  {journal} {\bibinfo  {journal} {Phys. Rev. Res.}\ }\textbf {\bibinfo {volume} {4}},\ \bibinfo {pages} {033140} (\bibinfo {year} {2022})}\BibitemShut {NoStop}%
\bibitem [{\citenamefont {Winick}\ \emph {et~al.}(2022)\citenamefont {Winick}, \citenamefont {Wallman}, \citenamefont {Dahlen}, \citenamefont {Hincks}, \citenamefont {Ospadov},\ and\ \citenamefont {Emerson}}]{winick2022concepts}%
  \BibitemOpen
  \bibfield  {author} {\bibinfo {author} {\bibfnamefont {A.}~\bibnamefont {Winick}}, \bibinfo {author} {\bibfnamefont {J.~J.}\ \bibnamefont {Wallman}}, \bibinfo {author} {\bibfnamefont {D.}~\bibnamefont {Dahlen}}, \bibinfo {author} {\bibfnamefont {I.}~\bibnamefont {Hincks}}, \bibinfo {author} {\bibfnamefont {E.}~\bibnamefont {Ospadov}},\ and\ \bibinfo {author} {\bibfnamefont {J.}~\bibnamefont {Emerson}},\ }\bibfield  {title} {\bibinfo {title} {Concepts and conditions for error suppression through randomized compiling},\ }\href {https://arxiv.org/abs/2212.07500} {\bibfield  {journal} {\bibinfo  {journal} {arXiv preprint arXiv:2212.07500}\ } (\bibinfo {year} {2022})}\BibitemShut {NoStop}%
\bibitem [{\citenamefont {Gu}\ \emph {et~al.}(2023)\citenamefont {Gu}, \citenamefont {Ma}, \citenamefont {Forcellini},\ and\ \citenamefont {Liu}}]{PhysRevLett.130.250601}%
  \BibitemOpen
  \bibfield  {author} {\bibinfo {author} {\bibfnamefont {Y.}~\bibnamefont {Gu}}, \bibinfo {author} {\bibfnamefont {Y.}~\bibnamefont {Ma}}, \bibinfo {author} {\bibfnamefont {N.}~\bibnamefont {Forcellini}},\ and\ \bibinfo {author} {\bibfnamefont {D.~E.}\ \bibnamefont {Liu}},\ }\bibfield  {title} {\bibinfo {title} {Noise-resilient phase estimation with randomized compiling},\ }\href {https://doi.org/10.1103/PhysRevLett.130.250601} {\bibfield  {journal} {\bibinfo  {journal} {Phys. Rev. Lett.}\ }\textbf {\bibinfo {volume} {130}},\ \bibinfo {pages} {250601} (\bibinfo {year} {2023})}\BibitemShut {NoStop}%
\bibitem [{\citenamefont {Urbanek}\ \emph {et~al.}(2021)\citenamefont {Urbanek}, \citenamefont {Nachman}, \citenamefont {Pascuzzi}, \citenamefont {He}, \citenamefont {Bauer},\ and\ \citenamefont {de~Jong}}]{urbanek2021mitigating}%
  \BibitemOpen
  \bibfield  {author} {\bibinfo {author} {\bibfnamefont {M.}~\bibnamefont {Urbanek}}, \bibinfo {author} {\bibfnamefont {B.}~\bibnamefont {Nachman}}, \bibinfo {author} {\bibfnamefont {V.~R.}\ \bibnamefont {Pascuzzi}}, \bibinfo {author} {\bibfnamefont {A.}~\bibnamefont {He}}, \bibinfo {author} {\bibfnamefont {C.~W.}\ \bibnamefont {Bauer}},\ and\ \bibinfo {author} {\bibfnamefont {W.~A.}\ \bibnamefont {de~Jong}},\ }\bibfield  {title} {\bibinfo {title} {Mitigating depolarizing noise on quantum computers with noise-estimation circuits},\ }\href {http://dx.doi.org/10.1103/PhysRevLett.127.270502} {\bibfield  {journal} {\bibinfo  {journal} {Physical review letters}\ }\textbf {\bibinfo {volume} {127}},\ \bibinfo {pages} {270502} (\bibinfo {year} {2021})}\BibitemShut {NoStop}%
\bibitem [{\citenamefont {Faehrmann}\ \emph {et~al.}(2022)\citenamefont {Faehrmann}, \citenamefont {Steudtner}, \citenamefont {Kueng}, \citenamefont {Kieferova},\ and\ \citenamefont {Eisert}}]{Faehrmann_2022}%
  \BibitemOpen
  \bibfield  {author} {\bibinfo {author} {\bibfnamefont {P.~K.}\ \bibnamefont {Faehrmann}}, \bibinfo {author} {\bibfnamefont {M.}~\bibnamefont {Steudtner}}, \bibinfo {author} {\bibfnamefont {R.}~\bibnamefont {Kueng}}, \bibinfo {author} {\bibfnamefont {M.}~\bibnamefont {Kieferova}},\ and\ \bibinfo {author} {\bibfnamefont {J.}~\bibnamefont {Eisert}},\ }\bibfield  {title} {\bibinfo {title} {Randomizing multi-product formulas for hamiltonian simulation},\ }\href {https://doi.org/10.22331/q-2022-09-19-806} {\bibfield  {journal} {\bibinfo  {journal} {Quantum}\ }\textbf {\bibinfo {volume} {6}},\ \bibinfo {pages} {806} (\bibinfo {year} {2022})}\BibitemShut {NoStop}%
\bibitem [{\citenamefont {Iyer}\ \emph {et~al.}(2022)\citenamefont {Iyer}, \citenamefont {Jain}, \citenamefont {Bartlett},\ and\ \citenamefont {Emerson}}]{Iyer2022}%
  \BibitemOpen
  \bibfield  {author} {\bibinfo {author} {\bibfnamefont {P.}~\bibnamefont {Iyer}}, \bibinfo {author} {\bibfnamefont {A.}~\bibnamefont {Jain}}, \bibinfo {author} {\bibfnamefont {S.~D.}\ \bibnamefont {Bartlett}},\ and\ \bibinfo {author} {\bibfnamefont {J.}~\bibnamefont {Emerson}},\ }\bibfield  {title} {\bibinfo {title} {Efficient diagnostics for quantum error correction},\ }\href {https://doi.org/10.1103/PhysRevResearch.4.043218} {\bibfield  {journal} {\bibinfo  {journal} {Phys. Rev. Res.}\ }\textbf {\bibinfo {volume} {4}},\ \bibinfo {pages} {043218} (\bibinfo {year} {2022})}\BibitemShut {NoStop}%
\bibitem [{\citenamefont {Jain}\ \emph {et~al.}(2023)\citenamefont {Jain}, \citenamefont {Iyer}, \citenamefont {Bartlett},\ and\ \citenamefont {Emerson}}]{Jain2023}%
  \BibitemOpen
  \bibfield  {author} {\bibinfo {author} {\bibfnamefont {A.}~\bibnamefont {Jain}}, \bibinfo {author} {\bibfnamefont {P.}~\bibnamefont {Iyer}}, \bibinfo {author} {\bibfnamefont {S.~D.}\ \bibnamefont {Bartlett}},\ and\ \bibinfo {author} {\bibfnamefont {J.}~\bibnamefont {Emerson}},\ }\href {https://doi.org/10.1103/PhysRevResearch.5.033049} {\bibinfo {title} {Improved quantum error correction with randomized compiling}} (\bibinfo {year} {2023})\BibitemShut {NoStop}%
\bibitem [{\citenamefont {Iyer}\ \emph {et~al.}(2025)\citenamefont {Iyer}, \citenamefont {Jain}, \citenamefont {Bartlett},\ and\ \citenamefont {Emerson}}]{iyer2025enhancing}%
  \BibitemOpen
  \bibfield  {author} {\bibinfo {author} {\bibfnamefont {P.}~\bibnamefont {Iyer}}, \bibinfo {author} {\bibfnamefont {A.}~\bibnamefont {Jain}}, \bibinfo {author} {\bibfnamefont {S.~D.}\ \bibnamefont {Bartlett}},\ and\ \bibinfo {author} {\bibfnamefont {J.}~\bibnamefont {Emerson}},\ }\bibfield  {title} {\bibinfo {title} {Enhancing decoding performance using efficient error learning},\ }\href {https://arxiv.org/abs/2507.08536} {\bibfield  {journal} {\bibinfo  {journal} {arXiv preprint arXiv:2507.08536}\ } (\bibinfo {year} {2025})}\BibitemShut {NoStop}%
\bibitem [{\citenamefont {Sannamoth}(2026)}]{my_repo}%
  \BibitemOpen
  \bibfield  {author} {\bibinfo {author} {\bibfnamefont {D.}~\bibnamefont {Sannamoth}},\ }\href {https://doi.org/10.5281/zenodo.21042850} {\bibinfo {title} {Codes for easier randomizing gate provide more accurate fidelity estimation}} (\bibinfo {year} {2026})\BibitemShut {NoStop}%
\end{thebibliography}%


\newpage
\appendix

\section{Construction of the Self-Consistent Gauge}
\label{Optimal gauge}

In numerical studies of randomized benchmarking, the simple and standard  approach is to assign an error to each gate, which can generally be gate-dependent. 
From this one can define a fidelity (or infidelity) associated with an average over the fidelities (or infidelities) associated with each of these noisy gates. This is the `error-on-the gate' approach to defining a ``theoretical infidelity" for the numerical simulation. However, the error rate measured by RB
is the average `error-between-the-gates' of the randomized sequence  \cite{Wallman_2018}, which can differ from the naive approach because it accounts for the possibility of coherent combination of the errors from sequential gates, i.e., between the error on any one gate and the error on the subsequent gate. Taking this into account resolves the concern raised in \cite{PhysRevLett.119.130502} about that RB actually measures.

For historical reasons, the above issues are normally discussed as a gauge freedom in defining the theoretical fidelity. For historical continuity, we follow this framework to ask and answer the following question: if we measure a fidelity from a numerical experiment such as RB, how can we compare it to a rigorous theoretical value associated with the error model when the relevant error is the average error-between-the gates? There is a rather simple way to answer this question, which we call the self-consistent gauge (SCG) fidelity.

Consider a set of noisy state representations $\set{\doubleket{\noisy{\rho}_i}}$, a set of noisy measurement representations $\set{\doublebra{\noisy{\mu}_i}}$, and a set of noisy superoperators $\set{\noisy{G}_i}$. With these vectors and matrices we can express the probability of outcome $k$ given any sequence of operators $\noisy{G}_{j_m} \cdots\noisy{G}_{j_1}$ applied on any state $\noisy{\rho}_i$:
\begin{align}
    &p( \text{outcome}~k | \text{operations} = [j_m, \cdots , j_i], \text{state}~i  )  \notag \\
     &= \doublebra{\noisy{\mu}_k} \noisy{G}_{j_m} \cdots\noisy{G}_{j_1} \doubleket{\noisy{\rho}_i}~.
\end{align}
These probabilities are the only physically measurable quantities, therefore the sets
$\set{S\doubleket{\noisy{\rho}_i}}$, $\set{\doublebra{\noisy{\mu}_i} S^{-1}}$, and $\set{S\noisy{G}_iS^{-1}}$ describe the same physical observations (here $S$ could be any $d^2 \times d^2$ invertible matrix).

In light of this gauge freedom, it has been proposed \cite{PhysRevLett.119.130502} that the process fidelity should actually be expressed in the following gauge-dependent form:
\begin{align}
    f(S \noisy{G}_i S^{-1},G_i) = \frac{{\rm{Tr}} (S \noisy{G}_i S^{-1}G_i^{-1})}{d^2}~,
\end{align}
which depends on the gauge matrix $S$. Without endorsing the correctness of this general claim, we adopt this  framework because it is general enough to accommodate the analysis of the difference between the `error-on-the-gate' model that defines the theoretical fidelity and the effective `error-between-the-gates' that determined the observed fidelity in numerical simulations.

Now we get to the crux of the issue: consider 
the simple case where noisy gates take the form
\begin{align}\label{eq:SimpleNoise}
    \noisy{G}_i = L G_i R~~~\forall~~\noisy{G}_i \in \set{\noisy{G}_i},
\end{align}
where $L$ and $R$ are left and right error channels respectively. In such scenario, the probability of outcome $k$ given any sequence of operators $\noisy{G}_{j_m} \cdots\noisy{G}_{j_1}$ applied on any state $\noisy{\rho}_i$ takes the form:
\begin{align}
    &p( \text{outcome}~k | \text{operations} = [j_m, \cdots , j_i], \text{state}~i  )  \notag \\
     &= \doublebra{\noisy{\mu}_k} L {G}_{j_m} RL {G}_{j_{m-1}} \cdots G_2 RL {G}_{j_1} R\doubleket{\noisy{\rho}_i}~.
\end{align}
Since the initial $R$ and last $L$ can be factored in state preparation and measurement errors, it becomes quite obvious that RB will provide an estimate of the process fidelity of $RL$ to the identity, $f(RL)$. On the other hand, 
the process fidelity averaged over the Cliffords is expressed as
\begin{align}
    \symbolset{E}_{\substack{G_i\\{\text{is Clifford}}}} \frac{{\rm{Tr}} (S L {G}_i RS^{-1}G_i^{-1})}{d^2} =&  f(SL)f(RS^{-1}) \notag \\
    &+O(r(SL)r(RS^{-1}))~.
\end{align}
Therefore, the RB fidelity gets in good agreement with the average fidelity obtained in the gauge $S$ as long as 
\begin{align} \label{eq:GaugeAgreement}
    f(RL)  = f(SL)f(RS^{-1}) + O(r(SL)r(RS^{-1}))~.
\end{align}
Two obvious choices are $S= R$ and $S = L^{-1}$, but let's explore options of $S$ that are constrained to physical unitaries. Suppose that $R$ is \emph{decoherent} as defined in \cite{Polar}.
That is, the leading Kraus operator of $R$ is positive semi-definite. 
As shown in \cite{Polar}, it 
follows that the composite fidelity is multiplicative: $f(RL) = f(R)f(L) + O(r(R)r(L))$. In such case, 
the choice $S = I$ (i.e. no gauge transformation) satisfies Eqn.~\ref{eq:GaugeAgreement}. Now, let's express $R = U_R D_R$ and $L = U_L D_L$ where $U_R$,$U_L$ are physical unitary channels and $D_R$,$D_L$ are physical decoherent channels. This polar decomposition is always possible for high-enough fidelity channels, as shown in \cite{Polar}.

\begin{figure*}[!htbp]
    \centering
    \includegraphics[width=0.9
    \textwidth]{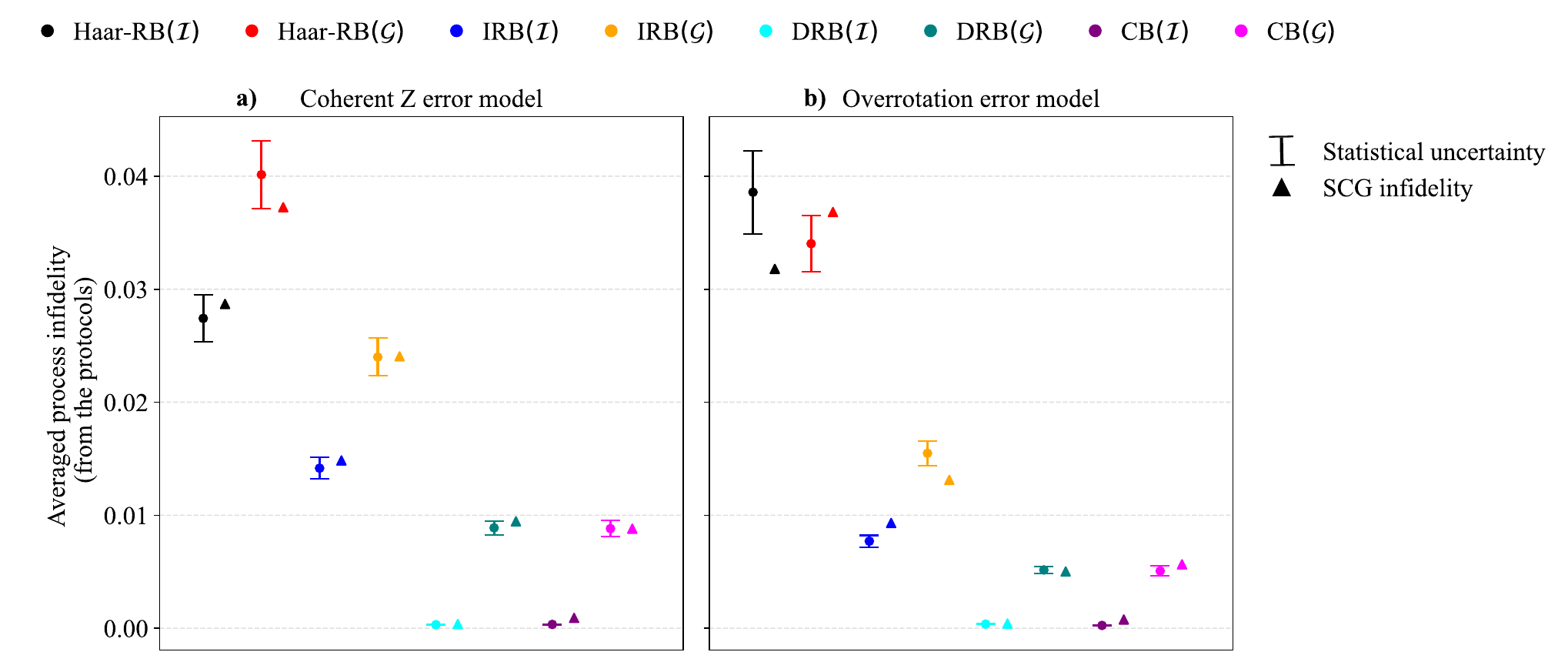} 
  \caption{\footnotesize Numerical plots showing that the estimate of the process infidelities obtained from the different protocols closely matches the self-consistent gauge infidelity values. \textbf{(a)} Error model with two-qubit gates  error of the form $e^{i\theta_2 ZZ}$ with $\theta_2=10^{\circ} $ and single qubit gates having fixed gate  error of the form $e^{i\theta_1 Z}$ with $\theta_1=1^{\circ} $.\textbf{(b)} Error model of the form $U^{1+\delta}$ where $U$ is the unitary on which noise is applied. For two qubit gates $\delta=0.05$ and for single qubit gates $\delta=0.01$ The triangles beside each estimate are obtained using the self-consistent gauge for that experiment as discussed in \ref{Optimal gauge}. The twirling group for finding the self-consistent gauge is similar to the twirling group for each protocol beside which it is plotted.}
  \label{fig:main_fig}
\end{figure*}

Moreover, there is a constructive way to find the polar decomposition of a channel $\Lambda$ close to the identity:
\begin{enumerate}
    \item Express $\Lambda$ in its Choi matrix representation $\Lambda_{\rm Choi}$.
    \item The eigenvalues of $\Lambda_{\rm Choi}$ can be interpreted as probabilities, and the eigenvectors are vectorized Kraus operators. Since $\Lambda$ is close to the identity, there is an eigenvalue of $\Lambda_{\rm Choi}$ that is close to $1$. Its corresponding Kraus operator is the leading Kraus operator $K_0$.
    \item Perform a standard polar decomposition $K_0 = U P$ where $U$ is a $d \times d$ unitary, and $P$ is $d \times d$ a positive semi-definite matrix. Let $\mathcal{U}$ be the superoperator version of $U$.
    \item $\Lambda = \mathcal{U}D$ where $D := \mathcal{U}^\dagger\Lambda$ is purely decoherent.
\end{enumerate}

Leveraging fidelity composition rules with decoherent channels, we get
\begin{align}
    f(RL) &= f(D_R)f(D_L) f(U_R U_L) \notag \\&+ O(r(D_R)r(D_LU_LU_R)+r(D_L)r(U_LU_R))~, \\
    f(SL) &= f(D_L)f(SU_L)  + O(r(SU_L)r(D_L))~, \\
    f(RS^{-1}) &= f(D_R) f(U_R S^{-1}) + O(r(U_RS^{-1})r(D_R))~.
\end{align}
Substituting those in Eqn.~\ref{eq:GaugeAgreement} and omitting the second order terms (s.o.t), we get the condition
\begin{align}
    f(U_R U_L) =   f(U_R S^{-1})f(SU_L) + {\rm s.o.t.}~,
\end{align}
meaning that $S=U_R$ or $S= U_L^{-1}$ satisfy Eqn.~\ref{eq:GaugeAgreement}. In other words, if we are given the simple noise model $\noisy{G}_i = U_L D_L G_i U_RD_R$, the average gate set fidelities of $\set{U_L^\dagger\noisy{G}_i U_L}$ and of 
$\set{U_R\noisy{G}_i U_R^\dagger}$ are in good agreement with the reported RB number.

It is now time to depart from the simple noise model $\noisy{G}_i = L G_i R$ (Eqn.~\ref{eq:SimpleNoise}). To do so, we will need to consider gate sets that are groups. As we shall see, we define a gauge-fixing method for any given group $\mathbb{G}$ as long as it forms a $t$-design for $t \geq 1$. First, it has been shown that given such group $\mathbb{G}$, there exist $L$ and $R$ (independent of $S$) such that;
\begin{align}\label{eq:RightEigen}
    \mathbb{E}_{G_i \in \mathbb{G}} G_i^{-1} (RS^{-1}) (S\noisy{G}_i S^{-1}) &= \Lambda^{\mathbb{G}}RS^{-1} ~, \\
    \mathbb{E}_{G_i \in \mathbb{G}}   (S\noisy{G}_iS^{-1}) (SL) G_i^{-1} &= SL \Lambda^{\mathbb{G}} ~,\label{eq:LeftEigen}
\end{align}
where $\Lambda^{\mathbb{G}}$ is a channel twirled by the group $\mathbb{G}$ and is independent of $S$. It is shown in \cite{Wallman_2018} that the decaying fidelities extracted by CB-like and RB-like protocols are the fidelities of $\Lambda^{\mathbb{G}}$. To understand what the fidelities of $\Lambda^{\mathbb{G}}$ correspond to, first notice that $L$ and $R$ can be approximated up to second order terms (s.o.t.) in the infidelities by the following physical channels \cite{Carignan_Dugas_2018}:
\begin{align}
    L &= \underbrace{\mathbb{E}_{G_i \in \mathbb{G}}\noisy{G}_4\noisy{G}_3\noisy{G}_2\noisy{G}_1 (G_4G_3G_2G_1)^{-1}}_{=: \mathcal{E}_L}+ {\rm s.o.t.} \\
    R &= \underbrace{\mathbb{E}_{G_i \in \mathbb{G}}(G_4G_3G_2G_1)^{-1}\noisy{G}_4\noisy{G}_3\noisy{G}_2\noisy{G}_1}_{=:\mathcal{E_R}} + {\rm s.o.t.}
\end{align}
Why $4$? Because the eigenvalues of the twirl super-duper operator $T_{\mathbb {G}}:= \mathbb{E}_{G_i \in \mathbb{G}}G_i^* \otimes \noisy{G}_i$ are either close to $1$ or at most $O(\sqrt{r_i})$, meaning that concatenating this twirl 4 times brings $O(\sqrt{r_i})$ to $O(r_i^2)$ (here $r_i$ refers to infidelities). 

Similarly as earlier, let's decompose $\mathcal{E}_L = U_L D_L$ and $\mathcal{E}_R = U_R D_R$ where $U_R$,$U_L$ are physical unitary channels and $D_R$,$D_L$ are physical decoherent channels.
Substituting these into Eqns.~\ref{eq:RightEigen} and \ref{eq:LeftEigen}, and taking the fidelity on each side, we obtain:
\begin{align}
    &f\left(\left[\mathbb{E}_{G_i \in \mathbb{G}}  S\noisy{G}_i S^{-1}G_i^{-1}\right] [U_RD_RS^{-1}] \right) =  ~ \notag \\
    & ~~~~~f(\Lambda^{\mathbb{G}})f(U_RD_RS^{-1}) + {\rm s.o.t} ~,\\
    &f\left(\left[\mathbb{E}_{G_i \in \mathbb{G}}  (G_i^{-1}S\noisy{G}_iS^{-1})\right] \left[SU_LD_L\right]\right)  =  \notag \\
    &~~~~~f(\Lambda^{\mathbb{G}}) f(SU_LD_L) + {\rm s.o.t.} ~.
\end{align}
This means that choosing either $S=U_R$ or $S=U_L^{-1}$ yield to a good agreement between the experimentally-obtained average decay and the average fidelity:
\begin{align}
    \mathbb{E}_{G_i \in \mathbb{G}} f\left( S\noisy{G}_iS^{-1}, G_i\right) =f(\Lambda^{\mathbb G}) + {\rm s.o.t.}~.
\end{align}

In summary, the gauge-fixing method that we propose to obtain an agreement between experimentally-obtained values and the average fidelity of a group $\mathbb{G}$ is:
\begin{enumerate}
    \item Find or approximate $\mathcal{E}_R$ and $\mathcal{E}_L$ as defined in Eqns.~\ref{eq:RightEigen} and \ref{eq:LeftEigen}. The approximation can be done by sampling rather than taking the entire average.
    \item Decompose $\mathcal{E}_L = U_L D_L$ and $\mathcal{E}_R = U_R D_R$ where $U_R$,$U_L$ are physical unitary channels and $D_R$,$D_L$ are physical decoherent channels. We explain how to do this polar decomposition earlier in this section.
    \item Choose either $S=U_R$ or $S=U_L^{-1}$.
\end{enumerate}
We followed this method to plot the triangles which denote the ``self-consistent gauge''(SCG) infidelity in Fig.~\ref{fig:main_fig}.

\section{Estimating the process infidelity of the interleaved gate}
\label{process infidelity calculation of intleaved gate}
When the twirling group is Clifford, the estimator for the process infidelity of the interleaved gate can be constructed from the single decay parameters for each of the interleaved sequence and the reference sequence \cite{PhysRevLett.109.080505}. In this case the estimator for the process infidelity of the error acting on the interleaved gate is given by
\begin{equation}
\hat{\epsilon}^{\mathcal{ F}}=\frac{d^2-1}{d^2}\Big(1-\frac{p^{\mathcal{E\circ F}}}{p^{\mathcal{E}}}\Big)
\end{equation}
 Using the relation of $p=1-\frac{d^2}{d^2-1}\epsilon$ \cite{Carignan_Dugas_2019}, the above equation can be rewritten as
\begin{equation}
\hat{\epsilon}^{\mathcal{F}}=\frac{\epsilon^{\mathcal{E\circ F}}-\epsilon^{\mathcal{E}}}{1-\frac{d^2}{d^2-1}\epsilon^{\mathcal{E}}}
\label{eq:expressing depolarizing in terms of infidelity}
\end{equation}
which under expansion up to second order terms gives
\begin{equation}
    \hat{\epsilon}^{\mathcal{ F}}=  \epsilon^{\mathcal{E\circ F}}- \epsilon^{\mathcal{ E}}+ \frac{d^2}{d^2-1}\epsilon^{\mathcal{ E\circ F}} \epsilon^{\mathcal{ E}}- \frac{d^2}{d^2-1}(\epsilon^{\mathcal{ E}})^2+O(\epsilon^3)
    \label{eq: depolarizing expansion}
\end{equation}
 
However not all interleaved benchmarking schemes do admit a single decay parameter `$p$' because the twirling group need not be a unitary 2-design \cite{PhysRevA.80.012304}. This motivates the need for a unified and protocol-independent estimator for the process infidelity of the error on the interleaved gate.
For such protocols, we begin with the observation that an estimator for the process fidelity of the interleaved gate is given by
\begin{equation}
  \hat{F}_p^{\mathcal{ F}}  =F_p^{\mathcal{E\circ F}} / F_p^{\mathcal{E}}
  \label{eqn:product of process fidelities}
\end{equation}
where $F_p^{\mathcal{E}}$ and $F_p^{\mathcal{ F}}$ are the observed process fidelities of the cycle of randomizing gates and interleaved gate respectively. 
We now show that this estimator is negligibly different from that proposed  originally in IRB.
Writing the estimator for process fidelity $\hat{F}_p=1-\hat{\epsilon}$, we obtain
\begin{equation}
   \hat{\epsilon}^{\mathcal{ F}}= 1- \frac{1-\epsilon^{\mathcal{E\circ F}}}{1-\epsilon^{\mathcal{E}}}
   \label{eq:ratio of infidelities}
\end{equation}
which reproduces Eqn.~\ref{eq:ratio of two fidelities}. Expanding Eqn.~\ref{eq:ratio of infidelities} expression up to second order terms gives

\begin{equation}
    \hat{\epsilon}^{\mathcal{ F}}= \epsilon^{\mathcal{E\circ F}}- \epsilon^{\mathcal{ E}}+ \epsilon^{\mathcal{ E\circ F}} \epsilon^{\mathcal{ E}}- (\epsilon^{\mathcal{ E}})^2+O(\epsilon^3)
    \label{eq: ratio expansion}
\end{equation}

Comparing Eqn.~\ref{eq: depolarizing expansion} and  Eqn.~\ref{eq: ratio expansion}, we observe that the two expressions are identical to each other up to the second order terms except the presence of the prefactor $\frac{d^2}{d^2-1}$. 

To quantify the difference when the twirling group is Clifford, we compare the two estimators for the error models used in our numerical studies. For the coherent Z error model, using Eqn.~\ref{eq:ratio of two fidelities} we get $\epsilon^{\mathcal{F}}=6.978(2.901)\times10^{-3}$ and using Eqn.~\ref{eq:expressing depolarizing in terms of infidelity}, we get $\epsilon^{\mathcal{F}}=6.985(2.904)\times10^{-3}$. Similarly, for the overrotation error model, using Eqn.~\ref{eq:ratio of two fidelities} we get $\epsilon^{\mathcal{F}}=8.450(2.397)\times10^{-3}$ and using Eqn.~\ref{eq:expressing depolarizing in terms of infidelity}, we get $\epsilon^{\mathcal{F}}=8.456(2.398)\times10^{-3}$. 
In both cases, the discrepancy between the two estimators is on the order of $O(10^{-6})$, which is negligible compared to the statistical uncertainty. This demonstrates that estimating the interleaved gate infidelity via Eqn.~\ref{eq:ratio of two fidelities} provides a consistent and accurate approach, while remaining applicable to a general class of interleaved benchmarking protocols.

\onecolumngrid
\section{{Numerical and experimental data}}
\label{Numerical and experimental data}

\begin{table*}[htbp]
\centering
\renewcommand{\arraystretch}{1.80}
\setlength{\tabcolsep}{9.0pt}
\footnotesize
\begin{tabular}{|l|l|c|c|c|c|}
\hline
Label
& \begin{tabular}{c}Interleaved\\protocol pair\end{tabular}
& \begin{tabular}{c}process inf.\\$(10^{-2})$\end{tabular}
& \begin{tabular}{c}CNOT err.\\$(10^{-2})$\end{tabular}
& \begin{tabular}{c}Sys. UB\\$(10^{-2})$\end{tabular}
& \begin{tabular}{c}Sys. LB\\$(10^{-2})$\end{tabular} \\
\hline
\multirow{2}{*}{Haar}
& Haar-RB($\mathcal{G}$)
& $4.01\pm0.30$
& \multirow{2}{*}{$1.30\pm0.37$}
& \multirow{2}{*}{$12.94$}
& \multirow{2}{*}{$0.12$} \\
\cline{2-3}
& Haar-RB($\mathcal{I}$)
& $2.74\pm0.20$
&
&
& \\
\hline

\multirow{2}{*}{Clifford}
& IRB($\mathcal{G}$)
& $2.39\pm0.16$
& \multirow{2}{*}{$0.99\pm 0.19$}
& \multirow{2}{*}{$7.36$}
& \multirow{2}{*}{$0.13+$} \\
\cline{2-3}
& IRB($\mathcal{I}$)
& $1.41\pm 0.09$
&
&
& \\
\hline

\multirow{2}{*}{Local Clifford}
& DRB($\mathcal{G}$)
& $0.88\pm0.06$
& \multirow{2}{*}{$0.85\pm0.06$}
& \multirow{2}{*}{$1.25$}
& \multirow{2}{*}{$0.58$} \\
\cline{2-3}
& DRB($\mathcal{I}$)
& $0.03\pm0.001$
&
&
& \\
\hline

\multirow{2}{*}{Pauli}
& CB($\mathcal{G}$)
& $0.88\pm0.07$
& \multirow{2}{*}{$0.84\pm0.07$}
& \multirow{2}{*}{$1.26$}
& \multirow{2}{*}{$0.57$} \\
\cline{2-3}
& CB($\mathcal{I}$)
& $0.03\pm0.003$
&
&
& \\
\hline
\end{tabular}
\caption{Numerical values underlying Fig.\ref{fig:sys uncertainty as function of twirling group}. Each $(\mathcal{G}),(\mathcal{I})$ pair gives a single
inferred CNOT error rate and corresponding systematic interval. Statistical
uncertainties denote one standard error from resampling. The columns ``Sys. UB'' and ``Sys. LB''
denote the upper  and lower systematic bounds using Eqn. \ref{eq:systematic-bound}. Finally, the theoretical value of CNOT process infidelity is $0.759 \times 10^{-2}$.  } 
\label{tab:fig2_data}
\end{table*}

\begin{table*}[htbp]
\centering

\renewcommand{\arraystretch}{1.80}
\setlength{\tabcolsep}{9.0pt}
\footnotesize
\begin{tabular}{|c|c|c|c|c|}
\hline
\begin{tabular}{c}
Avg. 1Q err.\\
$(10^{-3})$
\end{tabular}
&
\begin{tabular}{c}
Haar\\
$(10^{-2})$
\end{tabular}
&
\begin{tabular}{c}
Clifford\\
$(10^{-2})$
\end{tabular}
&
\begin{tabular}{c}
Local Clifford\\
$(10^{-2})$
\end{tabular}
&
\begin{tabular}{c}
Pauli\\
$(10^{-2})$
\end{tabular}
\\
\hline
$2.33$ & $28.98$ & $15.24$ & $4.23$ & $5.00$ \\
\hline
$1.61$ & $21.25$ & $12.13$ & $2.42$ & $2.55$ \\
\hline
$0.62$ & $15.05$ & $9.61$ & $1.52$ & $1.21$ \\
\hline
$0.28$ & $12.62$ & $8.04$ & $0.84$ & $1.04$ \\
\hline
$0.25$ & $11.46$& $7.44$ & $0.75$ & $0.85$\\
\hline
$0.13$ & $11.09$ & $7.38$ & $0.57$ &$0.64$  \\
\hline
$0.08$ & $11.20$ & $7.18$ & $0.42$ & $0.23$ \\
\hline
\end{tabular}
\caption{
Numerical values underlying Fig.\ref{fig:rel error vs ratio}. The first column gives the average
single-qubit gate error rate used on the horizontal axis. The remaining columns
give the size of the systematic uncertainty which is the difference between upper bound and lower bound of the systematic uncertainty, for each randomization label.
}
\label{tab:sys_width_vs_1q_error}
\end{table*}

\begin{table*}[t]
\centering

\renewcommand{\arraystretch}{1.70}
\setlength{\tabcolsep}{2.3pt}
\footnotesize
\begin{tabular}{|l|l|c|c|c|c|c|c|c|}
\hline
Label
& \begin{tabular}{c}Interleaved\\protocol pair\end{tabular}
& \begin{tabular}{c}process inf.\\$(10^{-2})$\end{tabular}
& \begin{tabular}{c}CNOT err.\\$(10^{-2})$\end{tabular}
& \begin{tabular}{c}Sys. LB\\$(10^{-2})$\end{tabular}
& \begin{tabular}{c}Sys. UB\\$(10^{-2})$\end{tabular}
& \begin{tabular}{c}SCG CNOT err.\\$(10^{-2})$\end{tabular}
& \begin{tabular}{c}SCG UB\\$(10^{-2})$\end{tabular}
& \begin{tabular}{c}SCG LB\\$(10^{-2})$\end{tabular} \\
\hline

\multirow{2}{*}{Haar}
& Haar-RB($\mathcal{G}$)
& $3.40\pm0.24$
& \multirow{2}{*}{$-0.47\pm0.46$}
& \multirow{2}{*}{$13.98$}
& \multirow{2}{*}{$0.01$}
& \multirow{2}{*}{$0.51$}
& \multirow{2}{*}{$13.23$}
& \multirow{2}{*}{$0.01$} \\
\cline{2-3}
& Haar-RB($\mathcal{I}$)
& $3.85\pm0.36$
&
&
&
&
&
& \\
\hline

\multirow{2}{*}{Clifford}
& IRB($\mathcal{G}$)
& $1.54\pm0.10$
& \multirow{2}{*}{$0.78\pm0.12$}
& \multirow{2}{*}{$4.45$}
& \multirow{2}{*}{$0.13$}
& \multirow{2}{*}{$0.38$}
& \multirow{2}{*}{$4.40$}
& \multirow{2}{*}{$0.03$} \\
\cline{2-3}
& IRB($\mathcal{I}$)
& $0.77\pm0.05$
&
&
&
&
&
& \\
\hline

\multirow{2}{*}{Local Clifford}
& DRB($\mathcal{G}$)
& $0.51\pm0.03$
& \multirow{2}{*}{$0.47\pm0.03$}
& \multirow{2}{*}{$0.82$}
& \multirow{2}{*}{$0.27$}
& \multirow{2}{*}{$0.46$}
& \multirow{2}{*}{$0.83$}
& \multirow{2}{*}{$0.25$} \\
\cline{2-3}
& DRB($\mathcal{I}$)
& $0.03\pm0.001$
&
&
&
&
&
& \\
\hline

\multirow{2}{*}{Pauli}
& CB($\mathcal{G}$)
& $0.50\pm0.04$
& \multirow{2}{*}{$0.48\pm0.04$}
& \multirow{2}{*}{$0.75$}
& \multirow{2}{*}{$0.30$}
& \multirow{2}{*}{$0.48$}
& \multirow{2}{*}{$1.06$}
& \multirow{2}{*}{$0.22$} \\
\cline{2-3}
& CB($\mathcal{I}$)
& $0.02\pm0.002$
&
&
&
&
&
& \\
\hline
\end{tabular}
\caption{
Numerical values underlying Fig.~\ref{fig:true and theoretical}.  The columns for SCG corresponds to the self-consistent gauge estimates and bounds as detailed in Appendix. \ref{Optimal gauge}.
The theoretical value of the CNOT process infidelity is $0.461 \times 10^{-2}$}
\label{tab:Fig.4 data}
\end{table*}

\begin{table*}[htbp]
\centering

\renewcommand{\arraystretch}{1.80}
\setlength{\tabcolsep}{6.5pt}
\footnotesize
\begin{tabular}{|l|l|l|c|c|c|c|}
\hline
Label
& Interference
& \begin{tabular}{c}Interleaved\\protocol pair\end{tabular}
& \begin{tabular}{c}process inf.\\$(10^{-2})$\end{tabular}
& \begin{tabular}{c}CNOT err.\\$(10^{-2})$\end{tabular}
& \begin{tabular}{c}Sys. UB\\$(10^{-2})$\end{tabular}
& \begin{tabular}{c}Sys. LB\\$(10^{-2})$\end{tabular} \\
\hline

\multirow{4}{*}{Clifford}
& \multirow{2}{*}{Constructive}
& IRB($\mathcal{G}$)
& $2.67\pm0.20$
& \multirow{2}{*}{$1.91\pm0.20$}
& \multirow{2}{*}{$6.22$}
& \multirow{2}{*}{$0.58$} \\
\cline{3-4}
&
& IRB($\mathcal{I}$)
& $0.77\pm0.04$
&
&
& \\
\cline{2-7}

& \multirow{2}{*}{Destructive}
& IRB($\mathcal{G}$)
& $0.00\pm0.00$
& \multirow{2}{*}{$-0.77\pm0.04$}
& \multirow{2}{*}{$0.77$}
& \multirow{2}{*}{$0.77$} \\
\cline{3-4}
&
& IRB($\mathcal{I}$)
& $0.77\pm0.04$
&
&
& \\
\hline

\end{tabular}
\caption{Numerical values underlying Fig ~\ref{fig:construction_destruction}. The theoretical value of process infidelity of CNOT is $0.759 \times 10^{-2}$. Notice that for the destructive case, the systematic upper bound and lower bound collapse to a single value which is consistent with the theoretical value}
\label{tab:fig 5 data}
\end{table*}

\begin{table*}[t]
\centering

\label{tab:xrb_device_numerics}
\renewcommand{\arraystretch}{1.70}
\setlength{\tabcolsep}{4.0pt}
\scriptsize
\begin{tabular}{|l|l|l|c|c|c|c|c|c|}
\hline
Device
& Label
& \begin{tabular}{c}Interleaved\\protocol pair\end{tabular}
& \begin{tabular}{c}process inf.\\$(10^{-2})$\end{tabular}
& \begin{tabular}{c}CNOT err.\\$(10^{-2})$\end{tabular}
& \begin{tabular}{c}Sys. UB\\$(10^{-2})$\end{tabular}
& \begin{tabular}{c}Sys. LB\\$(10^{-2})$\end{tabular}
& \begin{tabular}{c}XRB UB\\$(10^{-2})$\end{tabular}
& \begin{tabular}{c}XRB LB\\$(10^{-2})$\end{tabular} \\
\hline

\multirow{6}{*}{AQT(4,5)}
& \multirow{2}{*}{Haar}
& Haar-RB($\mathcal{G}$)
& $9.3\pm3.6$
& \multirow{2}{*}{$-1.22\pm5.18$}
& \multirow{2}{*}{$35.49$}
& \multirow{2}{*}{$0.03$}
& \multirow{2}{*}{$27.50$}
& \multirow{2}{*}{$0.04$} \\
\cline{3-4}
&
& Haar-RB($\mathcal{I}$)
& $10.4\pm2.9$
&
&
&
&
& \\
\cline{2-9}

& \multirow{2}{*}{Clifford}
& IRB($\mathcal{G}$)
& $6.5\pm2.1$
& \multirow{2}{*}{$1.87\pm 2.42$}
& \multirow{2}{*}{$21.04$}
& \multirow{2}{*}{$0.15$}
& \multirow{2}{*}{$17.06$}
& \multirow{2}{*}{$0.19$} \\
\cline{3-4}
&
& IRB($\mathcal{I}$)
& $4.71\pm 0.98$
&
&
&
&
& \\
\cline{2-9}

& \multirow{2}{*}{Pauli}
& CB($\mathcal{G}$)
& $3.16\pm0.28$
& \multirow{2}{*}{$1.96\pm 0.30$}
& \multirow{2}{*}{$8.14$}
& \multirow{2}{*}{$0.46$}
& \multirow{2}{*}{--}
& \multirow{2}{*}{--} \\
\cline{3-4}
&
& CB($\mathcal{I}$)
& $1.22\pm 0.1$
&
&
&
&
& \\
\hline

\multirow{6}{*}{AQT(5,6)}
& \multirow{2}{*}{Haar}
& Haar-RB($\mathcal{G}$)
& $5.68 \pm0.75$
& \multirow{2}{*}{$1.92\pm0.87$}
& \multirow{2}{*}{$17.95$}
& \multirow{2}{*}{$0.19$}
& \multirow{2}{*}{$6.63$}
& \multirow{2}{*}{$0.55$} \\
\cline{3-4}
&
& Haar-RB($\mathcal{I}$)
& $3.83\pm0.38$
&
&
&
&
& \\
\cline{2-9}

& \multirow{2}{*}{Clifford}
& IRB($\mathcal{G}$)
& $3.48\pm0.48$
& \multirow{2}{*}{$1.50\pm0.54$}
& \multirow{2}{*}{$10.49$}
& \multirow{2}{*}{$0.20$}
& \multirow{2}{*}{$4.09$}
& \multirow{2}{*}{$0.54$} \\
\cline{3-4}
&
& IRB($\mathcal{I}$)
& $2.01\pm0.23$
&
&
&
&
& \\
\cline{2-9}

& \multirow{2}{*}{Pauli}
& CB($\mathcal{G}$)
& $1.156\pm0.06$
& \multirow{2}{*}{$0.67\pm 0.06$}
& \multirow{2}{*}{$3.11$}
& \multirow{2}{*}{$0.14$}
& \multirow{2}{*}{--}
& \multirow{2}{*}{--} \\
\cline{3-4}
&
& CB($\mathcal{I}$)
& $0.486\pm0.025$
&
&
&
&
& \\
\hline

\multirow{4}{*}{H0(0,1)}
& \multirow{2}{*}{Haar}
& Haar-RB($\mathcal{G}$)
& $3.66\pm 0.81$
& \multirow{2}{*}{$1.26 \pm 0.87$}
& \multirow{2}{*}{$11.69$}
& \multirow{2}{*}{$0.12$}
& \multirow{2}{*}{$4.58$}
& \multirow{2}{*}{$0.34$} \\
\cline{3-4}
&
& Haar-RB($\mathcal{I}$)
& $2.43 \pm 0.27$
&
&
&
&
& \\
\cline{2-9}

& \multirow{2}{*}{Pauli}
& CB($\mathcal{G}$)
& $0.83 \pm 0.053$
& \multirow{2}{*}{$0.73\pm0.05$}
& \multirow{2}{*}{$1.49$}
& \multirow{2}{*}{$0.35$}
& \multirow{2}{*}{--}
& \multirow{2}{*}{--} \\
\cline{3-4}
&
& CB($\mathcal{I}$)
& $0.098\pm0.02$
&
&
&
&
& \\
\hline

\multirow{4}{*}{Yorktown(0,2)}
& \multirow{2}{*}{Haar}
& Haar-RB($\mathcal{G}$)
& $6.96\pm0.98$
& \multirow{2}{*}{$1.35\pm1.23$}
& \multirow{2}{*}{$23.62$}
& \multirow{2}{*}{$0.06$}
& \multirow{2}{*}{$9.20$}
& \multirow{2}{*}{$0.19$} \\
\cline{3-4}
&
& Haar-RB($\mathcal{I}$)
& $5.68\pm0.63$
&
&
&
&
& \\
\cline{2-9}

& \multirow{2}{*}{Pauli}
& CB($\mathcal{G}$)
& $1.615\pm0.064$
& \multirow{2}{*}{$1.41\pm0.07$}
& \multirow{2}{*}{$2.95$}
& \multirow{2}{*}{$0.67$}
& \multirow{2}{*}{--}
& \multirow{2}{*}{--} \\
\cline{3-4}
&
& CB($\mathcal{I}$)
& $0.205\pm0.04$
&
&
&
&
& \\
\hline

\multirow{4}{*}{Yorktown(3,4)}
& \multirow{2}{*}{Haar}
& Haar-RB($\mathcal{G}$)
& $8.5\pm1.7$
& \multirow{2}{*}{$2.29\pm1.95$}
& \multirow{2}{*}{$27.37$}
& \multirow{2}{*}{$0.16$}
& \multirow{2}{*}{$7.36$}
& \multirow{2}{*}{$0.70$} \\
\cline{3-4}
&
& Haar-RB($\mathcal{I}$)
& $6.35\pm0.69$
&
&
&
&
& \\
\cline{2-9}

& \multirow{2}{*}{Pauli}
& CB($\mathcal{G}$)
& $2.377\pm0.092$
& \multirow{2}{*}{$2.07\pm0.10$}
& \multirow{2}{*}{$4.37$}
& \multirow{2}{*}{$0.97$}
& \multirow{2}{*}{--}
& \multirow{2}{*}{--} \\
\cline{3-4}
&
& CB($\mathcal{I}$)
& $0.313\pm0.043$
&
&
&
&
& \\
\hline

\end{tabular}
\caption{
Experimental data underlying Fig.~\ref{fig:expeiments}. For each device, rows are grouped by
randomization label. The columns ``XRB LB'' and ``XRB UB'' denote
the XRB bounds where available, these bounds are reported for Haar and Clifford randomization. 
}
\label{Experimental data table}
\end{table*}

\end{document}